\documentclass[12pt, a4paper]{article}
\usepackage{float}
\usepackage{amscd}
\usepackage{amssymb}
\usepackage{layout,xcolor, multirow,textcomp}
\usepackage{graphicx}
\usepackage{longtable}
\usepackage{amsfonts,amssymb,amsmath,amsthm,mathrsfs,
latexsym,amsxtra,graphicx, appendix,setspace,fix-cm, cite,lscape,
bm}
\usepackage{mciteplus}
\usepackage{caption}

\usepackage{tabularx}
\usepackage[version = 3]{mhchem}
\usepackage{subcaption}
\usepackage{array,multirow}

\usepackage{adjustbox}
\usepackage{hyperref}
\usepackage{arydshln}

\newcommand{\bmat}{\left(\begin{array}}
\newcommand{\emat}{\end{array}\right)}
\def\gtrsim{\mathrel{\raise.3ex\hbox{$>$\kern-.75em\lower1ex\hbox{$\sim$}}
}
}

\def\-{\hphantom{-}}

\def\s2{\frac{1}{\sqrt2}}

\def\beq{\begin{equation}}
\def\eeq{\end{equation}}
\def\beqa{\begin{eqnarray}}
\def\eeqa{\end{eqnarray}}

\def\mg{m_{3/2}}
\def\mg2{m^2_{3/2}}

\def\Dsl{\,\raise.15ex\hbox{/}\mkern-13.5mu D} 

\def\be{\begin{equation}}
\def\ee{\end{equation}}
\def\bea{\begin{eqnarray}}
\def\eea{\end{eqnarray}}

\DeclareMathOperator{\Spin}{\mathit{Spin}}

\newcommand{\Mj}{M_{\bar{J}}}
\newcommand{\Hkj}{\mathcal{H}_{\mathcal{K}\bar{J}}}
\newcommand{\Mkj}{M_{\hat{K}\bar{J}}}

\definecolor{mOrange}{RGB}{255,166,0}
\definecolor{mBittersweet}{RGB}{255,99,97}
\definecolor{mMulberry}{RGB}{188,80,144}
\definecolor{mViolet}{RGB}{88,80,141}
\definecolor{mIndigo}{RGB}{0,63,92}
\definecolor{mGold}{RGB}{204,133,0}

\makeatletter
\@addtoreset{equation}{section}
\makeatother

\begin{document}
\pagestyle{plain}
\begin{titlepage}
\begin{center}
 \LARGE{ Stringy Gauge Structures
\\[6mm]}

\large{\bf  Gerardo Aldazabal${}^{a,b,c}$, Eduardo C. Andr\'es$^{a,c}$, 
Luc\'ia M. Cabrera${}^{a,c}$, Martín Mayo${}^{a,b,c, \dagger}$
 \\}
 
\small{ ${}^a$ {\em G. F\'isica CAB-CNEA, }\\{\em Centro At\'omico 
Bariloche, Av. Bustillo 9500, Bariloche, 
Argentina.}\\ 
${}^b${\em Consejo Nacional de Investigaciones Científicas y Técnicas 
(CONICET)} \\
${}^c${\em Instituto Balseiro, Universidad Nacional de Cuyo (UNCUYO) } \\[-0.3em]
{\em  Av. Bustillo 9500, R8402AGP,  Bariloche, 
Argentina.}\\[-0.3em]}
{\scriptsize
$^\dagger$Present address:
ELTEMATE, Karl-Scharnagl-Ring 5, 80539 Munich, Germany.\par}
\end{center}
{\scriptsize{{{E-mail: {gerardo.aldazabal@ib.edu.ar}, 
{eduardo.andres@ib.edu.ar}, {lucia.cabrera@ib.edu.ar}, {martin.mayo@ib.edu.ar} 
}}}}
\vspace{0.5cm}

\noindent
 {\bf Abstract}:
 A  field theory action, able to incorporate stringy gauge symmetry enhancement-breaking effects, which occur at particular points of moduli space in a heterotic string toroidal compactification framework, was previously constructed from three-point interactions and duality symmetries. Via doubled periodic coordinates and a non-commutative product, it encodes the non-Abelian structure in a background-independent form. 
 
 Moduli-space dependence becomes manifest upon performing a generalized mode expansion. Here we further elaborate on this construction. In particular, we propose field transformations for the retained sectors that make up the foundations of a potential symmetry of the action and that, when mode expanded at enhancement points in moduli space, contain the usual non-Abelian gauge transformations. We also examine the covariance of the corresponding scalar derivative structures. Although the proposal must be regarded as a truncated version of a full, presently unknown field-theory description, it already encodes representations of massless and massive states containing up to one oscillator excitation. Finally, we relate the construction to gauged DFT with an extended tangent space by associating generalized Kaluza--Klein modes considered here with additional generalized-frame directions. At enhancement, their brackets reproduce the enhanced gauge algebra; away from enhancement, the same algebraic structure is recovered by rotating the Cartan basis.

\vspace{.5cm}

\end{titlepage}

\begin{small}
\tableofcontents
\end{small}

\newpage\section{Introduction} \label{sec: Introduction}

So-called ``stringy features'' refer to phenomena associated with the fact that a string is an extended object. Therefore, these are not expected to manifest in a field theory of point particles. A neat example is the phenomenon of enhancement of the gauge symmetry that occurs in string compactifications at some points in moduli space (namely for certain values of background fields). At such points, compact momenta combine with stringy winding modes such that some vector bosons become massless. Another closely related example is T-duality, which connects the physics of strings on very different geometrical backgrounds.
Despite the crucial differences between point-like and extended objects, different proposals were advanced in order to incorporate these stringy effects into a field theory -- see for instance the early works \cite{Siegel:1993th,Hull:2009mi}.  
A key strategy, inspired in toroidal string compactifications, is to incorporate for each winding mode $\tilde p$ a dual conjugate coordinate $\tilde y$ on top of the internal coordinate $y$, conjugate to Kaluza-Klein compact momentum  $p$. This doubling of coordinates led to the name of Double Field Theory (DFT), see  \cite{Aldazabal:2013sca, Hohm:2013bwa} for some reviews, and \cite{Hohm:2011a} for the heterotic formulation of DFT.

In Ref.~\cite{Aldazabal:2018uzm}, by incorporating ideas from DFT, a field theory action able to describe stringy gauge symmetry enhancement-breaking effects, of  a heterotic string toroidal compactification in $d$-spacetime dimensions, was presented.

The construction is based on the computation of three-point interactions and duality symmetries.  
When the fields are written in terms of double periodic coordinates, as suggested by DFT formulations, the action has a background-independent form and non-Abelian structures appear  to be encoded in a non-commutative product. 
It is when a generalized Kaluza-Klein (GKK) mode expansion~\cite{Hohm:2013b} -- namely, in terms of KK momenta and windings -- is performed,  that a dependence on  moduli space manifests. 

 The formulation at hand is restricted to fields containing up to $N=1$  string
oscillator number and includes  the fields, massless and massive,  that participate  in the enhancement phenomena. 
Clearly a full, consistent description  would require the introduction  of an infinite number of fields of all possible oscillator numbers and spins. 
Nevertheless, even if  this action should be  taken as a truncated version of a full (unknown) field theory description,  it already encodes interesting stringy phenomena. 
When moving from one moduli point to another, the gauge group changes and the fields must reorganize into representations of the gauge group at this specific new point. In this process some vector fields, labeled  by definite  values of windings and compact momenta,  become massive whereas others, generically with different values of windings and KK momenta,  can become massless.  Interestingly enough, the proposed action encodes this information for representations containing up to one oscillator mode in string theory. Even if restricted to the low-oscillator number sectors, the formulation contains a generically infinite tower of GKK modes, which makes it reminiscent of string field theory, in which a string field packages an infinite tower of spacetime component fields \cite{Hull:2009mi,Kugo:1992md}.

In the present work  we further elaborate on this construction, with an overarching interest in identifying the background-independent field transformations that may -- before any mode expansion -- keep the action invariant. When mode expanded at a specific moduli point, they must  contain the usual gauge transformations. This  points towards identifying, at least part of, the full underlying symmetries of the theory that we observe as symmetries of a given vacuum when choosing a given point in moduli space.

Notice that this construction differs from a conventional low-energy effective theory formulated at a fixed point in moduli space. Such a theory integrates out the states that are massive at that point and describes the remaining light fields through an ordinary gauge-theory Lagrangian coupled to gravity, with no explicit reference to winding modes~\cite{Hohm:2015}.

Interestingly enough a connection with gauged-DFT theories  with extended tangent space -- see \cite{Aldazabal:2016,Aldazabal:2017,Aldazabal:2017b,Cagnacci:2017} -- can be established by identifying generalized Kaluza-Klein (GKK) modes with the extended directions.

We organize the article as follows: In Section \ref{sec: Senh}, we describe the main ingredients of  the action proposed in \cite{Aldazabal:2018uzm}. In Section \ref{sec: Symmetries} we study field transformations that, up to truncations, result in covariance of the derivative structures proposed in \cite{Aldazabal:2018uzm}. In particular we show that when these are mode expanded, and by keeping massless or slightly massive states around an enhancement point, they reduce to the usual gauge transformations. Section \ref{subsec: Examples} deals with specific examples. In Section  \ref{sec: hidden_algebra_GKK} we discuss the connections with some DFT proposals. A summary and a discussion of the limitations and possible extensions of the present work are presented in Section \ref{sec: Remarks_and_Outlook}. Technical aspects are reserved to the Appendices, where  a more detailed description of the $\star$-product is extended to incorporate the heterotic string gauge modes.

\section{The interpolating action}
\label{sec: Senh}

 In this section we set some notation and introduce  the basic ingredients of the proposed action, based on \cite{Aldazabal:2018uzm}.
 We are interested in heterotic string theory with $r$ dimensions compactified on a torus. 
 We encode the internal KK momenta $p^m$, winding modes $\tilde p_m $, with $m=1,\dots r$, and the 16 Left-moving heterotic momenta $P^I$ with $I=1,\dots 16$, into a generalized KK momentum vector $\mathbb P^{B}\equiv (P^I, p_m,\tilde p^m )$. The fields considered will depend on both $d$ space-time $x^{\mu}$ coordinates as well as on internal compact toroidal   $\mathbb Y^{B}\equiv (y^I, y^m,\tilde y_m )$ coordinates. Namely, besides the $y^I$ coordinates associated to the heterotic string degrees of freedom, we introduce $2r$ double coordinates $(y^m,\tilde y_m)$, conjugate to internal momenta and winding modes, respectively. 

A complete heterotic action should contain all the infinitely many fields arising from the string spectrum. We denote these generic fields by  $\Psi_{\cal N}(x,\mathbb Y)$, with a dependence  on both space-time   as well as  internal coordinates. The subindex ${\cal N}\equiv (\mu_1,\mu_2,\dots;{\cal I},{\cal J})$ encodes space-time indices (Greek characters) as well as internal indices (calligraphic  capital Latin characters) \footnote{We will be more specific below.} associated to oscillator numbers. Therefore, a $d$-dimensional action will, schematically, look like
\begin{equation}
 S_{\Psi}=\int d^dx {\cal L}_ {\psi}(x)=\int d^dx\int d\mathbb Y{\cal L}\left(\Psi_{\cal N}(x,\mathbb
Y),\partial_ {\mu}\Psi_{\cal N}(x,\mathbb
Y)\right)
\label{eq: accion_psi}
\end{equation}
with ${\cal L}_ {\psi}(x)=\int d\mathbb Y{\cal L}\left(\Psi_{\cal N}(x,\mathbb
Y),\partial_ {\mu}\Psi_{\cal N}(x,\mathbb
Y)\right)$.
The explicit form of the kinetic and interaction terms should be readable from string scattering amplitudes of string vertex operators associated to the fields $\Psi_{\cal N}$. Let us emphasize that, at this stage, the Lagrangian does not show an explicit dependence on string moduli. It is when the fields are mode-expanded that moduli become apparent.

Let us  denote a moduli point  by  $\Phi\equiv(g,b,A) $, encoding the background metric $g$, the $b$ field and Wilson line values $A$. String states are characterized by Left and Right momenta, dependent on moduli as well as winding and momentum modes -- see \eqref{eq: leftrightmomenta} in Appendix \ref{ap: HeteroticStringBasics} -- encoded into a vector on a Lorentzian self-dual Narain lattice

\begin{equation}
\label{eq: Kgral}
    \mathbb K(\mathbb P;\Phi)
    =
    \bigl(\kappa_L(\mathbb P;\Phi),k_R(\mathbb P;\Phi)\bigr),
\end{equation}
where the full left-moving momentum is
\begin{equation}
    \kappa_L^{\hat I}(\mathbb P;\Phi)
    =
    \bigl(K_{L}^I(\mathbb P;\Phi),k_L^m(\mathbb P;\Phi)\bigr),
    \qquad
    \hat I=(I,m)\, .
\end{equation}
Here $K_{L}^I$ denotes the heterotic gauge-lattice part written in the chosen simple root basis of $\mathfrak{so}(32)$, while $k_L^m$ and $k_R^{m}$ denote the compact left- and right-moving components. We include a more detailed account of our index choices below.

A  GKK expansion\footnote{In what follows, to simplify the notation, we will mostly omit the dependence of $\mathbb K$ on $\mathbb P$ and $\Phi$. } reads
\begin{equation}
 \Psi_{\cal N}(x, {\mathbb Y})
=\sum_{\mathbb
K}
\Psi_{\cal N}^{( {\mathbb K})}(x) e^{i \mathbb{K}_{\cal I} \mathbb{Y}^{\cal I}}\,
=\sum_{{\mathbb K}}
\Psi_{\cal N}^{({\mathbb K})}(x) e^{i( \kappa_L.y_L+ k_R.y_R)}\,
\label{eq: psiexp}
\end{equation}
The sum is over Narain's lattice vectors satisfying
\begin{equation}
 \frac12{\mathbb K}^2=\frac12 \kappa_L^2- \frac12 k_R^2=\tilde
p.p+\frac12P^2=(1-N+\bar N)\, .
\label{eq: LMCwidingstext}
\end{equation}
 where $  N=N_B$ is the bosonic L-oscillator number, and $\bar N=\bar N_B+\bar N_F+\bar E_0$, with $\bar N_B$ the bosonic R-oscillator number,  $\bar N_F$ is the R fermion
oscillator number and $\bar E_0=-\frac12(0) $ for the NS (R) sector.

In what follows, besides the space-time indices,  we will use  internal left ${\hat I}$ and right $\bar I$ indices encoded in a double index   ${\cal I}= (\hat I,\bar I)$. These are associated to the presence of oscillators in vertex operators. The left index splits as
\begin{equation}
 \hat I=(I,m),
    \qquad
    I=1,\ldots,16,
    \qquad
    m=1,\ldots,r\, ,   
\end{equation}
where $I$ labels the heterotic gauge-lattice directions and $m$ labels the compact left-moving torus directions. The right-moving index $\bar I=1,\ldots,r$ only labels the compact right-moving torus directions. Equivalently, the heterotic gauge-lattice momenta $P^I$ may be viewed as arising from sixteen chiral compact bosons, or formally from an auxiliary sixteen-dimensional torus with the right-moving momenta projected out by a chiral constraint. Thus $P^I$ contributes to the left-moving lattice data, but it does not introduce additional right-moving oscillator indices. For the zero GKK label, $\mathbb P=0$, these oscillator indices label the Cartan-sector gauge bosons: $\hat I=(I,m)$ labels the $16+r$ left Cartan directions, while $\bar I$ labels the $r$ right Abelian directions.


As previously mentioned, the fields $\Psi_{\cal N}^{( {\mathbb K})}(x)$ are inherited from string vertex operators. We will restrict our analysis to  fields with a small number of oscillators, namely $\bar N=0$ and $N=0,1$. These lead to the set of fields  quoted in Table \ref{tab: tablavertices}, where
\begin{equation}
 e^{i\mathbb K(\mathbb P;\Phi)\cdot\mathbb Y(z)} = e^{i\kappa_L(\mathbb P;\Phi)\cdot y_L(z) +i k_R(\mathbb P;\Phi)\cdot y_R(z)}\, ,
\end{equation}
and $\mathbb{K}$ satisfies
\begin{equation}
 \mathbb K^2 =\kappa_L^2-k_R^2 = 0,2 \qquad \text{for} \qquad \bar N=0,\quad N=1,0 \, .
 \label{eq: rlmc}
\end{equation}

\begin{table}[htbp]
\centering
\renewcommand{\arraystretch}{1.45}
\begin{adjustbox}{max width=\textwidth}
\begin{tabular}{
|>{\centering\arraybackslash}m{0.20\textwidth}
|>{\centering\arraybackslash}m{0.07\textwidth}
|>{\centering\arraybackslash}m{0.05\textwidth}
|>{\centering\arraybackslash}m{0.33\textwidth}|}
\hline
\textbf{Field modes} & $\boldsymbol{\mathbb K^2}$ & $\boldsymbol{N}$ & \textbf{Vertex operators} \\
\hline

$g_{\mu\nu}^{(\mathbb K)},\, b_{\mu\nu}^{(\mathbb K)},\, \phi^{(\mathbb K)}$
& $0$ & $1$
&
$\partial_z X^\mu \,\tilde\psi^\nu(z)\,
e^{i\mathbb K\cdot\mathbb Y(z)}e^{iq\cdot X(z)}$
\\
\hline

$A_{\mu}^{\bar I(\mathbb K)}$
& $0$ & $1$
&
$\partial_z X^\mu\,\tilde\psi^{\bar I}(z)\,
e^{i\mathbb K\cdot\mathbb Y(z)}e^{iq\cdot X(z)}$
\\
\hline

\multirow{2}{*}{$A_{\mu}^{\hat I(\mathbb K)}$}
& \multirow{2}{*}{$0$}
& \multirow{2}{*}{$1$}
&
$\partial_z Y^I\,\tilde\psi^\mu(z)\,
e^{i\mathbb K\cdot\mathbb Y(z)}e^{iq\cdot X(z)}$
\\
\cline{4-4}
&
&
&
$\partial_z Y^m\,\tilde\psi^\mu(z)\,
e^{i\mathbb K\cdot\mathbb Y(z)}e^{iq\cdot X(z)}$
\\
\hline

\multirow{2}{*}{$M_{\hat I\bar I}^{(\mathbb K)}$}
& \multirow{2}{*}{$0$}
& \multirow{2}{*}{$1$}
&
$\partial_z Y^I\,\tilde\psi^{\bar I}(z)\,
e^{i\mathbb K\cdot\mathbb Y(z)}e^{iq\cdot X(z)}$
\\
\cline{4-4}
&
&
&
$\partial_z Y^m\,\tilde\psi^{\bar I}(z)\,
e^{i\mathbb K\cdot\mathbb Y(z)}e^{iq\cdot X(z)}$
\\
\hline

$A_{\mu}^{(\mathbb K)}$
& $2$ & $0$
&
$\tilde\psi^\mu(z)\,
e^{i\mathbb K\cdot\mathbb Y(z)}e^{iq\cdot X(z)}$
\\
\hline

$M_{\bar I}^{(\mathbb K)}$
& $2$ & $0$
&
$\tilde\psi^{\bar I}(z)\,
e^{i\mathbb K\cdot\mathbb Y(z)}e^{iq\cdot X(z)}$
\\
\hline

\end{tabular}
\end{adjustbox}
\caption{\footnotesize Field modes, level-matching condition, left oscillator number and corresponding string vertex operators. In all cases considered here $\bar N_F=\frac12$, $\bar N_B=0$, and therefore $\bar N=\bar N_F+\bar N_B-\frac12=0$. The factor $e^{iq\cdot X(z)}$ is the ordinary spacetime plane wave, with $q_\mu$ the momentum along the non-compact spacetime directions.}
\label{tab: tablavertices}
\end{table}
Recall that the fields with $N=1$ originate from Kaluza-Klein reductions of ten dimensional massless fields whereas those with no left oscillators originate from charged fields associated with modes carrying nonzero quantum numbers in the compact momentum-winding lattice, the heterotic gauge lattice, or both. We will refer to the former as $N=1$ fields (called `Cartan' in \cite{Aldazabal:2018uzm}) and as $N=0$ fields to the latter (referred to as `charged' in \cite{Aldazabal:2018uzm}).

For instance, let us consider  the left vector  $A^{\hat{I}(\mathbb{K})}_{\mu}(q)$, $\mathbb{K}^2=0$, in the list above. This field will be generically massive. However, for $\mathbb P=0$, i.e. for vanishing momenta and windings, we have $\mathbb{K}\equiv \kappa_L=k_R=0$, and the state becomes massless, independently of moduli values, and  corresponds to a Left $N=1$ gauge vector $A_{\mu}^{\hat I}\equiv
A^{\hat{I}(0)}_{\mu} $. Notice that  this massless vector is included in the mode expansion of $ A_{\mu}^{\hat{I}}(x,\mathbb{Y})$ -- see \eqref{eq: psiexp}.
Similarly, we will have Right vector fields $A_{\mu}^{\bar I}\equiv A^{\bar I(0)}_{\mu} $ corresponding to the graviphotons.

On the other hand,  the vector field $A_{\mu}^{\mathbb{(K)}}$ with $\frac12{\mathbb{K}^2}=1$ will be generically massive. However, at an enhancement point $\Phi_0$, and for specific values of windings and momenta (i.e., for specific values of  $\mathbb P$), we could have
\begin{equation}
\label{eq: masslescondfp}
    k_R(\mathbb P;\Phi_0)=0,
    \qquad
    \kappa_L(\mathbb P;\Phi_0)=\alpha^{(\mathbb P)},
    \qquad
    \frac{1}{2}\left(\alpha^{(\mathbb P)}\right)^2=1\, .
\end{equation}

Namely, at the enhancement point $\Phi_0$, the full left-moving momentum
$\kappa_L(\mathbb P;\Phi_0)$ is identified with a root $\alpha^{(\mathbb P)}$ of the enhanced algebra, while the right-moving momentum vanishes. The corresponding vector modes therefore become massless, see \eqref{eq: LRstringmasses}, and are identified with the charged generators of the enhanced gauge algebra.\footnote{At the level of string vertex operators, the internal exponential associated with such a mode becomes the affine current corresponding to the root, $ e^{i\kappa_L(\mathbb P;\Phi_0)\cdot Y_L(z)} = e^{i\alpha^{(\mathbb P)}\cdot Y_L(z)} \longrightarrow J_{\alpha^{(\mathbb P)}}(z)$.
}

At a different enhancement point\footnote{The explicit enhancement points considered are also fixed points of appropriate discrete T-duality transformations. We nevertheless use the term enhancement point when referring to the appearance of additional massless states, reserving T-duality fixed point for statements concerning the duality action.}, a different set of $\mathbb{P}$s will ensure \eqref{eq: masslescondfp}, leading to a different enhanced gauge group. We will denote this set of GKK modes
by
\begin{equation}
\label{eq: enhancementsector}
  G(\Phi_0)
  =
  \left\{
  \mathbb P\equiv (P^I,p_m,\tilde p^{\,m})
  \;:\;
  k_R(\mathbb P;\Phi_0)=0,
  \quad
  \kappa_L(\mathbb P;\Phi_0)=\alpha^{(\mathbb P)},
  \quad
  m^2=0
  \right\}.
\end{equation}

Namely, $G(\Phi_0)$ is the set of GKK labels of the modes that become
massless at the enhancement point $\Phi_0$. We denote by $n_c$ the number of positive roots of the enhanced algebra. The complete set therefore has cardinality
$\left|G(\Phi_0)\right|=2n_c$. For each
$\mathbb P\in G(\Phi_0)$, the corresponding internal momentum satisfies
$\kappa_L(\mathbb P;\Phi_0)=\alpha^{(\mathbb P)}$, with
$\alpha^{(\mathbb P)}$ a root of the enhanced algebra. Thus, at $\Phi_0$, the
vector mode labelled by $\mathbb P$ is identified with the charged left gauge vector
field $A_\mu^{(\alpha^{(\mathbb P)})}(x)$. Together with the massless Left $N=1$ fields, they give rise to an enhanced gauge group $G_{\Phi_0}$ of rank $16+r$ and dimension $16+r+2n_c$. By including the $r$ Right $ A_{\mu}^{\bar I}$ gauge vector fields we obtain the full $G_{\Phi_0}\times U(1)^r_R$ gauge group at enhancement point $\Phi_0$. Similar  considerations are valid for the scalar fields. 
At $\Phi_0$, and for $\mathbb{P}\in  G(\Phi_0)$, the charged fields $M^{(\mathbb{K})}_{\bar J}$ and $M_{\hat{I}\bar J}^{(0)}$ organize into the adjoint representation of the enhanced gauge group. Massless fields include also the metric, Kalb-Ramond field and dilaton field, obtained from the first row of Table \ref{tab: tablavertices} for $\mathbb{K}=0$. The pattern of enhanced gauge groups over the heterotic Narain moduli space,
including the role of Wilson lines, has been analyzed in detail in the literature, see for instance \cite{Narain:1986,Narain:1987,Mohaupt:1992jf,Giveon:1994,
Fraiman:2018,Font:2020rsk}.

Let us stress that, even if we are considering a truncation of the whole theory, with a restricted set of fields, our construction contains  information about massive states:  the fields in Table \ref{tab: tablavertices} with $\mathbb{P} \notin G(\Phi_0)$ have nonzero masses, and organize into massive representations of the gauge group at $\Phi_0$. For high masses, these massive representations will contain states with several internal oscillators that we are not including, as accounted for in restriction \eqref{eq: rlmc}. However,  representations containing up to $N=1$ do fall within our description. Some examples were presented in \cite{Aldazabal:2018uzm} and we provide a more general discussion below.

Summarizing, from the full tower of string fields collectively denoted by $\Psi$, our construction retains the zero-GKK massless fields descending from ten-dimensional massless fields, together with the momentum-carrying modes that participate in symmetry enhancement. In the sector considered here, the former belong to the $N=1$, $\mathbb K^2=0$ sector, while the latter belong to the $N=0$, $\mathbb K^2=2$ sector and become massless at special points in moduli space.

Therefore, we split the action \eqref{eq: accion_psi}  as
\begin{equation}
 S({\Psi})=S_{enh}({\Psi})+S'(\Psi)
\label{eq: Senhs'}
\end{equation}
where $S_{enh}({\Psi})$ contains the fields $\Psi(x,\mathbb{Y})= A_{\mu}^{\cal {I}},A_{\mu},M_{\hat{I}\bar J},\dots$ that participate in the enhancement process\footnote{We omit the writing of the dependence on the space-time and internal coordinates $(x,\mathbb{Y})$ in order to lighten the notation.},  and $S'$ holds all of the rest. 
The splitting in \eqref{eq: Senhs'} should be distinguished from the low-energy split obtained after choosing a particular enhancement point. Given $\Phi_0$, $S_{enh}$ contains both the zero-GKK massless fields and the modes in $G(\Phi_0)$ that become
massless at $\Phi_0$, together with modes that remain heavy at that point. Thus, near
enhancement points $\Phi_i$, one may further decompose $S_{enh}$ as
\begin{equation}
\label{eq: localLightHeavySplit}
    S_{enh}(\Phi)
    =
    S_{\text{light at }\Phi_0}(\Phi)
    +
    S_{\text{heavy at }\Phi_0}(\Phi) = S_{\text{light at }\Phi_1}(\Phi)
    +
    S_{\text{heavy at }\Phi_1}(\Phi)= \cdots
\end{equation}
At low energies and for $\Phi\simeq\Phi_i$, the heavy part can be neglected, leaving
\begin{equation}
\label{eq: localEffectiveAction}
    S_{\text{eff}}(\Phi\simeq\Phi_i)
    =
    S_{\text{light at }\Phi_i}(\Phi\simeq\Phi_i)\, .
\end{equation}
At the enhancement point itself, the modes in $G(\Phi_i)$ become massless and
combine with the Cartan-sector zero modes into the enhanced gauge theory with gauge
group $G_{\Phi_i}\times U(1)^r_R$. When moving slightly away from $\Phi_i$, these
same modes generically acquire masses and the enhanced, non-abelian symmetry is either partially broken, or broken back to the
generic Abelian gauge group $U(1)^{r+16}_L\times U(1)^r_R$ . Thus $S_{\text{eff}}(\Phi\simeq\Phi_i)$ describes not only the enhanced theory at $\Phi_i$, but also its nearby broken phase, containing massless Abelian fields together with the slightly massive states that transform under the Abelian group. Around a different enhancement point $\Phi_j$, a different set of GKK labels becomes light and the corresponding light/heavy split is generally different.

The explicit form of  $S_{enh}({\Psi})$ is
\begin{equation} \label{eq: hetActionUplift}
\begin{aligned}
S_{enh}(\Psi)=&\int d^dxd\mathbb{Y}\sqrt{g} e^{-2\varphi}\Bigg[
(
R+4\partial^\mu\varphi\partial_\mu\varphi-\frac1{12}H_{\mu\nu\rho}H^{\mu\nu\rho
}) \\
& -\frac{1}{4}
\mathcal{H}_{\mathcal{I}\mathcal{J}}\star F^{\mathcal{I}}_{\mu\nu}\star 
F^{\mathcal{J} \mu\nu}  + 
\frac{1}{8}\mathcal{D}_{\mu}\mathcal{H}^{\mathcal{I}\mathcal{J}}\star\mathcal{D}
^{\mu}\mathcal{H}_{\mathcal{I}\mathcal{J}}\\
&-\frac{1}{4} F_{\mu\nu}\star F^{\mu\nu}+\frac{1}{4} 
\mathcal{D}_{\mu}M^{\mathcal{I}}\star\mathcal{D}^{\mu}M_{\mathcal{I}}-\frac{1}{2
} M_{\mathcal{I}} \star F^{\mu\nu}\star F^{\mathcal{I}}_{\mu\nu}  \\
& - 
\frac{1}{4} 
\partial_{\mathcal{J}}M^{\mathcal{I}}\star\partial_{\mathcal{K}}M_{\mathcal{I}}
\big(H^{\mathcal{J}\mathcal{K}} -\eta^{\mathcal{J}\mathcal{K}})+ i \frac{1}{2} 
\partial_{\mathcal{I}}M^{\mathcal{J}}\star 
M_{\mathcal{J}}\star M^{\mathcal{I}}
\Bigg]\, .
\end{aligned} 
\end{equation}
The  different terms in the action are defined as follows
\footnote{Relative to Eqs.~(2.31)--(2.32) of
\cite{Aldazabal:2018uzm}, $D_\mu M_{\mathcal J}$ is unchanged. The vector field strengths acquire additional terms with internal
derivatives. In $F_{\mu\nu}^{\mathcal I}$, the coefficient of
$A_{[\mu}\star A_{\nu]}^{\mathcal I}$ is halved and the sign of
$\partial^{\mathcal I}A_\mu\star A_\nu$ is reversed.
In $D_\mu\mathcal H_{\mathcal{KJ}}$, the coefficient of
$A_\mu\star\mathcal H_{\mathcal{KJ}}$ is halved,
a double-derivative term is added, and the transport term is
replaced by the operator $\mathcal L$ defined in
\eqref{eq: op_LieDerivative}. These changes allow for uniform treatment of arbitrary $\mathrm{SU(n)}$ enhancements.}. For the vector field-strengths,
\begin{equation} \label{eq: FmunuChargedUplift}
F_{\mu\nu} = 2 \partial_{[\mu}A_{\nu]} + g [[ A_{\mu}, A_{\nu} ]]_{\star}\, ,
\end{equation}
\begin{equation} \label{eq: FmunuCartanUplift}
F_{\mu\nu}^{\mathcal{I}} =  2\partial_{[\mu}A_{\nu]}^{\cal I} + g[[A_{\mu}, A_{\nu}]]_{\star}^{\mathcal{I}}\,
\end{equation}
where the star-brackets\footnote{Notice that $[[V, W]]_{\star}^{\mathcal{I}}$ is similar in structure to a C-bracket. However, our definition makes use of the $\star$-product, and contains an asymmetric term. } above are defined as 
\begin{equation}\label{eq: starBracket_Cargado}
[[ V_{\mu}, V_{\nu} ]]_{\star} = i V_{\mu}\star V_{\nu} + 2 V^{\cal K}_{[\mu}\star \partial_{\cal K}V_{\nu]} + 2 \partial_{\cal K}V_{[\mu}\star V^{\cal K}_{\nu]} \, ,
\end{equation}
\begin{equation}\label{eq: starBracket_Cartan}
    [[V, W]]_{\star}^{\mathcal{I}} = F(V)_{\mathcal{Q}}^{\mathcal{I}}W^{\mathcal{Q}} - F(W)_{\mathcal{Q}}^{\mathcal{I}}V^{\mathcal{Q}} - \partial^{\mathcal{I}} V \star W \, , 
\end{equation}
and the operator $F(V)^{\mathcal{I}}_{\mathcal{Q}}$ in \eqref{eq: starBracket_Cartan} reads
\begin{equation}
    F(V)^{\mathcal{I}}_{\mathcal{Q}}W^{\mathcal{Q}} =\left(V^{\mathcal{K}} \star \partial_{\mathcal{K}}\delta^{\mathcal{I}}_{\mathcal{Q}} + i V \star \delta^{I}_{\mathcal{Q}} + i \partial^{\mathcal{I}}\partial_{\mathcal{Q}}V \star  \right)W^{\mathcal{Q}} \, .
\end{equation}

The derivatives of scalar fields are
\begin{equation}\label{eq: DmuMjUplift}
    \mathcal{D}_{\mu}M_{\mathcal{J}} = \partial_{\mu}M_{\mathcal{J}} + gF(A_{\mu})M_{\mathcal{J}} - g \partial^{\mathcal{K}}A_{\mu}\star \left(\mathcal{H}_{\mathcal{KJ}} - \eta_{\mathcal{KJ}} \right)
\end{equation}
\begin{equation}\label{eq: DmuHkjUplift}
    D_{\mu}\mathcal{H}_{\mathcal{KJ}} = \partial_{\mu}\mathcal{H}_{\mathcal{KJ}} + g F(A_{\mu})^{\mathcal{QP}}_{\mathcal{KJ}}\mathcal{H}_{\mathcal{QP} } + g \partial_{\mathcal{K}}A_{\mu}\star M_{\mathcal{J}} 
\end{equation}
where we have defined operators
\begin{equation} \label{eq: opCargados}
    F(V) = iV \star \, + V_{\mathcal{K}  } \star\partial^{\mathcal{K}} \, ,
\end{equation}
\begin{equation}\label{eq: opCartanes}
    F(V)^{\mathcal{QP}}_{\mathcal{K}\mathcal{J}} = \mathcal{L}(V)_{\mathcal{K}\mathcal{J}}^{\mathcal{QP}}+ iV \star \delta^{\mathcal{Q}}_{\mathcal{K}}\delta^{\mathcal{P}}_{\mathcal{J}} + i \partial^{\mathcal{Q}}\partial_{\mathcal{K}}V \star \delta^{\mathcal{P}}_{\mathcal{J}} \, .
\end{equation}
In \eqref{eq: opCartanes} we have also included the operator associated to the Lie derivative,
\begin{equation}\label{eq: op_LieDerivative}
    \mathcal{L}(\lambda)_{\mathcal{K}\mathcal{J}}^{\mathcal{QP}} = \lambda^{\mathcal{R}} \star \partial_{\mathcal{R}}\,\,\delta^{\mathcal{Q}}_{\mathcal{K}}\delta^{\mathcal{P}}_{\mathcal{J}} + \partial_{(\mathcal{J}} \lambda^{\mathcal{P}} \star \delta^{\mathcal{Q}}_{\mathcal{K})}\, - \partial^{\mathcal{P}} \lambda_{(\mathcal{J}} \star \delta^{\mathcal{Q}}_{\mathcal{K})}  \, .
\end{equation}

Inspired by DFT constructions~\cite{Aldazabal:2018uzm}, \eqref{eq: hetActionUplift} includes an $O(r_l,r)$ metric ${\cal H}_{}^{\cal I \cal J}$. When expanded on fluctuations
around a flat background it reads
\begin{equation}
 {\cal H}_{}^{\cal I \cal J}= \delta^{\cal I \cal J}+ {\cal H}^{(1)\cal I \cal
J}+ \frac12{\cal
H}^{(2)\cal I \cal J}+\dots
\label{scalarfluct}
\end{equation}
where matrix elements vanish  unless
\begin{equation}
\label{eq: HkjFluctuations}
\begin{aligned}
{\cal H}^{(1)}_{\hat I\bar J}
&= -M_{\hat I\bar J},
&\qquad
{\cal H}^{(1)}_{\bar J\hat I}
&= -M^T_{\hat I\bar J},
\\
{\cal H}^{(2)}_{\hat I\hat J}
&= (MM^T)_{\hat I\hat J},
&\qquad
{\cal H}^{(2)}_{\bar I\bar J}
&= (M^T M)_{\bar I\bar J}\, .
\end{aligned}
\end{equation}

Finally, the three-form $H$ is defined as
\begin{equation}
H=dB + F^{\mathcal{I}}\star\wedge A_{\mathcal{I}} + F\star\wedge A\, .
\end{equation}

It is worth emphasizing that expressions \eqref{eq: FmunuChargedUplift}--\eqref{eq: HkjFluctuations} are written in the unexpanded doubled-coordinate formulation. They carry no explicit dependence on the moduli $\Phi$; the moduli enter only after
GKK expansion, through the map $\mathbb P\mapsto \mathbb K(\mathbb P;\Phi)$. For further details on this mapping, see Appendix \ref{ap: HeteroticStringBasics}. The general details of the action are discussed in \cite{Aldazabal:2018uzm}. Several considerations are in order:

\begin{itemize}\item
The field $\star$-product is generically non-commutative and associative, and some of its properties are discussed both in the next section, as well as in more detail on Appendix \ref{ap: star_product}. 
If we denote by $\phi_{N_\phi}$ a field originated from a vertex with $N_\phi$ oscillators, then it can be shown that when the product $\phi_{N_\phi}\star \psi_{N_\psi}$ is mode expanded, and a single mode $\mathbb K$ is isolated,
\begin{equation}
(\phi_{N_\phi}\star\psi_{N_\psi})^{(\mathbb{K})}=e^{i\pi 
(\frac12\mathbb{K}^2+N_\phi+N_\psi)}(\psi_{N_\psi}\star\phi_{N_\phi})^{(\mathbb{K})} \, .
\label{eq: noncommutative12}
\end{equation}

For a fixed output mode $\mathbb{K}$, with $\mathbb K^2=0$, the $\star$-product is commutative when $N_\phi=N_\psi$ and anti-commutative when $N_\phi\ne N_\psi$. For $\mathbb K^2=2$, these properties are reversed, i.e. the product is commutative (anti-commutative) for $N_\phi\neq N_\psi$ ($N_\phi= N_\psi$). In practice we see, for instance, that $ A_{\mu}\star A_{\nu}=-A_{\nu}\star A_{\mu}$, and $A^{\mathcal{I}}_{\mu}\star\partial_{\mathcal{I}}A_{\nu}=\partial_{\mathcal{I}}A_{\nu} \star A^{\mathcal{I}}_{\mu}$ in \eqref{eq: FmunuChargedUplift} for $F_{\mu\nu}$. The role of this product is instrumental in encoding the non-commutativity expected in non-Abelian gauge theories, and will supply the appropriate factors involved in derivative as well as transformation structures.
 
\item
When fields are mode expanded according to \eqref{eq: psiexp}, terms up to cubic order correctly reproduce the tree level cubic string scattering amplitudes. On the neighborhood of each specific enhancement moduli point, and when only the slightly  massive modes that become massless at this point are kept, then the usual,  commutative, effective gauge theory action is recovered  after integrating over the internal coordinates. The gauge symmetry gets enhanced exactly at the enhancement point.
\item
The usual heterotic low energy effective action, in terms of the ten dimensional massless metric, Kalb-Ramond and dilaton fields  is recovered if no compact dimensions are considered.
\end{itemize}
\subsection{Mode expansions}
Here we present some examples of the GKK mode expansions and demonstrate the role of the $\star$-product.  When considering the full GKK mode expansion of the two-field product $\phi_{N_\phi}\star \psi_{N_\psi}$, from which we isolated a single mode in \eqref{eq: noncommutative12}, we have
\begin{equation}
\label{eq: star2fieldsModes}
\begin{aligned}
(\phi_{N_\phi}\star\psi_{N_\psi})(x,\mathbb Y)
&=
\sum_{\mathbb K_1}
(\phi_{N_\phi}\star\psi_{N_\psi})^{(\mathbb K_1)}(x)\,
e^{i\mathbb K_1\cdot\mathbb Y}
\\
&=
\sum_{\mathbb K_2,\mathbb K_3}^{\prime}
e^{i\pi p_{2}\cdot \tilde{p}_{3}}
\phi_{N_\phi}^{(\mathbb K_2)}(x)
\psi_{N_\psi}^{(\mathbb K_3)}(x)
e^{i(\mathbb K_2 + \mathbb K_3)\cdot\mathbb Y}\\
&=
\sum_{\mathbb K_2,\mathbb K_3}^{\prime}
e^{i\pi (1-N_{1})}\tilde{f}_{\mathbb{K}_{1}\mathbb{K}_{2}\mathbb{K}_{3}}
\phi_{N_\phi}^{(\mathbb K_2)}(x)
\psi_{N_\psi}^{(\mathbb K_3)}(x)
e^{i(\mathbb K_2 + \mathbb K_3)\cdot\mathbb Y}\, ,
\end{aligned}
\end{equation}
where in the second and third lines, $\mathbb K_1=\mathbb K_2+\mathbb K_3$ and $N_1\equiv 1-\tfrac12\mathbb K_1^2$. The primes on the sums, omitted in subsequent mode expansions, restrict the input momenta to
$\tfrac12\mathbb K_2^2=1-N_\phi$ and
$\tfrac12\mathbb K_3^2=1-N_\psi$. A phase ${p}_2\cdot \tilde p _{3}={p}_{2 m}\tilde p^{3m}+{p}_{2I} \tilde p_{3}^I$, dependent on the KK momenta $p_{2}$ associated to the first field, carrying momentum  $\mathbb K_{2}(\mathbb P_2;\Phi)$, and the winding numbers $\tilde p_{3}$ of second mode $\mathbb K_{3}(\mathbb P_3;\Phi)$ is introduced\footnote{The first term corresponds to a sum over the internal  compactification lattice indices. The sum over heterotic directions is  constrained by a chiral projection that eliminates Right heterotic momenta. It  can be expressed as 
\begin{equation}\label{eq: so32phases}
 p _{1I}\tilde{p}_2^I= \frac12 P_1EP_2
\end{equation}
in terms of  $\Spin(32)$ or  $E_8\times E_8$ weights with 
$E_{IJ}=G_{IJ}+B_{IJ}$. For more details, see Appendix \ref{ap: star_product}.}.

The phase $\tilde{f}$ appearing in the third line of \eqref{eq: star2fieldsModes} is defined as
\begin{equation}
\label{eq: phaseGral}
 \tilde{f}_{\mathbb{K}
_1 
\mathbb {K}_2
\mathbb{K}_3}=e^{i{\pi } 
{p}_1\cdot \tilde p_2}e^{i{\pi } 
({p}_1+{p}_2)\cdot \tilde p_3}\equiv\pm 1\, ,
\end{equation}
and will play a crucial role throughout this work. At enhancement points these phases, encoding KK momenta and winding modes, will supply the Cartan--Weyl structure constants of the enhanced gauge algebra, while the Cartan--root structure constants will emerge from the corresponding left momenta. For further details surrounding definition \eqref{eq: phaseGral}, see Appendix \ref{ap: star_product}.

When cubic terms in the action are integrated over the internal coordinates, only terms with vanishing total GKK label survive, $   \mathbb P_1+\mathbb P_2+\mathbb P_3=0$. For the ordered product $(\phi_1\star\phi_2)\star\phi_3$, the corresponding phase is the product of two elementary two-field phases,
\begin{equation}
\label{eq: cubicStarPhase}
    {\cal C}_{123}
    =
    e^{i\pi p_1\cdot\tilde p_2}
    e^{i\pi (p_1+p_2)\cdot\tilde p_3}.
\end{equation}
Using $\mathbb P_1+\mathbb P_2=-\mathbb P_3$, this becomes
\begin{equation}
\label{eq: cubicStarPhaseZeroMode}
    {\cal C}_{123}
    =
    e^{i\pi p_1\cdot\tilde p_2}
    e^{-i\pi p_3\cdot\tilde p_3}.
\end{equation}
Thus the phases appearing in cubic interaction terms are not independent objects: they are products of the same elementary cocycle phases generated by the $\star$-product. Their permutation properties, and their relation to the Cartan--Weyl structure constants at enhancement points, are discussed in Appendix \ref{ap: star_product}.

Finally, at an enhancement point, the GKK-labelled modes must be related to the Cartan--Weyl basis of the enhanced algebra. We take one representative $\mathbb P_i$ for each positive root from $G(\Phi_{0})$, with
\[
    \kappa_L(\mathbb P_i;\Phi_0)=\alpha_i,
    \qquad
    \kappa_L(-\mathbb P_i;\Phi_0)=-\alpha_i .
\]
Following the convention in \cite{Aldazabal:2018uzm}, we identify the corresponding
charged vector modes as
\begin{equation}
\label{eq: modeRootPrescription}
    A_\mu^{(\mathbb K_i)}(x)
    \equiv
    A_\mu^{(\alpha_i)}(x) \, 
    ,
    \qquad
    -A_\mu^{(-\mathbb K_i)}(x)
    \equiv
    A_\mu^{(-\alpha_i)}(x)
    \qquad \text{for }\,\,
    \mathbb K_i=\mathbb K(\mathbb P_i;\Phi_0)\, .
\end{equation}
Equivalently, the mode with label $-\mathbb P_i$ is identified with the negative-root
field with an additional minus sign. This prescription fixes the relative orientation
between positive- and negative-root modes, and will be used below when comparing
GKK-expanded expressions with the standard Cartan--Weyl form of non-Abelian gauge
transformations. The same convention is used for the corresponding gauge parameters,
and analogously for charged scalar modes whenever they are written in a Cartan--Weyl
basis.

\section{Gauge-like transformations and covariance}
\label{sec: Symmetries}

As highlighted in Section \ref{sec: Senh}, the truncated action \eqref{eq: hetActionUplift}, although incomplete, does continuously interpolate among gauge theories with enhanced gauge symmetries that arise at different points in moduli space, and displays a background-independent form. As a mathematical object, \eqref{eq: hetActionUplift} is thus well-suited for exploring the question of what underlying structure makes this interpolation natural.

Just as the ordinary non-Abelian covariant derivatives at enhancement points guided the construction of the uplifted expressions \eqref{eq: DmuMjUplift} and \eqref{eq: DmuHkjUplift}, the corresponding enhancement-point gauge transformations guide our construction of $\star$-product transformations before mode expansion. In the comparison between the GKK-expanded transformations and ordinary non-Abelian gauge variations, the required algebraic data is already present in momentum factors, which reproduce the Cartan action, while the phases $\tilde f$ generated by the $\star$-product reproduce the signs of the charged-root structure constants.

In what follows, we propose a set of $\star$-product transformations for $N=0,1$ fields and show that, after GKK expansion at symmetry enhancement points, they reproduce standard non-Abelian variations for fields in the relevant representations. We illustrate this with two examples. Finally, we check that the same transformations make the derivative structures \eqref{eq: DmuMjUplift} and \eqref{eq: DmuHkjUplift} transform covariantly in the unexpanded $\star$-product formulation.

\subsection{Gauge transformations}\label{sec: GaugeTransformations}

Building up from the results in \cite{Aldazabal:2018uzm}, where expressions \eqref{eq: DmuMjUplift} and \eqref{eq: DmuHkjUplift} for the would-be covariant derivatives involved in the action \eqref{eq: hetActionUplift} were presented, we would now like to answer the question of whether it is possible  to propose would-be gauge transformations. That is, field transformations in terms of the $\star$-product that contain typical, non-Abelian gauge transformations that become manifest at symmetry-enhancement points in moduli space.

Gauge invariance of the symmetry-enhanced effective theories proves a useful guiding principle on which to base our construction. Given a scalar field $\Phi^{s}$, associated to oscillator numbers $N_{s}=0,1$ and $\bar{N}_{s} = 0$, we expect its transformation law to be

\begin{equation}\label{eq: deltaPhiTypical}
    \delta \Phi^{s} = i \left( \lambda^{a}T_{a}\right)^{s}\,_{r}\Phi^{r} \, ,
\end{equation}
where $\lambda^{a}$ is the gauge parameter, $T_{a}$ are the relevant generators, and $r$ runs over all possible states in the representation. We have omitted dependence on the set of spacetime coordinates $x$ to simplify notation. 

Vectors in the theory that do not take part in symmetry-enhancement phenomena will also transform according to \eqref{eq: deltaPhiTypical}, while for the gauge vector bosons of the theory, which transform in the adjoint representation, we expect

\begin{equation}\label{eq: deltaAmuTypical}
    \delta A_{\mu}^{c} = -\frac{1}{g} \partial_{\mu}\lambda^{c} + i f_{ab}\,^{c} \lambda^{a}A_{\mu}^{b} \, ,
\end{equation}
with $f_{ab}\,^{c}$ labeling the symmetry group's structure constants, and $a,b$ and $c$ are state indices.

We now propose a set of $\star$-product transformations that reproduce \eqref{eq: deltaPhiTypical} and \eqref{eq: deltaAmuTypical} at enhancement points in moduli space, which we illustrate with two examples in Section \ref{subsec: Examples}.

\subsubsection{Scalar fields}\label{sec: variationM}

\subsubsection*{$N=0$ scalars}

Our proposal for the transformation law of $N=0$ scalar fields is
\begin{equation} \label{eq: deltaMj}
\delta M_{\mathcal{J}} = F(\lambda)M_{\mathcal{J}} - \partial^{\mathcal{K}}\lambda \star\left( \mathcal{H}_{\mathcal{JK}} - \eta_{\mathcal{JK}} \right)\, ,
\end{equation}
where the gauge-parameter-dependent operator $F(\lambda)$ is the same we defined in \eqref{eq: opCargados}.

With the help of the mode-expansion prescription \eqref{eq: star2fieldsModes}, and choosing to keep only zero-modes of $N=1$ gauge parameters, we arrive at the mode-expanded version of \eqref{eq: deltaMj},
\begin{equation}\label{eq: deltaMjGKK}
\begin{aligned}
\delta M^{(\mathbb{K}_{1})}_{\bar{J}} =& -i \sum_{\mathbb{K}_{2}}\tilde{f}_{\mathbb{K}_{1}\mathbb{K}_{2}\mathbb{K}_{3}} \lambda^{(\mathbb{K}_{2})}  M_{\bar{J}}^{(\mathbb{K}_{3})}  + i \lambda_{\mathcal{K}}^{(0)}\left(\mathbb{K}_{1} \right)^{\mathcal{K}} M_{\bar{J}}^{(\mathbb{K}_{1})}\\
& -i \sum_{\mathbb{K}_{2}}\tilde{f}_{\mathbb{K}_{1}\mathbb{K}_{2}\mathbb{K}_{3}} M_{\hat{I}\bar{J}}^{(\mathbb{K}_{3})}\left(\mathbb{K}_{2} \right)^{\hat{I}}\lambda^{(\mathbb{K}_{2})} \, ,
\end{aligned}
\end{equation}
where momentum conservation is imposed, $\mathbb{K}_{1} = \mathbb{K}_{2} + \mathbb{K}_{3} $.

At first glance, and upon comparison with \eqref{eq: deltaPhiTypical}, it might seem that we are dealing with completely different expressions. However, when we locate ourselves at an enhanced-symmetry point in moduli space, equation \eqref{eq: deltaMjGKK} becomes considerably more transparent. On the one hand, the phases $\tilde{f}_{\mathbb{K}_{1}\mathbb{K}_{2}\mathbb{K}_{3}}$ will have well-defined relations with the group's structure constants (see Section \ref{subsec: gkk_bracket_algebra} and Appendix \ref{ap: star_product}). The generalized momentum $\mathbb{K}_{1}$ will yield the matrix elements of the Cartan subalgebra's generators, and the momentum $\mathbb{K}_{2}$ will reduce to one of the group's roots and combine with the scalar $M_{\mathcal{KJ}}^{(\mathbb{K}_{3})}$ to yield the physical fields of the theory associated with states carrying $N=1$ oscillators. In fact, it is possible to show that \eqref{eq: deltaPhiTypical} and \eqref{eq: deltaMjGKK} are completely equivalent. This is illustrated throughout the examples at the end of this section.

\subsubsection*{\texorpdfstring{$N=1$ scalars}{$N=1$ scalars}}

We propose the variation for $N=1$ scalar fields through that of the generalized metric $\mathcal{H}_{\mathcal{KJ}}$,
\begin{equation} \label{eq: deltaHkj}
\begin{aligned}
\delta \mathcal{H}_{\mathcal{KJ}} =  F(\lambda)^{\mathcal{QP}}_{\mathcal{K}\mathcal{J}} \mathcal{H}_{\mathcal{QP}} + \partial_{\mathcal{K}} \lambda \star M_{\mathcal{J}} \, 
\end{aligned}
\end{equation}
where, once more, the gauge-parameter dependent operator $F(\lambda)^{\mathcal{QP}}_{\mathcal{K}\mathcal{J}}$ is the same we defined in \eqref{eq: opCartanes}.
Since the $N = 1$ scalar fields we will be dealing with are $M_{\hat{I}\bar{J}}$, and these appear as the first order fluctuations of $\mathcal{H}_{\mathcal{IJ}}$ (see equation \eqref{eq: HkjFluctuations}), we can recast \eqref{eq: deltaHkj} for algebraic convenience as
\begin{equation} \label{eq: deltaMkj}
\delta M_{\hat{I}\bar{J}} = \lambda^{\mathcal{K}}\star\partial_{\mathcal{K}} M_{\hat{I}\bar{J}} + i\lambda \star M_{\hat{I}\bar{J}} + i\partial^{\mathcal{K}}\partial_{\hat{I}}\lambda \star M_{\mathcal{K}\bar{J}} - \partial_{\hat{I}}\lambda \star M_{\bar{J}}\, ,
\end{equation}
where we have suppressed terms involving derivatives of the $N=1$ gauge parameter $\lambda^{\mathcal{K}}$, as these turn out to be identically zero upon mode expansion, since we are restricting them to their zero modes. Both GKK mode-expansions, that of \eqref{eq: deltaHkj} and of \eqref{eq: deltaMkj}, lead to
\begin{equation}\label{eq: deltaMkj_GKK}
\begin{aligned}
\delta M_{\hat{I}\bar{J}}^{(\mathbb{K}_{1})} &= i\lambda^{\mathcal{K}(0)}\left(\mathbb{K}_{1}\right)_{\mathcal{K}}M_{\hat{I}\bar{J}}^{(\mathbb{K}_{1})} + i \sum_{\mathbb{K}_{2}}\tilde{f}_{\mathbb{K}_{1}\mathbb{K}_{2}\mathbb{K}_{3}} \lambda^{(\mathbb{K}_{2})} M_{\hat{I}\bar{J}} ^{(\mathbb{K}_{3})}\\
& \quad -i \sum_{\mathbb{K}_{2}}\tilde{f}_{\mathbb{K}_{1}\mathbb{K}_{2}\mathbb{K}_{3}} \left( \mathbb{K}_{2}\right)^{\mathcal{K}}\left(\mathbb{K}_{2}\right)_{\hat{I}}\lambda^{(\mathbb{K}_{2})} M_{\mathcal{K}\bar{J}} ^{(\mathbb{K}_{3})} - i\sum_{\mathbb{K}_{2}}\tilde{f}_{\mathbb{K}_{1}\mathbb{K}_{2}\mathbb{K}_{3}}  \lambda^{(\mathbb{K}_{2})}\left(\mathbb{K}_{2}\right)_{\hat{I}} M_{\bar{J}}^{(\mathbb{K}_{3})}\, ,
\end{aligned}
\end{equation}
where once more we impose $\mathbb{K}_{1} = \mathbb{K}_{2} + \mathbb{K}_{3}$. Just like with the $N=0$ scalars, expression \eqref{eq: deltaMkj_GKK} can be shown to exactly match \eqref{eq: deltaPhiTypical} when written in terms of the theory's physical fields.

\subsubsection{Gauge bosons}

\subsubsection*{$N=0$ vector fields}

Our proposal for $N=0$ gauge vector bosons again leverages the operator defined in \eqref{eq: opCargados} and takes the form
\begin{equation}\label{eq: deltaAmuGauge}
\delta A_{\mu} = - \frac{1}{g} \partial_{\mu}\lambda +  F(\lambda)A_{\mu} -  \partial_{\mathcal{K}}\lambda \star A_{\mu}^{\mathcal{K}} \, .
\end{equation}
Its GKK mode expansion yields
\begin{equation}\label{eq: deltaAmuGaugeGKK}
\delta A_{\mu}^{(\mathbb{K}_{1})} = - \frac{1}{g}\partial_{\mu}\lambda^{(\mathbb{K}_{1})} -  i\sum_{\mathbb{K}_{2}} \tilde{f}_{\mathbb{K}_{1}\mathbb{K}_{2}\mathbb{K}_{3}} \lambda^{(\mathbb{K}_{2})} A_{\mu}^{(\mathbb{K}_{3})} + i \lambda^{\mathcal{K}(0)} \mathbb{K}_{1\mathcal{K}} A_{\mu}^{(\mathbb{K}_{1})} - i \mathbb{K}_{1\mathcal{K}} \lambda^{(\mathbb{K}_{1})}  A_{\mu}^{\mathcal{K}(0)} \, ,
\end{equation} 
where $\mathbb K_1=\mathbb K_2+\mathbb K_3$, as in the scalar mode expansions above.

In contrast to what happens with the scalar fields, it is straightforward to check that \eqref{eq: deltaAmuGaugeGKK} exactly matches \eqref{eq: deltaAmuTypical} when located at an enhancement moduli point, by a mere rewriting of the involved phases in terms of the group's structure constants as detailed in \eqref{eq: ctesFaseTildef}, and without any further redefinitions of physical fields. Without loss of generality, let $\alpha_{s}$ be a positive root of the enhanced group $G_{L}$. From \eqref{eq: deltaAmuGaugeGKK},
\begin{equation}\label{eq: deltaAmuGaugeGKKExample_a}
\delta A_{\mu}^{(\alpha_{s})} = - \frac{1}{g}\partial_{\mu}\lambda^{(\alpha_{s})} -  i\sum_{\alpha_{l},\,  \alpha_{r}} \tilde{f}_{\alpha_{s}\alpha_{l}\alpha_{r}} \lambda^{(\alpha_{l})} A_{\mu}^{(\alpha_{r})} + i \alpha_{s\mathcal{K}} \lambda^{\mathcal{K}(0)} A_{\mu}^{(\alpha_{s})} - i \alpha_{s\mathcal{K}} \lambda^{(\alpha_{s})}  A_{\mu}^{\mathcal{K}(0)} \, ,
\end{equation} 
where $\alpha_{l}$ and $\alpha_{r}$ are any two roots of $G_{L}$ that satisfy $\alpha_{s} = \alpha_{l} + \alpha_{r}$. Using the phase convention reviewed in Appendix \ref{ap: star_product} together with the mode-to-root prescription \eqref{eq: modeRootPrescription}, this is re-written as
\begin{equation}\label{eq: deltaAmuGaugeGKKExample_b}
\begin{aligned}
\delta A_{\mu}^{(\alpha_{s})} =& - \frac{1}{g}\partial_{\mu}\lambda^{(\alpha_{s})} +  i \sum_{\alpha_{i},\,  \alpha_{j}} -f_{\alpha_{i}(-\alpha_{j})}\,^{\alpha_{s}} \left( \lambda^{(-\alpha_{j})} A_{\mu}^{(\alpha_{i})} - \lambda^{(\alpha_{i})} A_{\mu}^{(-\alpha_{j})} \right) \\
&- i \sum_{\alpha_{x}, \, \alpha_{z}} f_{\alpha_{x}\alpha_{z}}\,^{\alpha_{s}} \lambda^{(\alpha_{z})}A_{\mu}^{(\alpha_{x})} + i f_{\mathcal{K}\alpha_{s}}\,^{\alpha_{s}} \left( \lambda^{\mathcal{K}(0)} A_{\mu}^{(\alpha_{s})} - \lambda^{(\alpha_{s})}  A_{\mu}^{\mathcal{K}(0)} \right)\, ,
\end{aligned}
\end{equation} 
which immediately yields \eqref{eq: deltaAmuTypical} by application of antisymmetry in the indices of structure constants of $G_{L}$. Here we assume $\lbrace \alpha_{i},\alpha_{j}, \alpha_{x}, \alpha_{z}\rbrace$ are positive roots of $G_{L}$  such that $\alpha_{i} - \alpha_{j} = \alpha_{x} + \alpha_{z} = \alpha_{s}$. Notice several labels $\lbrace i,j,x,z \rbrace$ might satisfy these relations for a given $s$.

\subsubsection*{$N=1$ vector fields}

The $\star$-product expression that reproduces \eqref{eq: deltaAmuTypical} upon mode-expansion is
\begin{equation}\label{eq: deltaAmuCartanGauge}
\delta A_{\mu}^{\mathcal{I}} = - \frac{1}{g} \partial_{\mu}\lambda^{\mathcal{I}} + \partial^{\mathcal{I}}\lambda\star A_{\mu} \, .
\end{equation}
The variation of the zeroth mode is then
\begin{equation}\label{eq: deltaAmuCartanGaugeGKK}
\delta A_{\mu}^{\hat{I}(0)} = - \frac{1}{g}\partial_{\mu}\lambda^{\hat{I}(0)} + i \sum_{\mathbb{K}} \mathbb{K}^{\hat{I}}\left( \lambda^{(\mathbb{K})} A_{\mu}^{(-\mathbb{K})} - \lambda^{(-\mathbb{K})} A_{\mu}^{(\mathbb{K})} \right) \, ,
\end{equation}
where we have already applied identification \eqref{eq: modeRootPrescription}, pre-identifying momenta $\mathbb{K}$ with the will-be positive roots of $G_{L}$ to avoid confusion. Once more, using the results in Appendix \ref{ap: star_product}, recovery of \eqref{eq: deltaAmuTypical} is immediate.

\subsection{Examples}
\label{subsec: Examples}

Throughout the examples in this section, we set the Wilson lines to zero,
\(A^{I}{}_{m}=0\), and adopt the conventions
\(\widetilde R=\alpha'/R\) and \(R=\sqrt{\alpha'}\), where $R$ is the common
radius parameter and $\widetilde R$ its dual, as summarized in
Appendix~\ref{ap: HeteroticStringBasics}. The choice $R=\sqrt{\alpha'}$ and vanishing Wilson lines simplify these examples but neither is required by the preceding construction.

\subsubsection{Compactification on \texorpdfstring{$T^{2}$}{T2} - \texorpdfstring{$\mathrm{SU(3)}_{L}$}{SU(3)L} enhancement point} \label{sec: su3example}

Following up on the example presented in \cite{Aldazabal:2018uzm}, we consider a 2-torus compactification. The generalized momentum encoding KK and winding modes is $\mathbb{P} = \left(P^{I}, p_{1}, p_{2}; \tilde{p}^{1}, \tilde{p}^{2} \right)$. At a generic moduli point $\Phi = \left(g, b, A\right)$, nonzero momenta give rise to massive states. Massless vectors arise from zero modes, $A_{\mu}^{\hat{I}(0)} \equiv A_{\mu}^{I(0)}, A_{\mu}^{1(0)}, A_{\mu}^{2(0)}$, and lead to the generic $U(1)^{18}_{L}\times U(1)_{R}^{2} $ gauge group. Enhancements will occur at specific moduli points. For instance, on the zero Wilson line locus $\Phi \equiv (g,b,0)$, with generic $g$ and $b$, $S_{enh}\left(\Phi \equiv (g,b,0)\right)$ corresponds to a well-defined $\mathrm{SO(32)}_{L} \times \mathrm{U(1)^{2}}_{L} \times \mathrm{U(1)}_{R}^{2}$ gauge theory \cite{Aldazabal:2018uzm}.

Moduli points $\Phi \equiv (g,b,0)$ can lead to further enhancements for specific values of $g$ and $b$ on the compactification 2-torus. In particular, an $\mathrm{\mathrm{SU(3)}}$ gauge symmetry enhancement of the $\mathrm{U(1)^{2}}_{L}$ factor occurs for $g$ and $b$ the Cartan matrix and antisymmetric tensor of $\mathrm{SU(3)}$, respectively.

We will focus here on a massive representation to illustrate the scope of our construction, and study the $\bm{6}_{(-\frac{2}{3}, -\frac{1}{3})}$ representation of $\mathrm{SU(3)}$, the full details of which are in Appendix \ref{ap: SU3facts}.  
In order for expressions \eqref{eq: DmuMjUplift} and \eqref{eq: DmuHkjUplift} to reproduce a well-defined covariant derivative for scalar fields $M_{\bar{J}}^{(\Lambda_{s})}$ and $M_{m\bar{J}}^{(\Lambda_{s})}$ in this representation, the latter need to be redefined \cite{Aldazabal:2018uzm}. The normalized fields are

\begin{equation}\label{eq: finalScFieldsSU3}
\begin{aligned}
\Phi^{s}_{\bar{J}} &= M^{(\Lambda_{s})}_{\bar{J}} \quad \text{for} \quad s=1,3,6 \, , \\
\Phi^{2}_{\bar{J}} &= \frac{-\alpha_{1}^{m}}{\sqrt{2}} M^{(\Lambda_{2})}_{m\bar{J}} \, , \\
\Phi^{4}_{\bar{J}} &= \frac{\alpha_{3}^{m}}{\sqrt{2}} M^{(\Lambda_{4})}_{m\bar{J}} \, , \\
\Phi^{5}_{\bar{J}} &= \frac{\alpha_{2}^{m}}{\sqrt{2}} M^{(\Lambda_{5})}_{m\bar{J}} \, , 
\end{aligned}
\end{equation}
where the $\alpha_{i}$, $i=1,2,3$ are the positive roots of $\mathrm{\mathrm{SU(3)}}$. For states $\Lambda_{s}$ with $s=2,4,5$, the two fields $\left(m = 1,2\right)$  $M_{m\bar{J}}^{(\Lambda_{s})}$ must combine into the physical state $\alpha_{i}^{m}M_{m\bar{J}}^{(\Lambda_{s})}$, where $i$ depends on $s$, while the orthogonal state $\Lambda_{s}^{m}M_{m\bar{J}}^{(\Lambda_{s})}$ must decouple.
   
The specific roots $\alpha_{i}$ that appear contracted with $M_{m\bar{J}}^{(\Lambda_{s})}$ satisfy, and are uniquely determined by, the condition $\alpha_{i}^{m}\Lambda_{sm}= 0$. Namely, they select the degrees of freedom of $M_{m\bar{J}}^{(\Lambda_{s})}$ orthogonal to $\Lambda_{s}$ as the physical ones.

Since massive vector bosons have the same weights as the scalars, the same line of reasoning leads to a consistent covariant derivative of the massive vector fields. The physical vector bosons are
\begin{equation}\label{eq: finalVecFields}
\begin{aligned}
A_{\mu}^{s} &= A_{\mu}^{(\Lambda_{s})} \quad \text{for} \quad s=1,3,6 \, , \\
A_{\mu}^{2} &= \frac{-\alpha_{1m}}{\sqrt{2}} A_{\mu}^{m(\Lambda_{2})} \, , \\
A_{\mu}^{4} &= \frac{\alpha_{3m}}{\sqrt{2}} A_{\mu}^{m(\Lambda_{4})} \, , \\
A_{\mu}^{5} &= \frac{\alpha_{2m}}{\sqrt{2}} A_{\mu}^{m(\Lambda_{5})} \, . 
\end{aligned}
\end{equation}
It is also possible to write these physical linear combinations in a more general fashion, both for scalar and vector fields, linking the normalization constants to the phases defined in \eqref{eq: phaseGral} and entries of the relevant generators $T_{a}$. Explicitly, and taking scalar fields as an example, for $s = 2, 4, 5$ we have
\begin{equation}\label{eq: gralPhysFields_6SU3}
\Phi^{s}_{\bar{J}} = \frac{-\tilde{f}_{x_{1}\alpha_{l}s}}{\left(T_{\alpha_{l}}\right)_{x_{1}s}}\alpha_{l}^{m}M_{m\bar{J}} ^{(\Lambda_{s})} = \frac{\tilde{f}_{x_{2}(-\alpha_{l})s}}{\left(T_{\alpha_{l}}\right)_{sx_{2}}}\left( - \alpha_{l}\right)^{m}M_{m\bar{J}} ^{(\Lambda_{s})} \, ,
\end{equation}
where, as before, each $\alpha_{i}$ is uniquely determined by the condition $\alpha_{i}^{m}\Lambda_{sm}= 0$, as are the labels $x_{1}$ and $x_{2}$, since for phases to be nonzero we require $\Lambda_{x_{1}} - \alpha_{i} = \Lambda_{s}$ or $\Lambda_{x_{2}} - \left(-\alpha_{i}\right) = \Lambda_{s}$. In the last line of \eqref{eq: gralPhysFields_6SU3} we have used that $T_{-\alpha_{i}} = \left(T_{\alpha_{i}}\right)^{t}$. 

The current setup is the perfect testing ground for our transformation proposals. Recall that a well-defined gauge transformation for fields $\Phi^{s}_{\bar{J}}$ on this multiplet is given by \eqref{eq: deltaPhiTypical}. Splitting indices for transparency, it reads
\begin{equation} \label{eq: deltaPhiTypical_Split}
\begin{aligned}
\delta \Phi^{s}_{\bar{J}} &= i \left(  \lambda^{(\alpha_{l})}\left(T_{\alpha_{l}}\right)^{s}\,_{r_{1}}\Phi^{r_{1}}_{\bar{J}} +  \lambda^{m(0)}\left(T_{m}\right)^{s}\,_{s}\Phi^{s}_{\bar{J}} +   \lambda^{(\alpha_{l})}\left(T_{\alpha_{l}}\right)^{s}\,_{r_{2}}\Phi^{r_{2}}_{\bar{J}}\right) \, ,
\end{aligned}
\end{equation} 
where $s$ is any of the 6 states in the representation, $r_{1} \in \lbrace 1,3,6 \rbrace $, $r_{2} \in \lbrace 2,4,5 \rbrace $, and $r_{1}, r_{2} \neq s$. Summation is implied over all indices except $s$. 

When considering $N=0$ scalar fields corresponding to states with $N_{s}=0$, $s=1,3,6$, our transformation proposal is given by \eqref{eq: deltaMj}. For fields in the symmetric representation, this can be recast as
\begin{equation} \label{eq: deltaMj_GKK_SU3}
\delta M^{(\Lambda_{s})}_{\bar{J}} =  -i \tilde{f}_{s\alpha_{l}r_{1}} \lambda^{(\alpha_{l})}  M_{\bar{J}}^{(\Lambda_{r_{1}})} + i \lambda^{m(0)}\left(\Lambda_{s}\right)_{m}  M^{(\Lambda_{s})}_{\bar{J}}   -i \tilde{f}_{s\alpha_{l}r_{2}}\lambda^{(\alpha_{l})}\alpha^{m}_{l} M_{m\bar{J}}^{(\Lambda_{r_{2}})} \, .
\end{equation}
 If we take for example $s=1$, \eqref{eq: deltaMj_GKK_SU3} can immediately be shown to coincide with the expected structure of \eqref{eq: deltaPhiTypical_Split},
\begin{equation}\label{eq: deltaMj_GKK_SU3_Lambda1}
\begin{aligned}
\delta M^{(\Lambda_{1})}_{\bar{J}} &= i \left(  \lambda^{m(0)}\Lambda_{1m}M^{(\Lambda_{1})}_{\bar{J}} + \lambda^{(\alpha_{1})}(-\tilde{f}_{1\alpha_{1}2})\alpha_{1}^{m}M_{m\bar{J}} ^{(\Lambda_{2})} + \lambda^{(\alpha_{3})}(-\tilde{f}_{1\alpha_{3}4})\alpha_{3}^{m}M_{m\bar{J}} ^{(\Lambda_{4})} \right) \,
\end{aligned}
\end{equation}
 by using  either \eqref{eq: finalScFieldsSU3} or \eqref{eq: gralPhysFields_6SU3}, and the basis described in Appendix \ref{ap: SU3facts}. For $s=3,6$ the situation is completely analogous.

 For the $N=1$ fields the transformation proposal is, locating \eqref{eq: deltaMkj_GKK} at the $\mathrm{SU(3)}$ point, 
 \begin{equation}\label{eq: deltaMkj_GKK_SU3}
\begin{aligned}
\delta M_{m\bar{J}}^{(\Lambda_{s})} &= i\lambda^{\mathcal{K}(0)}\left(\Lambda_{s}\right)_{\mathcal{K}}M_{m\bar{J}}^{(\Lambda_{s})} + i \sum_{\alpha_{l}}\tilde{f}_{s\alpha_{l}r_{2}} \lambda^{(\alpha_{l})} M_{m\bar{J}} ^{(\Lambda_{r_{2}})}\\
& \quad -i \sum_{\alpha_{l}}\tilde{f}_{s\alpha_{l}r_{2}} \alpha_{l}^{\mathcal{K}}\alpha_{lm}\lambda^{(\alpha_{l})} M_{\mathcal{K}\bar{J}} ^{(\Lambda_{r_{2}})} - i \sum_{\alpha_{l}} \tilde{f}_{s\alpha_{l}r_{1}} \lambda^{(\alpha_{l})}\left(\alpha_{l}\right)_{m} M_{\bar{J}}^{(\Lambda_{r_{1}})}\, .
\end{aligned}
\end{equation}

Taking $s=2$ in \eqref{eq: deltaMkj_GKK_SU3} as an example, and projecting the expression with $\frac{-\alpha_{1}^{m}}{\sqrt{2}}$ according to \eqref{eq: finalScFieldsSU3}, leads to
\begin{equation} \label{eq: delta Mkj_GKK_SU3_Lambda2}
\begin{aligned}
\delta \Phi^{2}_{\bar{J}} &= i \left(   \lambda^{m(0)}\Lambda_{2m}\Phi^{2}_{\bar{J}}+  \lambda^{(\alpha_{1})}\sqrt{2}\Phi^{3}_{\bar{J}} + \lambda^{(-\alpha_{1})}\sqrt{2}\Phi^{1}_{\bar{J}} +\lambda^{(\alpha_{2})}\Phi^{4}_{\bar{J}} + \lambda^{(\alpha_{3})}\Phi^{5}_{\bar{J}}  \right)\, ,
\end{aligned}
\end{equation}
which is exactly the prediction from \eqref{eq: deltaPhiTypical_Split}. Again, analogous results hold for $s=4,5$. Just as was the case for the derivatives of massive vector bosons, the same logic leads to consistent transformation laws for these fields.

\subsubsection{Compactification on \texorpdfstring{$T^{3}$}{T3} - \texorpdfstring{$\mathrm{SU(4)}_{L}$}{SU(4)L} enhancement point} \label{sec: su4example}

We now consider a $3$-torus compactification. In particular, the moduli point $\Phi = (g,b,0)$, with $g$ and $b$ the Cartan matrix and antisymmetric tensor of $\mathrm{SU(4)}$ respectively, promotes an enhancement to $\mathrm{SU(4)}$ of the abelian factor $\mathrm{U(1)^{3}}_{L}
$. We can apply the results from \cite{Aldazabal:2018uzm} to this example to further demonstrate the robustness of \eqref{eq: DmuMjUplift} and \eqref{eq: DmuHkjUplift}, as well as the transformation proposals \eqref{eq: deltaMj} and \eqref{eq: deltaMkj}.

In appendix \ref{ap: SU4facts} we summarise the information about $\mathrm{SU(4)}$ necessary to follow the current example.

The set of KK momenta and winding numbers in \eqref{eq: KKyW_SU4}, and also detailed in Table \ref{tab: LambdasKKyW_SU4_15},  label the twelve charged massless vectors of $SU(4)_{L}$. When combined with the $\mathrm{SO}(32)$ modes \cite{Aldazabal:2018uzm}, these define a heterotic low energy effective theory $S_{eff}(\Psi)$, with gauge group $\mathrm{SO}(32)_{L} \times \mathrm{SU}(4)_{L} \times \mathrm{U}(1)^{3}_{R}$.
On the following example, we focus on a massive representation of the enhanced gauge group to demonstrate the full scope of our construction. We consider a $10$-dimensional representation of $\mathrm{SU(4)}$. The massive modes $\Lambda_{s}, \,\, s=1 \cdots 10$, corresponding to vectors and scalars, and filling this multiplet, are organised in Table \ref{tab: LambdasKKyW_SU4_10} and have masses $\alpha' m^{2} =1$.
We are free to choose the associated right charge as long as the LMC
\begin{equation}\label{eq: lmcSU4}
    \left(l^{m}_{L}\alpha_{m}\right)^{2} - \left(l^{m}_{R}\alpha_{m}\right)^{2} =  2 \left(1 - N \right) \, 
\end{equation}
is satisfied for each state in the multiplet. We choose $\left(l^{1}_{R}, l^{2}_{R}, l^{3}_{R} \right) = \left(-\frac{1}{2},0,\frac{1}{2} \right)$. By following \cite{Aldazabal:2018uzm}, it is straightforward to calculate the covariant derivatives of scalar fields in this multiplet and arrive at the appropriate physical fields involved. Here, we show the analogous procedure, starting from our proposed variations.
From \eqref{eq: deltaMjGKK}, and for $s=1$, we find
\begin{equation} \label{eq: deltaMj_10SU4_L1}
\begin{aligned}
\delta M_{\bar{J}}^{(\Lambda_{1})} &= - i\left( \tilde{f}_{1\alpha_{1} 2}\alpha_{1}^{m} \lambda^{(\alpha_{1})}M_{m\bar{J}}^{(\Lambda_{2})} + \tilde{f}_{1\alpha_{4} 4}\alpha_{4}^{m} \lambda^{(\alpha_{4})}M_{m\bar{J}}^{(\Lambda_{4})} \right. \\
&\quad  \left. + \tilde{f}_{1\alpha_{6} 7}\alpha_{6}^{m} \lambda^{(\alpha_{6})}M_{m\bar{J}}^{(\Lambda_{7})} \right) + i \Lambda_{1m}\lambda^{m(0)}M_{\bar{J}}^{(\Lambda_{1})}\\
& =i\left( -\alpha_{1}^{m}\lambda^{(\alpha_{1})}M_{m\bar{J}}^{(\Lambda_{2})} +\alpha_{4}^{m} \lambda^{(\alpha_{4})}M_{m\bar{J}}^{(\Lambda_{4})} -\alpha_{6}^{m} \lambda^{(\alpha_{6})} M_{m\bar{J}}^{(\Lambda_{7})} \right) \\
&\quad + i \left( \sqrt{2}\lambda^{1(0)} + \sqrt{\frac{2}{3}}\lambda^{2(0)} + \frac{1}{\sqrt{3}} \lambda^{3(0)} \right)M_{\bar{J}}^{(\Lambda_{1})} \, .
 \end{aligned}
\end{equation}
For this to take the form of \eqref{eq: deltaPhiTypical}, and by using the generators described in \eqref{eq: Generators10SU4}, we find we must take
\begin{equation}\label{eq: 10SU4newFields1}
\Phi_{\bar{J}}^{1} = M_{\bar{J}}^{(\Lambda_{1})} \, , \quad \Phi_{\bar{J}}^{2} = \frac{-\alpha_{1}^{m}}{\sqrt{2}}M_{m\bar{J}}^{(\Lambda_{2})} \, , \quad {\Phi_{\bar{J}}^{4} = \frac{\alpha_{4}^{m}}{\sqrt{2}}M_{m\bar{J}}^{(\Lambda_{4})}}\, , \quad
{\Phi_{\bar{J}}^{7} = \frac{-\alpha_{6}^{m}}{\sqrt{2}}M_{m\bar{J}}^{(\Lambda_{7})}}\, .
\end{equation}
In fact, and in a perfectly analogous manner to what we observed in the case of $\mathrm{SU}(3)$, the physical, non-decoupling linear combinations of $N=1$ fields are obtained in a straightforward way by enforcing the form of a covariant derivative to \eqref{eq: DmuMjUplift} (or, equivalently, the transformation \eqref{eq: deltaMj}) once it has been mode-expanded and written down for all $N=0$ states of the representation, i.e. $\Lambda_{s} \, | \, s= 1,3,6,10$. The collected definitions of physical fields for the representation under study are 
\begin{equation}\label{eq: finalN=1Fields10SU4}
\begin{alignedat}{3}
\Phi^{s}_{\bar{J}} &= M^{(\Lambda_{s})}_{\bar{J}} \qquad \text{for} 
 && s=1,3,6,10, &&  \\[0.5em]
\Phi^{2}_{\bar{J}} &= \frac{-\alpha_{1}^{m}}{\sqrt{2}} M^{(\Lambda_{2})}_{m\bar{J}},
\qquad&
\Phi^{4}_{\bar{J}} &= \frac{\alpha_{4}^{m}}{\sqrt{2}} M^{(\Lambda_{4})}_{m\bar{J}},
\qquad&
\Phi^{5}_{\bar{J}} &= \frac{\alpha_{2}^{m}}{\sqrt{2}} M^{(\Lambda_{5})}_{m\bar{J}},
\\[0.5em]
\Phi^{7}_{\bar{J}} &= \frac{-\alpha_{6}^{m}}{\sqrt{2}} M^{(\Lambda_{7})}_{m\bar{J}},
\qquad&
\Phi^{8}_{\bar{J}} &= \frac{-\alpha_{5}^{m}}{\sqrt{2}} M^{(\Lambda_{8})}_{m\bar{J}},
\qquad&
\Phi^{9}_{\bar{J}} &= \frac{\alpha_{3}^{m}}{\sqrt{2}} M^{(\Lambda_{9})}_{m\bar{J}} .
\end{alignedat}
\end{equation}

Calculating now the variation for the first weight associated to $N=1$ fields, from \eqref{eq: deltaMkj_GKK}, we obtain
\begin{equation} \label{eq: deltaMkj_10SU4_L2}
\begin{aligned}
\delta M_{m\bar{J}}^{(\Lambda_{2})} &= - i \left( \tilde{f}_{2\alpha_{1}3} \lambda^{(\alpha_{1})}\alpha_{1m} M_{\bar{J}}^{(\Lambda_{3})} + \tilde{f}_{2(-\alpha_{1})1}(- \lambda^{(-\alpha_{1})})(-\alpha_{1m}) M_{\bar{J}}^{(\Lambda_{1})} \right) \\
& \quad+ i \left( \tilde{f}_{2\alpha_{2}4}\lambda^{(\alpha_{2})}(\delta^{\mathcal{K}}_{m} - \alpha^{\mathcal{K}}_{2}\alpha_{2m})M_{\mathcal{K}\bar{J}}^{(\Lambda_{4})} + \tilde{f}_{2\alpha_{4}5}\lambda^{(\alpha_{4})}(\delta^{\mathcal{K}}_{m} - \alpha^{\mathcal{K}}_{4}\alpha_{4m})M_{\mathcal{K}\bar{J}}^{(\Lambda_{5})} \right. \\
& \quad \left. +\tilde{f}_{2\alpha_{5}7}\lambda^{(\alpha_{5})} (\delta^{\mathcal{K}}_{m} - \alpha^{\mathcal{K}}_{5}\alpha_{5m}) M_{\mathcal{K}\bar{J}}^{(\Lambda_{7})} + \tilde{f}_{2\alpha_{6}8}\lambda^{(\alpha_{6})}(\delta^{\mathcal{K}}_{m} - \alpha^{\mathcal{K}}_{6}\alpha_{6m})M_{\mathcal{K}\bar{J}}^{(\Lambda_{8})}  \right) \\
&\quad +i \left( \sqrt{\frac{2}{3}}\lambda^{2(0)} + \frac{1}{\sqrt{3}}\lambda^{3(0)}  \right) M_{m\bar{J}}^{(\Lambda_{2})}   \, \\
&= i \left(  \lambda^{(\alpha_{1})}(-\alpha_{1m}) M_{\bar{J}}^{(\Lambda_{3})} + \lambda^{(-\alpha_{1})}(-\alpha_{1m}) M_{\bar{J}}^{(\Lambda_{1})} \right) \\
& \quad + i \left( \lambda^{(\alpha_{2})}(-\delta^{\mathcal{K}}_{m} + \alpha^{\mathcal{K}}_{2}\alpha_{2m})M_{\mathcal{K}\bar{J}}^{(\Lambda_{4})} + \lambda^{(\alpha_{4})}(\delta^{\mathcal{K}}_{m} - \alpha^{\mathcal{K}}_{4}\alpha_{4m})M_{\mathcal{K}\bar{J}}^{(\Lambda_{5})} \right.\\
& \quad \left. +\lambda^{(\alpha_{5})}(\delta^{\mathcal{K}}_{m} - \alpha^{\mathcal{K}}_{5}\alpha_{5m})M_{\mathcal{K}\bar{J}}^{(\Lambda_{7})} + \lambda^{(\alpha_{6})}(-\delta^{\mathcal{K}}_{m} + \alpha^{\mathcal{K}}_{6}\alpha_{6m})M_{\mathcal{K}\bar{J}}^{(\Lambda_{8})}  \right) \\
&\quad +i \left( \sqrt{\frac{2}{3}}\lambda^{2(0)} + \frac{1}{\sqrt{3}}\lambda^{3(0)}  \right) M_{m\bar{J}}^{(\Lambda_{2})}   \, \\
\end{aligned}
\end{equation}
Prompted by \eqref{eq: finalN=1Fields10SU4}, we project this with $-\frac{\alpha_{1}^{m}}{\sqrt{2}}$, leading to
\begin{equation} \label{eq: deltaMkj_10SU4_L2_bis}
    \begin{aligned}
        \delta \Phi^{2}_{\bar{J}} &= i \left(  \sqrt{2} \lambda^{(\alpha_{1})} M_{\bar{J}}^{(\Lambda_{3})} +\sqrt{2} \lambda^{(-\alpha_{1})} M_{\bar{J}}^{(\Lambda_{1})} \right) + i \left( \lambda^{(\alpha_{2})}\left( \frac{\alpha_{4}^{m}}{\sqrt{2}} M_{m\bar{J}}^{(\Lambda_{4})}\right) \right. \\
        & \quad \left.+ \lambda^{(\alpha_{4})}\left(\frac{\alpha_{2}^{m}}{\sqrt{2}} M_{m\bar{J}}^{(\Lambda_{5})} \right) +\lambda^{(\alpha_{5})}\left( \frac{-\alpha_{6}^{m}}{\sqrt{2}} M_{m\bar{J}}^{(\Lambda_{7})}\right) + \lambda^{(\alpha_{6})}\left( \frac{-\alpha_{5}^{m}}{\sqrt{2}} M_{m\bar{J}}^{(\Lambda_{8})}\right)  \right) \\
        &\quad +i \left( \sqrt{\frac{2}{3}}\lambda^{2(0)} + \frac{1}{\sqrt{3}}\lambda^{3(0)}  \right) \Phi_{\bar{J}}^{2}   \, \\
        &= i \left(  \sqrt{2} \lambda^{(\alpha_{1})} \Phi^{3}_{\bar{J}} +\sqrt{2} \lambda^{(-\alpha_{1})} \Phi^{1}_{\bar{J}} \right) + i \left( \lambda^{(\alpha_{2})}\Phi^{4}_{\bar{J}} + \lambda^{(\alpha_{4})}\Phi^{5}_{\bar{J}} +\lambda^{(\alpha_{5})}\Phi^{7}_{\bar{J}}\right.\\
        &\quad \left.+ \lambda^{(\alpha_{6})}\Phi^{8}_{\bar{J}}\right) +i \left( \sqrt{\frac{2}{3}}\lambda^{2(0)} + \frac{1}{\sqrt{3}}\lambda^{3(0)}  \right) \Phi_{\bar{J}}^{2}   \, ,
    \end{aligned}
\end{equation}
thus yielding an expression that is immediately consistent with the desired \eqref{eq: deltaPhiTypical}. Both the $\mathrm{SU(3)}$  example we previously looked at, as well as the present one, suggest there always exists a redefinition of the KK fields such that the theory retains the portion of $\Mkj^{(\Lambda_{s})}$ that results from projecting by the root satisfying $\alpha_{i}^{m}\Lambda_{sm} = 0$, so that the decoupling part lies on a hyperplane orthogonal to $\alpha_{i}$. These definitions lead to consistent gauge transformations as inherited from \eqref{eq: deltaMj} and \eqref{eq: deltaMkj}, with analogous reasoning applying for massive vector fields.

\paragraph{A note on a universal normalization of the physical fields}
For all the representations we have so far studied, both for $\mathrm{~SU}(3)$ as well as $\mathrm{~SU}(4)$, a certain normalization written in terms of the phases \eqref{eq: phaseGral} and the representation generators seems to hold,
\begin{equation}\label{eq: universalN=1FieldDef}
\Phi^{r}_{\bar{J}} = \frac{-\tilde{f}_{x\alpha_{l}r}}{\left(T_{\alpha_{l}}\right)_{xr}}\alpha_{l}^{m}M_{m\bar{J}} ^{(\Lambda_{r})} = \frac{\tilde{f}_{x(-\alpha_{l})r}}{\left(T_{\alpha_{l}}\right)_{rx}}\left( - \alpha_{l}\right)^{m}M_{m\bar{J}} ^{(\Lambda_{r})} \, ,
\end{equation}
where the $x$ will be uniquely determined since we must have either $\Lambda_{x} - \alpha_{l} = \Lambda_{r}$ or $\Lambda_{x} - \left( - \alpha_{l} \right) = \Lambda_{r}$ . In all cases, the weights and roots involved also satisfy $\alpha^{m}_{l}\Lambda_{rm} = 0$. The proposed derivatives and transformations extend to the analogous rank two symmetric representations of $\mathrm{SU(n)}$, since the root $\alpha_l$ polarizing each $N=1$ descendant vertex operator is precisely the root satisfying $\alpha_l\cdot\Lambda_r=0$ that selects its physical field combination, and thus accounts for the universal normalization in \eqref{eq: universalN=1FieldDef}.

\subsection{Covariance of derivatives}
\label{sec: covariance}
As previously shown in \cite{Aldazabal:2018uzm}, the general expressions for 
derivatives \eqref{eq: DmuMjUplift} and \eqref{eq: DmuHkjUplift} reduce, upon GKK mode expansion, to that of a standard covariant derivative in a non-Abelian gauge theory. The same holds true for the variations proposed in Section \ref{sec: GaugeTransformations} in relation to gauge transformations. A natural question follows: what happens when one does \emph{not} mode-expand? In this section we study the robustness of our proposals as tentative symmetries of the truncated, interpolating action \eqref{eq: hetActionUplift}. In other words, we demonstrate that \eqref{eq: DmuMjUplift} and \eqref{eq: DmuHkjUplift} transform in a way that is fully consistent with the transformations of the scalar fields they act upon, even before performing a mode expansion.\footnote{It is worth stressing that proposals \eqref{eq: DmuMjUplift} and \eqref{eq: DmuHkjUplift} are covariant in mode-expanded form by construction. However, it is not straightforward how this property translates for $\star$-product expressions.}

\subsubsection{Strategy}
The instinctive course of action to evaluate how the derivatives transform is to make a term-by-term comparison: extract transformation laws from expressions \eqref{eq: DmuMjUplift} and \eqref{eq: DmuHkjUplift} by varying each of their terms individually, using the results from Section \ref{sec: GaugeTransformations}, and compare these to the respective scalar's variations. While this indeed captures the essence of the computation, it is not sufficient to rigorously establish covariance. Studying only the final structure of the resulting terms overlooks crucial information encoded in the intermediate steps. To properly establish covariance, one must account for several relevant factors: momentum constraints that emerge after mode expansion, and which determine  the physically allowed products; and the possibility that terms which appear distinct before expansion ultimately give rise to identical mode-expanded structures when located at a symmetry enhancement point, once projected by the appropriate factors that determine physical fields. These nuances are not visible in a naïve term-by-term comparison but are indispensable for matching the transformation laws.

A crucial subtlety is that one must account for the `origin' of each factor appearing in a given $\star$-product. This `origin' — that is, the field content and transformation rule from which the term descends — determines the set of allowed $\mathbb{K}$ modes that can appear once a GKK mode expansion is performed. Momentum conservation in the $\star$-product demands that, after mode expansion, the total momentum $\mathbb{K}$ of the product equals that of the field being varied. In other words, the contributing combinations of momenta are not arbitrary; they are constrained by the requirement that the full variation must match the $\mathbb{K}$ structure of the term from which it arises. This is especially important when a symmetry-enhancement effect occurs, as it will constrain the combinations of states within a certain representation that can give rise to another weight of the theory. These constraints apply not only to the total momentum of the full product — which must match that of the field being varied — but also to the intermediate pairwise products that appear as building blocks of the variation.  
As a result, only specific combinations of modes are allowed to appear. Careful treatment of this consideration will, in fact, allow for the identification of seemingly different terms. 

As an illustration, consider the variation of $D_{\mu}M_{\mathcal{J}}$. The assumption that this derivative is covariant (which we encode in the calculation as a.c.; i.e. assuming covariance), and thus transforms according to \eqref{eq: deltaMj} results, among others, in the term
\begin{equation}\label{eq: hsExample1}
    \delta D_{\mu}M_{\mathcal{J}} \big|_{\mathrm{a.c.}} \, \supset i\lambda \star D_{\mu}M_{\mathcal{J}} \supset -g \lambda \star \left( A_{\mu} \star  M_{\mathcal{J}}\right)\, .
\end{equation}
Here and below, the symbol $\supset$ indicates that the expression on the right collects only selected contributions from the expression on the left, after expanding the relevant definitions as needed. Choosing a symmetry enhancement point and performing a mode expansion of \eqref{eq: hsExample1} reveals that, if the derivative being varied carries weight $\Lambda_{s}$, then constraints arise at each intermediate step in the computation. For instance, equation \eqref{eq: hsExample1} becomes:
\begin{equation}\label{eq: hsExample1_GKK}
\begin{aligned}
    \delta D_{\mu}M_{\bar{J}}^{(\Lambda_{s})} \big|_{\mathrm{a.c.}} \, &\supset -i\sum_{\Lambda_{s} = \alpha_{l} + \Lambda_{r}} \tilde{f}_{s\alpha_{l}r} \, \lambda^{(\alpha_{l})}  D_{\mu}M_{\bar{J}}^{(\Lambda_{r})} \\
    &\supset -g \sum_{\Lambda_{s} = \alpha_{l} + \Lambda_{r}} \sum_{\Lambda_{r} = \alpha_{m} + \Lambda_{t}} \tilde{f}_{s\alpha_{l}r} \, \tilde{f}_{r\alpha_{m}t} \, \lambda^{(\alpha_{l})}  A_{\mu}^{(\alpha_{m})}   M_{\bar{J}}^{(\Lambda_{t})}\, .
\end{aligned}
\end{equation}
Here, momentum conservation already constrains the possible values of the roots $\alpha_{l}$ and intermediate weights $\Lambda_{r}$ in the first step. Moreover, only those $\Lambda_{r}$ corresponding to physical, $N=0$ states in the multiplet can contribute.  The second line in \eqref{eq: hsExample1_GKK} introduces additional structure: expanding the covariant derivative yields further constraints on the combinations of roots and weights. In particular, the allowed values of $\alpha_{m}$ and $\Lambda_{t}$ must satisfy $\Lambda_{r} = \alpha_{m} + \Lambda_{t}$, and $\Lambda_{t}$ again corresponds to an $N=0$ scalar state. Thus, the allowed terms are determined recursively, with each intermediate step inheriting and refining the constraints from the previous one. 

In contrast, a term-by-term variation (which we denote by t.b.t.)\ of \eqref{eq: DmuMjUplift} does not immediately yield the $\star$-product term of three fields that is manifest in \eqref{eq: hsExample1}. Instead, we find
\begin{equation}
\begin{aligned}
    \delta D_{\mu}M_{\mathcal{J}} \big|_{\mathrm{t.b.t.}} \, &\supset g \delta A_{\mu}^{\mathcal{K}} \star \partial_{\mathcal{K}} M_{\mathcal{J}} -g \partial^{\mathcal{K}}A_{\mu} \star \delta \left(\mathcal{H}_{\mathcal{KJ}} - \eta_{\mathcal{KJ}} \right)\\
    & \supset g \left(\partial^{\mathcal{K}}\lambda \star A_{\mu}\right) \star \partial_{\mathcal{K}} M_{\mathcal{J}} - g \partial^{\mathcal{K}} A_{\mu} \star \left( \partial_{\mathcal{K}}\lambda  \star  M_{\mathcal{J}}\right) + \cdots 
\end{aligned}    
\end{equation}
Although these expressions appear distinct from \eqref{eq: hsExample1}, a mode expansion reveals their combination contains precisely the same terms as \eqref{eq: hsExample1_GKK}, 
\begin{equation}
\begin{aligned}
    \delta D_{\mu}M_{\bar{J}}^{(\Lambda_{s})} \big|_{\mathrm{t.b.t.}} \, &\supset ig \delta A_{\mu}^{\mathcal{K}(0)} \Lambda_{s\mathcal{K}} M_{\bar{J}}^{(\Lambda_{s})} +ig \sum_{\Lambda_{s} = \alpha_{l} + \Lambda_{r}} \alpha_{l}^{\mathcal{K}}A_{\mu}^{(\alpha_{l})}\,  \delta \left(\mathcal{H}_{\mathcal{K}\bar{J}} - \eta_{\mathcal{K}\bar{J}} \right)^{(\Lambda_{r})}\\
    & \supset -g \sum_{\alpha_{l}>0}\alpha_{l}^{\mathcal{K}}\Lambda_{s\mathcal{K}} \left(\lambda^{(-\alpha_{l})} A_{\mu}^{(\alpha_{l})} - \lambda^{(\alpha_{l})} A_{\mu}^{(-\alpha_{l})}\right) M_{\bar{J}}^{(\Lambda_{s})}\\& - g \sum_{\Lambda_{s} = \alpha_{l} + \Lambda_{r}}\sum_{\Lambda_{r} = \alpha_{m} + \Lambda_{t}}\tilde{f}_{s\alpha_{l}r}\tilde{f}_{r\alpha_{m}t}\alpha_{l}^{\mathcal{K}}\alpha_{m\mathcal{K}} A_{\mu}^{(\alpha_{l})}\lambda^{(\alpha_{m})}M_{\bar{J}}^{(\Lambda_{t})} + \cdots 
\end{aligned}    
\end{equation}
with additional contributions that cancel against the rest of the variation. This will become clearer through the explicit examples in the following sections.  This equivalence depends crucially on both momentum conservation at all steps, and on the inner products between roots of the same algebra: since $\alpha \cdot \beta \in \lbrace 0,\pm 1,\pm 2 \rbrace$, double derivatives $\partial_{\mathcal{K}}...\partial^{\mathcal{K}}$ acting on fields with definite weights may act as projectors—effectively selecting or discarding terms. These contractions are essential in establishing the matching of terms and the overall covariance of the derivative.

This simple example is illustrative enough to motivate the workflow outlined in Figure \ref{fig: workflow}, which has both downwards and upwards directions. The step-by-step process to prove covariance is as follows:

\begin{figure}[htb!]
\centering
\includegraphics[width=0.9\textwidth]{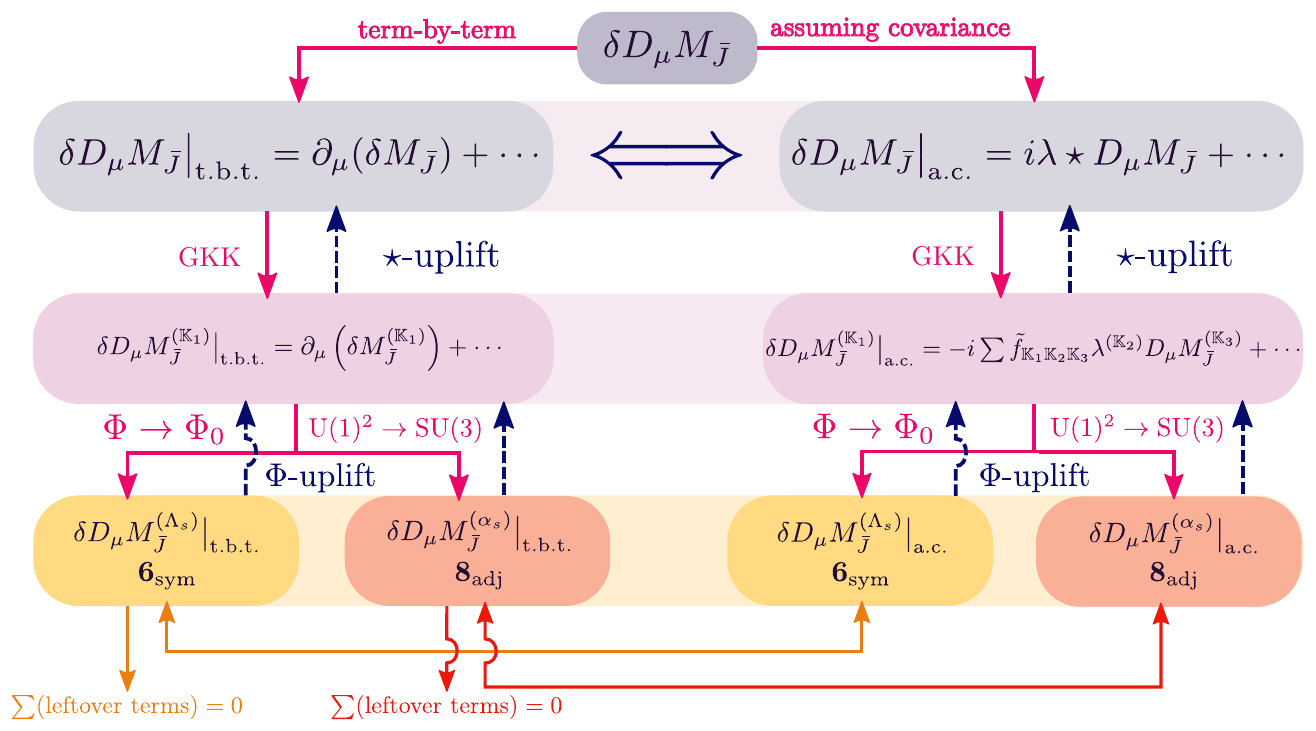}
\caption{Workflow for testing the covariance of $D_\mu M_{\bar J}$; the same procedure applies to $D_\mu M_{\mathcal K\bar J}$. The t.b.t.\ and a.c.\ variations are GKK-expanded, specialized to an $\mathrm{SU(3)}_L$ enhancement point, and projected onto the $\mathbf{6}$ and $\mathbf{8}$. Representation-level term matching is straightforward; apparent
leftover terms cancel after combining contributions. Solid downward arrows denote expansion and specialization, dotted upward arrows lift the matched correspondence to generic moduli and the unexpanded $\star$-product expressions. Horizontal double arrows connect matching terms.}
\label{fig: workflow}
\end{figure}

\begin{enumerate}
    \item \textbf{Calculate variations}

    We evaluate transformations using two methods: (i) \emph{term-by-term} (t.b.t.), where each term in \eqref{eq: DmuMjUplift} and \eqref{eq: DmuHkjUplift} is varied individually, and (ii) \emph{assuming covariance} (a.c.), where the whole derivative is treated as a covariant object, transforming according to either law \eqref{eq: deltaMj} or \eqref{eq: deltaHkj} .

The goal of this step is to generate two independent expressions for each variation, separately for \eqref{eq: DmuMjUplift} and \eqref{eq: DmuHkjUplift}. In later steps, we will compare these results to check for consistency between the two approaches.

In figure \ref{fig: workflow}, these calculations respectively correspond to the left (t.b.t.)\ and right (a.c.)\ branches at the top level, which we will follow throughout the comparison process.

    \item \textbf{Perform a GKK mode expansion}

    We now expand the results of the t.b.t.\ and a.c.\ calculations using the generalized Kaluza-Klein (GKK) expansion. This involves the obtained variations in terms of their individual momentum components across the Narain lattice.

Performing the GKK mode expansion is crucial because it allows us to track how each term transforms at the level of individual modes. This precision will later enable us to compare and match terms between the two approaches.

In Figure \ref{fig: workflow}, this step corresponds to moving downwards from the initial calculations (top level) to the expanded form (middle level), where terms from both approaches are broken into their mode components for comparison.

     \item \textbf{Choose an enhancement point}

     This step requires we choose an enhancement point as a proxy. At such points, gauge symmetry is fully manifest, making it easier to match terms from the t.b.t.\ and a.c.\ approaches.

     In principle, the choice could be any point in moduli space where the effective theory's symmetry group is enhanced on the left to $G_{L}$. For clarity, we work with the now familiar example of enhancement to $\mathrm{SO(32)}_{L} \times \mathrm{SU(3)}_{L}$, focusing on the $\mathrm{SU(3)}$ factor. However, we stress this is a mere tool, and does not render the results enhancement-point dependent.\

    Once the enhancement point is chosen, we use the expressions obtained in the previous step to write down the gauge transformations for the derivatives of scalars in multiplets of $G_{L}$ from both the t.b.t.\ and a.c.\ perspectives. Through this process, it becomes evident that not all terms that emerge in the second step of this workflow yield nonzero contributions for a given representation. Thus, we contemplate two reps for our enhancement choice: the adjoint and the symmetric representations of $\mathrm{SU(3)}_{L}$. These choices are sufficient to capture all essential features of the gauge structure.

    In figure \ref{fig: workflow}, this corresponds to the lowest level, where we organize terms from the mode expansions under specific representations of $G_{L}$.

    \item \textbf{Match terms on t.b.t.\ and a.c.\ calculations for the chosen enhancement}

    As covariance is guaranteed by construction in a point of enhanced gauge symmetry, it is straightforward to put transformation terms from the t.b.t.\ and a.c.\ calculations into one-to-one correspondence for each representation. As shown in the example at the beginning of this section, some terms arising from the a.c.\ assumption are not directly produced by the t.b.t.\ approach, but can be shown to emerge indirectly by combining contributions.

    This comparison corresponds to the lowest level of Figure \ref{fig: workflow}, where double-ended arrows illustrate the matching process, and arrows exiting the diagram illustrate cancellation of any leftover terms.

    \item \textbf{Match terms on t.b.t.\ and a.c.\ calculations for a generic moduli point}

     In this step, which commences the `upward' flow from the lowest level in the chart, we extend the term matching to a generic point in moduli space. We systematically reconstruct the a.c.\ expression for $\delta D_{\mu}M_{\bar{J}}^{(\Lambda_{s})}$ with the terms arising from the t.b.t.\ calculation. The process is analogous for $\delta D_{\mu}\mathcal{H}_{\hat{K}\bar{J}}^{(\Lambda_{s})} = - \delta D_{\mu}M_{\hat{K}\bar{J}}^{(\Lambda_{s})}$. With a precise understanding of the origin of each term in the representation-specific variations, we work backward through the GKK expansion. This process allows us to construct a systematic mapping—a dictionary—between the terms on both sides of the middle level of Figure \ref{fig: workflow}, ensuring that our transformation laws hold beyond specific enhancement points.

    \item \textbf{Match terms on t.b.t.\ and a.c.\ $\star$-product calculations} 

    Finally, we reconstruct $\delta D_{\mu}M_{\mathcal{J}}\big|_{\mathrm{a.c.}}$ using the terms derived from $\delta D_{\mu}M_{\mathcal{J}}\big|_{\mathrm{t.b.t.}}$. As we have a complete understanding of how each term arises from the GKK expansion, represented by the upward arrows reaching the topmost level of the comparison workflow, we can systematically verify that every term from the a.c.\ expression is accounted for in the t.b.t.\ calculation. This process establishes an exhaustive correspondence between both approaches, confirming that no terms are left unmatched. This demonstrates the internal consistency of our proposed transformations when interpreted within the full $\star$-product framework.

This process again corresponds to the topmost level of Figure \ref{fig: workflow}, indicated by horizontal arrows linking both branches of the figure, closing the loop of our consistency check.

\end{enumerate}

\subsubsection{Covariance of \texorpdfstring{ $D_{\mu}M_{\mathcal{J}}$}{DMj}}
Here we present some characteristic instances of the workflow outlined previously for the derivative of $N=0$ scalar fields, and prove it transforms covariantly. In Table \ref{tab: DmuMj} we summarize the results for term correspondence.\ 
To illustrate the first step of the process, we present the summarized structure of the term-by-term calculation,
\begin{equation}\label{eq: deltaDmuMjtbt}
\begin{aligned}
\delta D_{\mu}M_{\mathcal{J}}\big|_{\mathrm{t.b.t.}} = &\, \partial_{\mu}\left( \delta M_{\mathcal{J}} \right)+ ig\left( \delta A_{\mu} \star M_{\mathcal{J}} + A_{\mu} \star \delta M_{\mathcal{J}} \right) \\
&+ g \left( \delta A^{\mathcal{K}}_{\mu} \star \partial_{\mathcal{K}}M_{\mathcal{J}} + A^{\mathcal{K}}_{\mu} \star \partial_{\mathcal{K}}\left( \delta M_{\mathcal{J}} \right) \right)\\
& - g \left[  \partial^{\mathcal{K}}\left(\delta A_{\mu} \right) \star \left(\mathcal{H}_{\mathcal{K}\mathcal{J}} - \eta_{\mathcal{K}\mathcal{J}} \right) + \partial^{\mathcal{K}} A_{\mu}  \star \left( \delta \left(\mathcal{H}_{\mathcal{K}\mathcal{J}} - \eta_{\mathcal{K}\mathcal{J}} \right)\right) \right] \, ,
\end{aligned}
\end{equation} 
and the approach assuming covariance,
\begin{equation}\label{eq: deltaDmuMjac}
\delta D_{\mu}M_{\mathcal{J}}\big|_{\mathrm{a.c.}} = i\lambda \star D_{\mu}M_{\mathcal{J}} + \lambda_{\mathcal{K}}\star \partial^{\mathcal{K}}D_{\mu}M_{\mathcal{J}} - D_{\mu}\mathcal{H}_{\mathcal{KJ}} \star \partial^{\mathcal{K}}\lambda \, .
\end{equation}
Performing a GKK mode expansion of each is straightforward by systematically applying \eqref{eq: star2fieldsModes}  to  \eqref{eq: deltaDmuMjtbt} and \eqref{eq: deltaDmuMjac}, and so is the evaluation for the $\bm{6}_{(-\frac{2}{3}, -\frac{1}{3})}$ and $\bm{8}_{(0,0)}$ representations of $\mathrm{SU(3)}_{L}$, whose details are in Appendix \ref{ap: SU3facts}. We now analyse examples of indirect correspondence between terms in both approaches.

\paragraph{Symmetric representation}

To begin the comparison and reconstruct $\delta D_{\mu}M_{\mathcal{J}}\big|_{\mathrm{a.c.}}$, we first note that the first term in \eqref{eq: deltaDmuMjac} does not yield any contributions when looked at from the point of view of this representation. The reason is momentum conservation. For the chosen point, mode expansion of the first term in \eqref{eq: deltaDmuMjac} results in
\begin{equation}\label{eq: 1stTerm_deltaDmuMjac_GKK}
    \delta D_{\mu}M_{\bar{J}}^{(\Lambda_{s})}\big|_{\mathrm{a.c.}} \supset -i \sum_{\alpha_{l}} \tilde{f}_{\Lambda_{s}\alpha_{l}\Lambda_{r}} \lambda^{(\alpha_{l})} D_{\mu}M_{\bar{J}}^{(\Lambda_{r})}
\end{equation}
where $\Lambda_{r}$ must correspond to another state in the representation with $N_{r} = 0$, and $\alpha_{l}$ is a root of $\mathrm{SU(3)}_{L}$ such that $\Lambda_{s} = \Lambda_{r} + \alpha_{l}$. But there is no pair $(\Lambda_{r}, \alpha_{l})$, for any $s$, that satisfies these conditions in the symmetric representation, and so the term does not contribute. We will further study this term through the lens of the adjoint, where it does play an important role.

Let us analyze the interesting case of contributions from the t.b.t.\ calculation to the third term of \eqref{eq: deltaDmuMjac}. We will use the weight state $\Lambda_{1}$ as an example, but the same holds for all six charged states in the representation. By assuming covariance, the term
\begin{equation}
    \delta D_{\mu}M_{\bar{J}}^{(\Lambda_{1})}\big|_{\mathrm{a.c.}} \supset -\sqrt{2} g \lambda^{(\alpha_{1})} A_{\mu}^{\mathcal{K}(0)} \Lambda_{2\mathcal{K}} \Phi_{\bar{J}}^{2}\, ,
\end{equation}
where we have already used definitions \eqref{eq: finalScFieldsSU3}, emerges from expanding the third term in \eqref{eq: deltaDmuMjac},
\begin{equation} \label{eq: deltaDmuMjac_term3a}
    \delta D_{\mu}M_{\bar{J}}\big|_{\mathrm{a.c.}} \supset  - D_{\mu}\Hkj \star \partial^{\mathcal{K}}\lambda \supset -g \left(A_{\mu}^{\mathcal{P}} \star \partial_{\mathcal{P}}\Hkj\right) \star \partial^{\mathcal{K}}\lambda \, .
\end{equation}
On the contrary, it comes up in parts from the t.b.t.\ approach,
\begin{equation}
\begin{aligned}
     \delta D_{\mu}M_{\bar{J}}^{(\Lambda_{1})} \,\big|_{\mathrm{t.b.t.}} \, & \supset  -\sqrt{2} g \lambda^{(\alpha_{1})} A_{\mu}^{\mathcal{K}(0)} \Lambda_{1\mathcal{K}} \Phi_{\bar{J}}^{2} + \sqrt{2} g \lambda^{(\alpha_{1})} A_{\mu}^{\mathcal{K}(0)} \alpha_{1\mathcal{K}} \Phi_{\bar{J}}^{2}\\
     &=  -\sqrt{2} g \lambda^{(\alpha_{1})} A_{\mu}^{\mathcal{K}(0)} \Lambda_{2\mathcal{K}} \Phi_{\bar{J}}^{2} \, .
     \end{aligned}
\end{equation}
These emerge from the fifth and sixth terms in \eqref{eq: deltaDmuMjtbt},
\begin{equation}\label{eq: deltaDmuMjtbt_term5_6}
    \begin{aligned}
        \delta D_{\mu}M_{\mathcal{J}} \, \big|_{\mathrm{t.b.t.}} &\supset  g  A^{\mathcal{K}}_{\mu} \star \partial_{\mathcal{K}}\left( \delta M_{\mathcal{J}} \right)  - g \partial^{\mathcal{K}}\left(\delta A_{\mu} \right) \star \left(\mathcal{H}_{\mathcal{K}\mathcal{J}} - \eta_{\mathcal{K}\mathcal{J}} \right) \\
        & \supset -g A_{\mu}^{\mathcal{K}} \star \left( \partial_{\mathcal{K}}\left(\mathcal{H}_{\mathcal{P}\mathcal{J}} - \eta_{\mathcal{P}\mathcal{J}} \right) \star \partial^{\mathcal{P}}\lambda\right)\\
        &\quad - g \left( A_{\mu}^{\mathcal{K}} \star \left(\mathcal{H}_{\mathcal{P}\mathcal{J}} - \eta_{\mathcal{P}\mathcal{J}} \right) \right)\star \partial_{\mathcal{K}}\partial^{\mathcal{P}}\lambda \\
        & \quad + \cdots + g \left(\partial^{\mathcal{K}}\partial_{\mathcal{P}}\lambda \star A_{\mu}^{\mathcal{P}}\right) \star \left(\mathcal{H}_{\mathcal{K}\mathcal{J}} - \eta_{\mathcal{K}\mathcal{J}} \right) \, . 
    \end{aligned}
\end{equation}
From \eqref{eq: deltaDmuMjac_term3a} and \eqref{eq: deltaDmuMjtbt_term5_6}, we arrive at the partial correspondence described on the tenth row of Table \ref{tab: DmuMj}.

As highlighted before, it is not sufficient to use a single representation to put terms into correspondence and fill up Table \ref{tab: DmuMj}. We further analyse this situation from the adjoint point of view.

\paragraph{Adjoint representation}

The procedure for studying term correspondence from the perspective of this representation is completely analogous to that of the symmetric representation. Here we mainly focus on completing the construction by matching terms that did not yield any contributions for the latter.

We initially place our attention on the first term of \eqref{eq: deltaDmuMjac}, which does yield contributions for the adjoint, as opposed to the symmetric. Notice that since the sum of two roots of $\mathrm{SU(3)}_{L}$ can in principle result in another root, this term is allowed, in contrast to what we saw in equation \eqref{eq: 1stTerm_deltaDmuMjac_GKK}. We choose the state associated to root $\alpha_{3}$ as an example.

It is interesting to consider once more the term in \eqref{eq: hsExample1},
\begin{equation}
    \delta D_{\mu}M_{\mathcal{J}}\big|_{\mathrm{a.c.}} \supset i\lambda \star D_{\mu}M_{\mathcal{J}} \supset -g \lambda \star \left( A_{\mu} \star M_{\mathcal{J}} \right) \, ,
\end{equation}
which appears on the third row of Table \ref{tab: DmuMj}. In the previous section, we already mentioned \eqref{eq: hsExample1} arises as a combination of t.b.t.\ pieces that are shown in color in row 2 of Table \ref{tab: DmuMj}, to highlight their contribution to more than one a.c.\ term. They repeat in row 11 of the same Table. The same logic applies to other colored terms. For our present case, this example results in enhancement-point contributions such as
\begin{equation}
    \delta D_{\mu}\Mj^{(\alpha_{3})}\big|_{\mathrm{a.c.}} \supset -g \lambda^{(\alpha_{1})} A_{\mu}^{(-\alpha_{1})}\Mj^{(\alpha_{3})} + g \lambda^{(\alpha_{1})} A_{\mu}^{(\alpha_{3})}\Mj^{(-\alpha_{1})} + \cdots
\end{equation}
which on the t.b.t.\ approach emerge from combining the expansions of terms 4 and 7 in \eqref{eq: deltaDmuMjtbt}, respectively,
\begin{equation}\label{eq: deltaDmuMjtbt_term4_7}
    \begin{aligned}
        \delta D_{\mu}M_{\mathcal{J}} \, \big|_{\mathrm{t.b.t.}} &\supset  g  \delta A^{\mathcal{K}}_{\mu} \star \partial_{\mathcal{K}} M_{\mathcal{J}}  - g \partial^{\mathcal{\mathcal{K}}}A_{\mu} \star \delta \left(\mathcal{H}_{\mathcal{KJ}} - \eta_{\mathcal{KJ}}  \right) \\
        & \supset \textcolor{mGold}{g  \left(\partial^{\mathcal{K}}\lambda \star A_{\mu} \right) \star \partial_{\mathcal{K}} M_{\mathcal{J}}}  - \textcolor{mViolet}{ g \partial^{\mathcal{K}}A_{\mu} \star \left(  \partial_{\mathcal{K}}\lambda \star M_{\mathcal{J}}\right)} \, .
    \end{aligned}
\end{equation}

Taken together, the symmetric and adjoint analyses exhaust the distinct term-matching mechanisms required: direct correspondence, the recombination of several t.b.t.\ contributions into a single a.c.\ structure, and the cancellation of the remaining terms. The complete correspondence is summarized in Table~\ref{tab: DmuMj}.

With no terms left unmatched, we conclude proposals \eqref{eq: deltaMj}, \eqref{eq: deltaMkj}, \eqref{eq: deltaAmuGauge} and \eqref{eq: deltaAmuCartanGauge} lead the derivative for $N=0$ scalars \eqref{eq: DmuMjUplift} to a fully covariant transformation law in its uplifted form, which is what we set out to prove.

\begin{center}
\setlength{\LTleft}{0pt plus 1fill minus 1fill}
\setlength{\LTright}{0pt plus 1fill minus 1fill}
\begin{longtable}{|c|c|c|}


\caption{Covariance of $\delta D_{\mu}M_{\mathcal{J}}$. This table summarises how \emph{all} terms arising from $\delta D_{\mu}M_{\mathcal{J}} \big|_{\mathrm{t.b.t.}}$ combine to form $\delta D_{\mu}M_{\mathcal{J}} \big|_{\mathrm{a.c.}}$. There are indeed no unmatched terms in either calculation, and so we claim covariance of \eqref{eq: DmuMjUplift}. Notice that certain `t.b.t.'\ terms contribute to more than one `a.c.'\ term. These are shown in color.}\phantomsection\label{tab: DmuMj}\\
\hline
 $\bm{\delta D_{\mu}M_{\mathcal{J}}\big|_{\mathrm{t.b.t.}}}$  & \multicolumn{2}{|c|}{$\bm{\delta D_{\mu}M_{\mathcal{J}}}\big|_{\mathrm{a.c.}}$}  \\ \hline
\endfirsthead
  
 \multicolumn{3}{c}%
{{ \tablename\ \thetable{} -- continued from previous page}} \\
\hline $\bm{\delta D_{\mu}M_{\mathcal{J}}}\big|_{\mathrm{t.b.t.}}$  &  \multicolumn{2}{|c|}{$\bm{\delta D_{\mu}M_{\mathcal{J}}}\big|_{\mathrm{a.c.}}$}   \\ \hline
\endhead
 
\hline \multicolumn{3}{|r|}{{Continued on next page}} \\ \hline
\endfoot

\hline \hline
\endlastfoot

   $i\lambda \star \partial_{\mu}M_{\mathcal{J}}$ &  $i\lambda \star \partial_{\mu}M_{\mathcal{J}}$ & \multirow{6}{*}{$i\lambda \star D_{\mu}M_{\mathcal{J}}$} \\ \cline{1-2}

    \color{mGold}$g \left(\partial^{\mathcal{K}}\lambda \star A_{\mu}\right) \star \partial_{\mathcal{K}}M_{\mathcal{J}} \, +$ & \multirow{2}{*}{$-g \lambda \star \left( A_{\mu} \star M_{\mathcal{J}} \right)$} & \\
    \color{mViolet}$-g \partial^{\mathcal{K}}A_{\mu} \star \left(\partial_{\mathcal{K}}\lambda \star M_{\mathcal{J}} \right)$ & & \\ \cline{1-2}
    
    $ig A_{\mu}^{\mathcal{K}} \star \left(\partial_{\mathcal{K}}\lambda  \star M_{\mathcal{J}} \right) \, +$ & \multirow{3}{*}{$ig\lambda \star \left(A_{\mu}^{\mathcal{K}} \star \partial_{\mathcal{K}}M_{\mathcal{J}} \right)$} & \\
    $ig A_{\mu}^{\mathcal{K}} \star \left( \lambda  \star \partial_{\mathcal{K}}M_{\mathcal{J}} \right)\, +$ & & \\
    $-ig \left( \partial_{\mathcal{K}}\lambda \star A_{\mu}^{\mathcal{K}}\right) \star M_{\mathcal{J}}$ &  &  \\ 

\pagebreak[4]
     
    $-ig A_{\mu} \star \left(\left(\mathcal{H}_{\mathcal{KJ}} - \eta_{\mathcal{KJ}}  \right)\star \partial^{\mathcal{K}}\lambda  \right)\, +$ & \multirow{3}{*}{$-ig\lambda \star \left( \partial^{\mathcal{K}} A_{\mu} \star \left(\mathcal{H}_{\mathcal{KJ}} - \eta_{\mathcal{KJ}}  \right)\right)$} & \multirow{3}{*}{$i\lambda \star D_{\mu}M_{\mathcal{J}}$} \\ 
    \color{mMulberry}$-ig \left(\partial^{\mathcal{K}}\lambda \star A_{\mu} \right)\star \left(\mathcal{H}_{\mathcal{KJ}} - \eta_{\mathcal{KJ}}  \right)     \, +$ & & \\
    \color{mMulberry}$-ig \left(\lambda \star \partial^{\mathcal{K}}A_{\mu}\right) \star \left(\mathcal{H}_{\mathcal{KJ}} - \eta_{\mathcal{KJ}}  \right)$ & & \\ \hline

    $\lambda_{\mathcal{K}} \star \partial^{\mathcal{K}}\partial_{\mu} M_{\mathcal{J}} $ & $\lambda_{\mathcal{K}} \star \partial^{\mathcal{K}}\partial_{\mu} M_{\mathcal{J}} $ & \multirow{6}{*}{$\lambda_{\mathcal{K}} \star \partial^{\mathcal{K}} D_{\mu}M_{\mathcal{J}}$} \\ \cline{1-2}

   $ig \left(\lambda^{\mathcal{K}} \star \partial_{\mathcal{K}}A_{\mu} \right) \star M_{\mathcal{J}} \, +$ & $ig \lambda_{\mathcal{K}} \star \left( \partial^{\mathcal{K}}A_{\mu} \star M_{\mathcal{J}}  \right) \, + $ & \\ 
   $ig A_{\mu} \star \left( \lambda_{\mathcal{K}}  \star \partial^{\mathcal{K}} M_{\mathcal{J}} \right)$ &  $ig \lambda_{\mathcal{K}} \star \left( A_{\mu} \star \partial^{\mathcal{K}}M_{\mathcal{J}} \right) $ & \\ \cline{1-2}

  $g A_{\mu}^{\mathcal{K}} \star \left(\lambda_{\mathcal{P}} \star \partial_{\mathcal{K}} \partial^{\mathcal{P}}M_{\mathcal{J}}\right)$ &  $g \lambda_{\mathcal{K}} \star \left( A_{\mu}^{\mathcal{P}} \star \partial^{\mathcal{K}} \partial_{\mathcal{P}}M_{\mathcal{J}}\right)$ & \\ \cline{1-2}

  $-g \left( \lambda^{\mathcal{P}} \star \partial^{\mathcal{K}} \partial_{\mathcal{P}}A_{\mu}\right) \star \left(\mathcal{H}_{\mathcal{KJ}} - \eta_{\mathcal{KJ}}  \right) \, +$ & $-g \lambda_{\mathcal{K}} \star \left( \partial^{\mathcal{K}} \partial_{\mathcal{P}}A_{\mu} \star \left(\mathcal{H}_{\mathcal{P}\mathcal{J}} - \eta_{\mathcal{P}\mathcal{J}}\right) \right)\, + $ & \\ 
  $-g  \partial^{\mathcal{K}} A_{\mu} \star \left( \lambda^{\mathcal{P}} \star \partial_{\mathcal{P}}\left(\mathcal{H}_{\mathcal{KJ}} - \eta_{\mathcal{KJ}}  \right) \right) $ & $-g\lambda_{\mathcal{K}} \star \left(\partial_{\mathcal{P}}A_{\mu} \star \partial^{\mathcal{K}}\left(\mathcal{H}_{\mathcal{P}\mathcal{J}} - \eta_{\mathcal{P}\mathcal{J}} \right) \right) $ & \\ \hline

  $-\partial_{\mu} \left(\mathcal{H}_{\mathcal{KJ}} - \eta_{\mathcal{KJ}}  \right) \star \partial^{\mathcal{K}}\lambda $ & $-\partial_{\mu} \mathcal{H}_{\mathcal{K}\mathcal{J}}  \star \partial^{\mathcal{K}}\lambda $ & \multirow{12}{*}{$-D_{\mu}\mathcal{H}_{\mathcal{K}\mathcal{J}}  \star \partial^{\mathcal{K}}\lambda $} \\ \cline{1-2}

  $-g A_{\mu}^{\mathcal{K}} \star \left( \partial_{\mathcal{K}} \left(\mathcal{H}_{\mathcal{P}\mathcal{J}} - \eta_{\mathcal{P}\mathcal{J}} \right) \star \partial^{\mathcal{P}}\lambda \right) \, + $ &  \multirow{3}{*}{$-g \left( A_{\mu}^{\mathcal{P}} \star \partial_{\mathcal{P}} \mathcal{H}_{\mathcal{K}\mathcal{J}}  \right) \star \partial^{\mathcal{K}}\lambda $} & \\
  $-g A_{\mu}^{\mathcal{K}} \star \left(  \left(\mathcal{H}_{\mathcal{P}\mathcal{J}} - \eta_{\mathcal{P}\mathcal{J}} \right) \star \partial_{\mathcal{K}} \partial^{\mathcal{P}}\lambda \right) \, + $ & & \\ 
  $g \left(\partial^{\mathcal{K}}\partial_{\mathcal{P}} \lambda \star A_{\mu}^{\mathcal{P}} \right) \star \left(\mathcal{H}_{\mathcal{KJ}} - \eta_{\mathcal{KJ}}  \right)   $ &  & \\ \cline{1-2}
  
  $-g \left(\lambda \star A_{\mu} \right) \star M_{\mathcal{J}} \, +$ &  \multirow{4}{*}{$-g \left( \partial_{\mathcal{K}}A_{\mu} \star M_{\mathcal{J}} \right) \star \partial^{\mathcal{K}}\lambda $} & \\ 
  $-g A_{\mu} \star \left( \lambda \star M_{\mathcal{J}} \right) \, +$ & & \\
  \color{mGold} $g \left(\partial^{\mathcal{K}}\lambda \star A_{\mu} \right) \star \partial_{\mathcal{K}}M_{\mathcal{J}} \, +$ & & \\
  \color{mViolet} $-g \partial^{\mathcal{K}}A_{\mu} \star \left(\partial_{\mathcal{K}}\lambda  \star M_{\mathcal{J}} \right)$ & &\\ \cline{1-2}

  \color{mMulberry}$-ig \left(\partial^{\mathcal{K}}\lambda \star A_{\mu} \right)\star \left(\mathcal{H}_{\mathcal{KJ}} - \eta_{\mathcal{KJ}}  \right) \, + $ & \multirow{2}{*}{$-ig \left( A_{\mu} \star \mathcal{H}_{\mathcal{K}\mathcal{J}}  \right) \star \partial^{\mathcal{K}}\lambda \, + $} & \\
  \color{mMulberry}$-ig \left( \lambda \star \partial^{\mathcal{K}} A_{\mu} \right)\star \left(\mathcal{H}_{\mathcal{KJ}} - \eta_{\mathcal{KJ}}  \right) \, + $ & & \\ 
  $-ig \partial^{\mathcal{K}}A_{\mu} \star \left( \lambda \star \mathcal{H}_{\mathcal{K}\mathcal{J}}  \right) $ & \multirow{2}{*}{$-ig \left( \partial^{\mathcal{P}}\partial_{\mathcal{K}}A_{\mu} \star \left(\mathcal{H}_{\mathcal{PJ}} - \eta_{\mathcal{PJ}}  \right) \right) \star \partial^{\mathcal{K}}\lambda $}  & \\ 
  $-ig \partial^{\mathcal{K}}A_{\mu} \star \left(\partial^{\mathcal{P}}\partial_{\mathcal{K}} \lambda \star \left(\mathcal{H}_{\mathcal{PJ}} - \eta_{\mathcal{PJ}}  \right) \right) $ & & \\ 
    
\hline

\end{longtable}

\end{center}

\subsubsection{Covariance of \texorpdfstring{
$D_{\mu}M_{\mathcal{KJ}}$}{DHkj}}

As we have emphasized, term-matching techniques for derivatives of $N=1$ scalars are, for the most part, analogous to those exemplified for $N=0$ fields. In this case, the term-by-term perspective produces
\begin{equation}\label{eq: deltaDmuMkjtbt}
\begin{aligned}
    \delta D_{\mu} M_{\mathcal{K}\mathcal{J}}  \big|_{\mathrm{t.b.t.}} = & \partial_{\mu} \left( \delta M_{\mathcal{K}\mathcal{J}}  \right) + g \left( \delta A_{\mu}^{\mathcal{P}} \star \partial_{\mathcal{P}}M_{\mathcal{K}\mathcal{J}}  + A_{\mu}^{\mathcal{P}} \star \partial_{\mathcal{P}}\left( \delta M_{\mathcal{K}\mathcal{J}} \right) \right)\\
    & - g \left( \partial_{\mathcal{K}} \left(\delta A_{\mu} \right) \star M_{\mathcal{J}} + \partial_{\mathcal{K}} A_{\mu} \star \delta M_{\mathcal{J}} \right)\\
    & + ig \left( \delta A_{\mu} \star M_{\mathcal{K}\mathcal{J}}  + A_{\mu} \star \delta M_{\mathcal{K}\mathcal{J}}  \right)\\
    &+ ig\left( \partial^{\mathcal{P}}\partial_{\mathcal{K}}(\delta A_{\mu}) \star M_{\mathcal{P}\mathcal{J}}  + \partial^{\mathcal{P}}\partial_{\mathcal{K}}A_{\mu} \star \delta M_{\mathcal{P}\mathcal{J}} \right)
\end{aligned}    
\end{equation}
while assuming covariance leads to
\begin{equation} \label{eq: deltaDmuMkjac}
\begin{aligned}
    \delta D_{\mu} M_{\mathcal{K}\mathcal{J}} \big|_{\mathrm{a.c.}} =& \lambda^{\mathcal{P}} \star \partial_{\mathcal{P}} \left(D_{\mu} M_{\mathcal{K}\mathcal{J}}  \right) + i \lambda \star D_{\mu} M_{\mathcal{K}\mathcal{J}}  \\
    & + i \partial^{\mathcal{P}}\partial_{\mathcal{K}}\lambda \star D_{\mu} M_{\mathcal{P}\mathcal{J}}  
    - \partial_{\mathcal{K}}\lambda \star D_{\mu}M_{\mathcal{J}} \, .
\end{aligned}    
\end{equation}
Notice here we are already dealing with the derivative of scalars $M_{\mathcal{K}\mathcal{J}} $, which is the part of \eqref{eq: DmuHkjUplift} involved in reproducing symmetry enhancements. Once more, the GKK expansion of these expressions is straightforward by \eqref{eq: star2fieldsModes}. We present representative, involved examples for the proxy enhancement point under consideration.

\paragraph{Symmetric Representation}
This representation encompasses some of the most insightful instances of term matching. The totality of \eqref{eq: deltaDmuMkjac} yields contributions in this representation. Many of these can either be found to have a direct correspondence with a term from the t.b.t.\ approach, or said correspondence can be achieved through the methods already discussed. This logic applies, for example, to the first eight rows of Table \ref{tab: DmuMkj}, which make up the entire first term in \eqref{eq: deltaDmuMkjac}.

The following example illustrates why some rows of Table 3 cannot be read as one-to-one correspondences: the correct matching may require combining different a.c.\ subterms before comparing with the t.b.t.\ expansion, and conversely a single t.b.t.\ structure may contribute to more than one a.c.\ block.

Let us now consider $\delta D_{\mu}\Mkj^{(\Lambda_{2})}$ for the mode-expanded, symmetry-point examples. As anticipated, an interesting scenario arises. Consider for example the a.c.\ term
\begin{equation}\label{eq: hsExample3_a}
    \delta D_{\mu}M_{\mathcal{K}\mathcal{J}} \big|_{\mathrm{a.c.}} \supset i\lambda \star D_{\mu}M_{\mathcal{K}\mathcal{J}}  \supset -ig \lambda \star \left(\partial_{\mathcal{K}}A_{\mu} \star M_{\mathcal{J}}\right)
\end{equation}
which after mode-expansion produces
\begin{equation}\label{eq: hsExample3_b}
    \delta D_{\mu}\Mkj^{(\Lambda_{2})}\big|_{\mathrm{a.c.}} \supset g \lambda^{(\alpha_{2})}A_{\mu}^{(-\alpha_{3})}\alpha_{3\hat{K}}\Mj^{(\Lambda_{1})} + g \lambda^{(\alpha_{2})}A_{\mu}^{(\alpha_{3})}\alpha_{3\hat{K}}\Mj^{(\Lambda_{6})} + \cdots
\end{equation}
Contrary to the cases we analysed for the $N=0$ scalars, here there is no combination of terms in the t.b.t.\ calculation that produces, for instance, the first 3-field product in \eqref{eq: hsExample3_b}. The closest we find is the product
\begin{equation}\label{eq: hsExample3_c}
    \delta D_{\mu}\Mkj^{(\Lambda_{2})}\big|_{\mathrm{t.b.t.}} \supset g \lambda^{(\alpha_{2})}A_{\mu}^{(-\alpha_{3})}\alpha_{1\hat{K}}\Mj^{(\Lambda_{1})} + \cdots
\end{equation}
which arises from the t.b.t.\ combination
\begin{equation}\label{eq: hsExample3_d}
    \delta D_{\mu}M_{\mathcal{K}\mathcal{J}} \big|_{\mathrm{t.b.t.}} \supset -g \partial_{\mathcal{K}} (\delta A_{\mu})\star M_{\mathcal{J}} \supset -ig \left(\partial_{\mathcal{K}}\lambda \star A_{\mu} + \lambda \star \partial_{\mathcal{K}}A_{\mu} \right) \star M_{\mathcal{J}}  \, .
\end{equation}
However, the combination of the a.c.\ term in  \eqref{eq: hsExample3_a} and another, arising from the same approach, 
\begin{equation}\label{eq: hsExample3_e}
\begin{aligned}
    \delta D_{\mu}M_{\mathcal{K}\mathcal{J}} \big|_{\mathrm{a.c.}} &\supset i\lambda \star D_{\mu}M_{\mathcal{K}\mathcal{J}}  + i\partial^{\mathcal{P}}\partial_{\mathcal{K}}\lambda \star D_{\mu}M_{\mathcal{P}\mathcal{J}}  \\
    & \supset -ig \lambda \star \left(\partial_{\mathcal{K}}A_{\mu} \star M_{\mathcal{J}} \right) -ig \partial^{\mathcal{P}}\partial_{\mathcal{K}}\lambda \star \left(\partial_{\mathcal{P}}A_{\mu} \star M_{\mathcal{J}} \right)
    \end{aligned}
\end{equation}
yields \eqref{eq: hsExample3_c} upon mode expansion. Nevertheless, the second 3-field product in \eqref{eq: hsExample3_b} does not arise from the t.b.t.\ combination in \eqref{eq: hsExample3_d}. At the t.b.t.\ level, it instead comes from yet another term,
\begin{equation}\label{eq: hsExample3_f}
    \delta D_{\mu}M_{\mathcal{K}\mathcal{J}} \big|_{\mathrm{t.b.t.}} \supset ig \partial^{\mathcal{P}}\partial_{\mathcal{K}}A_{\mu} \star \delta M_{\mathcal{P}\mathcal{J}}  \supset -ig \partial^{\mathcal{P}}\partial_{\mathcal{K}}A_{\mu} \star \left(\partial_{\mathcal{P}}\lambda \star M_{\mathcal{J}}  \right)  \, .
\end{equation}
So we find that the full correspondence is that of row 11 in Table \ref{tab: DmuMkj}. The dotted lines indicate $\star$-products arising from different terms in \eqref{eq: deltaDmuMkjac}. Colors once more indicate repeated terms yielding mixed contributions.

\paragraph{Adjoint Representation} Analysis of this representation in the case of derivatives of $N=1$ fields does not yield new information, as the mapping of terms from both approaches, as presented in Table \ref{tab: DmuMkj}, can be carried out via the symmetric alone. A key observation about this representation is that only the last term in \eqref{eq: deltaDmuMkjac} will yield contributions, due to momentum conservation. 

Taken together, these analyses account for all structures appearing in the
a.c.\ variation and identify their counterparts in the t.b.t.\ calculation, and vice versa, including those correspondences that require combining several terms before
mode expansion.  We conclude the transformation law for derivatives of $N=1$ fields as proposed in \eqref{eq: DmuHkjUplift} to be fully covariant and consistent with \eqref{eq: deltaMj}, \eqref{eq: deltaMkj}, \eqref{eq: deltaAmuGauge} and \eqref{eq: deltaAmuCartanGauge}, with the full dictionary of terms presented in Table \ref{tab: DmuMkj}.

\begin{center}
\begin{longtable}{|c|c|c|}
\caption{Covariance of $\delta D_{\mu}M_{\mathcal{KJ}} $. This table summarises how \emph{all} terms arising from $\delta D_{\mu}M_{\mathcal{KJ}} \big|_{\mathrm{t.b.t.}}$ combine to form $\delta D_{\mu}M_{\mathcal{KJ}}  \big|_{\mathrm{a.c.}}$. There are indeed no unmatched terms on either calculation, and so we claim covariance of \eqref{eq: DmuHkjUplift}. Notice that certain `t.b.t.'\ terms contribute to more than one `a.c.'\ term and vice versa. These are shown in color.}
\phantomsection\label{tab: DmuMkj}\\
\hline
 $\bm{\delta D_{\mu}M_{\mathcal{KJ}} }\big|_{\mathrm{t.b.t.}}$  & \multicolumn{2}{|c|}{$\bm{\delta D_{\mu}M_{\mathcal{KJ}} }\big|_{\mathrm{a.c.}}$  } \\ \hline
\endfirsthead
  
 \multicolumn{3}{c}%
{{ \tablename\ \thetable{} -- continued from previous page}} \\
\hline $\bm{\delta D_{\mu}M_{\mathcal{KJ}} }\big|_{\mathrm{t.b.t.}}$ &  \multicolumn{2}{|c|}{$\bm{\delta D_{\mu}M_{\mathcal{KJ}} }\big|_{\mathrm{a.c.}}$}   \\ \hline
\endhead
\hline \multicolumn{3}{|r|}{{Continued on next page}} \\ \hline
\endfoot

\hline \hline
\endlastfoot


$\lambda^{\mathcal{P}} \star \partial_{\mathcal{P}} \partial_{\mu} M_{\mathcal{KJ}} $ & $\lambda^{\mathcal{P}} \star \partial_{\mathcal{P}} \partial_{\mu} M_{\mathcal{KJ}} $ & \multirow{8}{*}{$\lambda^{\mathcal{P}}\star\partial_{\mathcal{P}}\left( D_{\mu}M_{\mathcal{KJ}} \right)$}  \\ \cline{1-2}

$g A_{\mu}^{\mathcal{P}} \star \left( \lambda^{\mathcal{Q}} \star \partial_{\mathcal{P}}\partial_{\mathcal{Q}} M_{\mathcal{KJ}}  \right)$ & $g \lambda^{\mathcal{P}} \star \left( A_{\mu}^{\mathcal{Q}} \star \partial_{\mathcal{P}}\partial_{\mathcal{Q}} M_{\mathcal{KJ}}  \right)$ &  \\ \cline{1-2}

$-g \left(\lambda^{\mathcal{P}} \star \partial_{\mathcal{K}}\partial_{\mathcal{P}}A_{\mu} \right)\star M_{\mathcal{J}}  $ &  $-g \lambda^{\mathcal{P}} \star \left( \partial_{\mathcal{P}}\partial_{\mathcal{K}}A_{\mu}\star M_{\mathcal{J}}  \right)$ &  \\ \cline{1-2}

$-g \partial_{\mathcal{K}}A_{\mu} \star \left( \lambda_{\mathcal{P}} \star \partial^{\mathcal{P}} M_{\mathcal{J}}  \right)$ &  $-g \lambda^{\mathcal{P}} \star \left( \partial_{\mathcal{K}}A_{\mu}\star \partial_{\mathcal{P}}M_{\mathcal{J}}  \right)$ &   \\ \cline{1-2}

$ig \left( \lambda^{\mathcal{P}} \star \partial_{\mathcal{P}}A_{\mu}\right)\star M_{\mathcal{KJ}}  $ & $ig \lambda^{\mathcal{P}} \star  \left( \partial_{\mathcal{P}}A_{\mu}\star M_{\mathcal{KJ}}  \right) $ & \\ \cline{1-2}

$ig A_{\mu} \star \left( \lambda^{\mathcal{P}}\star \partial_{\mathcal{P}} M_{\mathcal{KJ}} \right)$ & $ig \lambda^{\mathcal{P}}\star \left( A_{\mu} \star \partial_{\mathcal{P}} M_{\mathcal{KJ}}  \right)$ &  \\ \cline{1-2}

$ig \left( \lambda^{\mathcal{Q}} \star \partial^{\mathcal{P}}\partial_{\mathcal{K}}\partial_{\mathcal{Q}}A_{\mu}\right)\star M_{\mathcal{P}\mathcal{J}}  $ &  $ig \lambda^{\mathcal{P}} \star  \left( \partial_{\mathcal{P}}\partial^{\mathcal{Q}}\partial_{\mathcal{K}}A_{\mu}\star M_{\mathcal{QJ}} \right) $ &  \\ \cline{1-2}

$ig \partial^{\mathcal{P}}\partial_{\mathcal{K}}A_{\mu} \star \left( \lambda^{\mathcal{Q}}\star \partial_{\mathcal{Q}} M_{\mathcal{P}\mathcal{J}} \right)$ & $ig \lambda^{\mathcal{P}}\star \left( \partial^{\mathcal{Q}}\partial_{\mathcal{K}}A_{\mu} \star \partial_{\mathcal{P}} M_{\mathcal{QJ}} \right)$ &  \\ \cline{1-3}

$ i\lambda \star \partial_{\mu}M_{\mathcal{KJ}}  $ & $ i\lambda \star \partial_{\mu}M_{\mathcal{KJ}}  $ & \multirow{6}{*}{$i \lambda \star D_{\mu}M_{\mathcal{KJ}} $}  \\  \cline{1-2}

$ig A_{\mu}^{\mathcal{P}} \star \left( \partial_{\mathcal{P}}\lambda \star M_{\mathcal{KJ}} \right) \, +$ & \multirow{3}{*}{$ig \lambda \star  \left( A_{\mu}^{\mathcal{P}} \star \partial_{\mathcal{P}} M_{\mathcal{KJ}}  \right)$} &  \\

$ig A_{\mu}^{\mathcal{P}} \star \left( \lambda \star \partial_{\mathcal{P}}M_{\mathcal{KJ}} \right) \, +$  & &  \\ 

$-ig \left(\partial_{\mathcal{P}}\lambda \star A_{\mu}^{\mathcal{P}}\right) \star M_{\mathcal{KJ}}  $  & &  \\ \cline{1-2}

\color{mMulberry}{$-ig \left( \partial_{\mathcal{K}} \lambda \star A_{\mu}\right) \star M_{\mathcal{J}} $} $\, + $ & \multirow{2}{*}{\color{purple}{$-ig \lambda \star \left( \partial_{\mathcal{K}}A_{\mu} \star M_{\mathcal{J}} \right) \, +$}} &    \\

\color{mMulberry}{$-ig \left( \lambda \star \partial_{\mathcal{K}}A_{\mu}\right) \star M_{\mathcal{J}} $} $\, + $ & &  \\ \cdashline{2-3}

$-ig \partial^{\mathcal{P}}\partial_{\mathcal{K}} A_{\mu}  \star \left( \partial_{\mathcal{P}}\lambda   \star M_{\mathcal{J}} \right)$ &  \color{purple}{$-ig \partial^{\mathcal{P}}\partial_{\mathcal{K}}\lambda \star \left( \partial_{\mathcal{P}}A_{\mu}\star M_{\mathcal{J}} \right)$} & $i \partial^{\mathcal{P}}\partial_{\mathcal{K}}\lambda \star \left( D_{\mu} M_{\mathcal{P}\mathcal{J}} \right)$  \\ \cline{1-2}\cdashline{3-3}


$g \left(\partial^{\mathcal{P}}\lambda \star A_{\mu} \right)\star \partial_{\mathcal{P}} M_{\mathcal{KJ}}  \,\, +$ &  \multirow{2}{*}{$-g \lambda \star \left( A_{\mu} \star M_{\mathcal{KJ}}  \right) $} & \multirow{3}{*}{$i \lambda \star D_{\mu}M_{\mathcal{KJ}} $} \\

$-g \left(  \lambda \star A_{\mu} \right) \star M_{\mathcal{KJ}}  $  & &  \\ \cline{1-2}


\color{mGold}{$g \partial_{\mathcal{K}}A_{\mu} \star \left( (\mathcal{H}_{\mathcal{PJ}}  - \eta_{\mathcal{PJ}}) \star \partial^\mathcal{P}\lambda\right) \,\, +$}  & $-g \lambda \star \left(\partial^{\mathcal{P}}\partial_{\mathcal{K}}A_{\mu} \star M_{\mathcal{P}\mathcal{J}}  \right) \,\, +$ &   \\ \cdashline{2-3}

\color{mGold}{$-g \left(\partial^{\mathcal{P}}\partial_{\mathcal{K}}\lambda \star A_{\mu}\right) \star M_{\mathcal{P}\mathcal{J}}  \,\, +$}& \multirow{2}{*}{$-g \partial^{\mathcal{P}}\partial_{\mathcal{K}}\lambda \star \left(A_{\mu} \star M_{\mathcal{P}\mathcal{J}}  \right) \,\, +$} & \multirow{4}{*}{$i \partial^{\mathcal{P}}\partial_{\mathcal{K}}\lambda \star \left( D_{\mu} M_{\mathcal{P}\mathcal{J}} \right)$}  \\

\color{mGold}{$-g \left(\partial_{\mathcal{K}}\lambda \star \partial^{\mathcal{P}}A_{\mu}\right) \star M_{\mathcal{P}\mathcal{J}}  \,\, +$}  &  &  \\ 

\color{mGold}{$-g \left(\partial^{\mathcal{P}}\lambda \star \partial_{\mathcal{K}}A_{\mu}\right) \star M_{\mathcal{P}\mathcal{J}}  \,\, +$} &\multirow{2}{*}{ $-g \partial^{\mathcal{P}}\partial_{\mathcal{K}}\lambda \star \left(\partial^{\mathcal{Q}}\partial_{\mathcal{P}}A_{\mu} \star M_{\mathcal{QJ}} \right) $} &   \\

\color{mGold}{$-g \left(\lambda \star \partial^{\mathcal{P}}\partial_{\mathcal{K}}A_{\mu}\right) \star M_{\mathcal{P}\mathcal{J}}  $} & &  \\ \hline


$-\partial_{\mathcal{K}} \lambda \star \partial_{\mu}M_{\mathcal{J}}  $  & $-\partial_{\mathcal{K}} \lambda \star \partial_{\mu}M_{\mathcal{J}}  $ & \multirow{15}{*}{$-\partial_{\mathcal{K}}\lambda \star D_{\mu}M_{\mathcal{J}} $} \\ \cline{1-2}

\color{mMulberry}{$-ig \left(\partial_{\mathcal{K}}\lambda \star A_{\mu}\right) \star M_{\mathcal{J}}  \, +$} &  \multirow{3}{*}{$-ig \partial_{\mathcal{K}}\lambda \star \left( A_{\mu} \star M_{\mathcal{J}}  \right)  $} &  \\

\color{mMulberry}{$-ig \left(\lambda \star \partial_{\mathcal{K}}A_{\mu}\right) \star M_{\mathcal{J}}  \, +$}  & &  \\

$-ig \partial_{\mathcal{K}} A_{\mu} \star \left( \lambda \star M_{\mathcal{J}}  \right) $  & &  \\ \cline{1-2}

$-g A_{\mu}^{\mathcal{P}} \star \left(  \partial_{\mathcal{P}}\partial_{\mathcal{K}} \lambda \star  M_{\mathcal{J}} \right) \,\, + $  & \multirow{3}{*}{$-g \partial_{\mathcal{K}}\lambda \star \left( A_{\mu}^{\mathcal{P}} \star \partial_{\mathcal{P}}M_{\mathcal{J}}  \right) $} &  \\

$-g A_{\mu}^{\mathcal{P}} \star \left(  \partial_{\mathcal{K}} \lambda \star \partial_{\mathcal{P}}  M_{\mathcal{J}}  \right) \, + $  & &  \\

$g  \left(\partial_{\mathcal{K}}  \partial_{\mathcal{P}} \lambda \star   A_{\mu}^{\mathcal{P}} \right) \star   M_{\mathcal{J}} $  & &  \\ \cline{1-2}

{\color{mGold}$g \partial_{\mathcal{K}}A_{\mu} \star  (\left(\mathcal{H}_{\mathcal{PJ}} - \eta_{\mathcal{PJ}} ) \star \partial^{\mathcal{P}}\lambda\right) \,\, + $} & \multirow{8}{*}{$g \partial_{\mathcal{K}}\lambda \star \left( \partial^{\mathcal{P}}A_{\mu} \star \left(\mathcal{H}_{\mathcal{PJ}} - \eta_{\mathcal{PJ}} \right) \right)$} &  \\

$-g A_{\mu} \star \left(\partial^{\mathcal{P}}\partial_{\mathcal{K}}\lambda \star M_{\mathcal{P}\mathcal{J}}  \right) \,\, +$  & &  \\

\color{mGold}{$-g \left(\partial^{\mathcal{P}}\partial_{\mathcal{K}}\lambda \star A_{\mu} \right) \star M_{\mathcal{P}\mathcal{J}}  \,\, + $}  & &  \\

\color{mGold}{$-g \left(\partial_{\mathcal{K}}\lambda \star \partial^{\mathcal{P}}A_{\mu} \right) \star M_{\mathcal{P}\mathcal{J}}  \,\, + $}  & &  \\
\color{mGold}{$-g \left(\partial^{\mathcal{P}}\lambda \star \partial_{\mathcal{K}}A_{\mu} \right) \star M_{\mathcal{P}\mathcal{J}}  \,\, + $} & &  \\
\color{mGold}{$-g \left(\lambda \star \partial^{\mathcal{P}}\partial_{\mathcal{K}}A_{\mu} \right) \star M_{\mathcal{P}\mathcal{J}}  \,\, + $}  & &  \\
$-g \partial^{\mathcal{P}}\partial_{\mathcal{K}}A_{\mu} \star \left(\lambda\star M_{\mathcal{P}\mathcal{J}}  \right) \,\, +$  & &  \\
$-g \partial^{\mathcal{P}}\partial_{\mathcal{K}}A_{\mu} \star \left(\partial^{\mathcal{Q}}\partial_{\mathcal{P}}\lambda\star M_{\mathcal{QJ}} \right)$  & &  \\ \cline{1-3}


$i\partial^{\mathcal{P}}\partial_{\mathcal{K}}\lambda \star \partial_{\mu}M_{\mathcal{P}\mathcal{J}} $ & $i\partial^{\mathcal{P}}\partial_{\mathcal{K}}\lambda \star \partial_{\mu}M_{\mathcal{P}\mathcal{J}} $ & \multirow{5}{*}{$i \partial^{\mathcal{P}}\partial_{\mathcal{K}}\lambda\star \left( D_{\mu}M_{\mathcal{P}\mathcal{J}} \right)$} \\ \cline{1-2}

$ig A_{\mu}^{\mathcal{P}} \star \left( \partial_{\mathcal{P}}\partial^{\mathcal{Q}}\partial_{\mathcal{K}}\lambda \star M_{\mathcal{QJ}}\right) \,\, +$ &  \multirow{4}{*}{$ig \partial^{\mathcal{P}}\partial_{\mathcal{K}} \lambda \star \left(A_{\mu}^{\mathcal{Q} } \star \partial_{\mathcal{Q}}M_{\mathcal{P}\mathcal{J}}  \right)$} &  \\

$ig A_{\mu}^{\mathcal{P}} \star \left( \partial^{\mathcal{Q}}\partial_{\mathcal{K}}\lambda \star \partial_{\mathcal{P}}M_{\mathcal{QJ}}\right)\,\, +$  & &  \\

$-ig \left(\partial^{\mathcal{P}} \partial_{\mathcal{K}}\partial_{\mathcal{Q}}\lambda \star A_{\mu}^{\mathcal{Q}}\right) \star M_{\mathcal{P}\mathcal{J}}  \,\, +$  & &  \\

$-ig \left( \partial_{\mathcal{K}}\partial_{\mathcal{Q}}\lambda \star \partial^{\mathcal{P}}A_{\mu}^{\mathcal{Q}}\right) \star M_{\mathcal{P}\mathcal{J}}  \,\, +$  & &  \\

$-ig \left( \partial^{\mathcal{P}}\partial_{\mathcal{Q}}\lambda \star \partial_{\mathcal{K}}A_{\mu}^{\mathcal{Q}}\right) \star M_{\mathcal{P}\mathcal{J}}  \,\, +$  & \multirow{2}{*}{$ig \partial^{\mathcal{P}}\partial_{\mathcal{K}} \lambda \star \left(A_{\mu}^{\mathcal{Q} } \star \partial_{\mathcal{Q}}M_{\mathcal{P}\mathcal{J}}  \right)$} & \multirow{2}{*}{$i \partial^{\mathcal{P}}\partial_{\mathcal{K}}\lambda\star \left( D_{\mu}M_{\mathcal{P}\mathcal{J}} \right)$}  \\

$-ig \left( \partial_{\mathcal{Q}}\lambda \star \partial^{\mathcal{P}}\partial_{\mathcal{K}}A_{\mu}^{\mathcal{Q}}\right) \star M_{\mathcal{P}\mathcal{J}} $  & &  \\ \cline{1-2}\cdashline{3-3}

\color{mMulberry}{$-ig \left(\partial_{\mathcal{K}}\lambda \star A_{\mu} \right) \star M_{\mathcal{J}}  \,\, +$}  & \multirow{2}{*}{\color{purple}{$-ig \lambda \star \left(\partial_{\mathcal{K}}A_{\mu} \star M_{\mathcal{J}}  \right)$}} & \multirow{2}{*}{$i \lambda \star D_{\mu}M_{\mathcal{KJ}} $}  \\
\color{mMulberry}{$-ig \left(\lambda \star \partial_{\mathcal{K}}A_{\mu} \right) \star M_{\mathcal{J}} \, \, +$}  & &  \\ \cdashline{1-3}

$-ig A_{\mu} \star \left(\partial_{\mathcal{K}}\lambda \star M_{\mathcal{J}}  \right)$  & \color{purple}{$-ig \partial^{\mathcal{P}}\partial_{\mathcal{K}}\lambda \star \left(\partial_{\mathcal{P}}A_{\mu} \star M_{\mathcal{J}}  \right)$} & $i \partial^{\mathcal{P}}\partial_{\mathcal{K}}\lambda\star \left( D_{\mu}M_{\mathcal{P}\mathcal{J}} \right)$ \\
\hline
\end{longtable}
\end{center}

\section{Hidden algebra of GKK modes}\label{sec: hidden_algebra_GKK}

The previous section showed that the proposed $\star$-product transformations reproduce the expected non-Abelian gauge transformations after GKK expansion at enhancement points, and that they act covariantly on the derivative structures entering the truncated action. We now isolate the algebraic structure carried by the GKK labels themselves. These labels are precisely those mapped to roots of the enhanced gauge algebra at the chosen enhancement point; away from that point the corresponding fields are generically massive, but the underlying momentum and winding data remain well defined. This makes it possible to ask whether these modes still carry an algebraic remnant of the enhanced symmetry, and how this structure can be expressed in a DFT-like language.

The point is that the relevant momentum-carrying modes behave, algebraically, as additional frame directions associated with the enhanced generators. This provides a natural bridge to gauged DFT \cite{Aldazabal:2011,Geissbuhler:2013,Grana:2012}: the effective enlargement of the generalized tangent space is realized here by Fourier modes of the doubled fields rather than postulated as independent tangent directions. We make this statement precise by constructing generalized-frame elements associated with Cartan and momentum-carrying modes, computing their bracket, and showing that the resulting structure constants are controlled by lattice data and $\star$-product phases, with a Jacobi identity that can be followed away from the enhancement locus. This parallels the enhanced tangent-space constructions of
\cite{Aldazabal:2016,Aldazabal:2017,Aldazabal:2017b,Cagnacci:2017}, and more broadly, generalized-geometric extensions of the heterotic tangent bundle by gauge degrees of freedom \cite{Coimbra:2014qaa}, although here the additional directions are realized as retained GKK Fourier modes.

\subsection{Generalized frames from GKK modes}\label{subsec: gkk_frames}

We first introduce the generalized frame elements associated with the Cartan directions and with the GKK modes that become charged generators at an enhancement point. We denote by
\begin{equation}
    r_L = r+16
\end{equation}
the number of left Cartan directions. We follow the index conventions established in section \ref{sec: Senh}, namely, Left indices are written as
\begin{equation}
    \hat I = (I, m), \qquad m=1,\ldots,r, \qquad I=1,\ldots,16 ,
\end{equation}
while right indices are denoted by
\begin{equation}
    \bar I = 1,\ldots,r .
\end{equation}

At an enhancement point $\Phi_0$, the GKK modes $\mathbb P_c$ that become massless satisfy
\begin{equation}
\label{eq: kLEqualsRoot}
    k_R(\mathbb P_c;\Phi_0)=0, 
    \qquad 
    \kappa_L(\mathbb P_c;\Phi_0)=\alpha_c ,
\end{equation}
where $\alpha_c$ is a root of the enhanced algebra. Away from the enhancement point, we keep the same lattice labels $\mathbb P_c$, but in general $k_R(\mathbb P_c;\Phi)\neq 0$. In what follows, $\alpha_c$ is used as a label inherited from the enhancement point.

We consider an extended frame space with index
\begin{equation}
    M=(\hat I,c1,c2,\bar I),
    \qquad c=1,\ldots,n_c.
\end{equation}
where $c$ labels the momentum-carrying directions associated with the retained GKK modes corresponding to positive roots of the enhanced algebra. Thus a vector in this frame has the schematic form
\begin{equation}
    V^M=(V^{\hat I};V^{c1};V^{c2};V^{\bar I}),
\end{equation}
We denote by
\begin{equation}
    D_M=(D_{\hat I},D_{c1},D_{c2},D_{\bar I})
\end{equation}
the corresponding basis elements. Since the physical doubled coordinates are only those associated with the left and right compact directions, the derivative has no component along the $2n_{c}$ directions:
\begin{equation}
    \partial_M = \left(\partial_{\hat I},0,0,\partial_{\bar I}\right) .
    \label{eq: frame_derivative}
\end{equation}

The Cartan frame elements are
\begin{align}
    H_{\hat I} &= H_{\hat I}{}^M D_M
    = -i\,\delta_{\hat I}{}^M D_M ,
    \label{eq: left_cartan_frame}
    \\
    H_{\bar I} &= H_{\bar I}{}^M D_M
    = -i\,\delta_{\bar I}{}^M D_M .
    \label{eq: right_cartan_frame}
\end{align}
The momentum-carrying frame elements are defined by
\begin{equation}
     E_{\pm\alpha_c}(\pm\mathbb P_c)
    =\frac{i}{\sqrt2}
      e^{\pm i(\kappa_L(\mathbb P_c)\cdot y_L
                    +k_R(\mathbb P_c)\cdot y_R)}
      (D_{c1}\pm iD_{c2})\, .
    \label{eq: charged_frame_element}
\end{equation}
To lighten notation, we will often suppress the explicit $\mathbb P_c$ dependence and write simply $E_{\pm\alpha_c}$. In the frame basis introduced above, we take the metric to be
\begin{equation}
 \eta_{MN}=
    \begin{pmatrix}
        \delta_{\hat I\hat J} & 0 & 0 & 0 \\
        0 & \delta_{cd} & 0 & 0 \\
        0 & 0 & \delta_{cd} & 0 \\
        0 & 0 & 0 & -\delta_{\bar I\bar J}
    \end{pmatrix}.
\label{eq: eta_LR}
\end{equation}
corresponding to an extension of the usual \(O(r_L,r)\) metric by a positive-definite block for the retained momentum-carrying directions, while aligned with the standard convention in which left directions have positive norm and right directions have negative norm. The momentum-carrying directions are treated as left directions, since at the enhancement point they become the charged generators of the enhanced left algebra.

\subsection{Generalized bracket and algebra away from enhancement points}
\label{subsec: gkk_bracket_algebra}

We now compute the algebraic relations between the frame elements introduced above. The bracket is defined through a generalized Lie derivative supplemented by an extra term acting on the momentum-carrying directions,
\begin{equation}\label{eq: gkk_LieDerivativeAlgebra}
\left(\tilde{\mathcal{L}}_{E_A}E_B \right)^M
=
E_A{}^N\partial_N E_B{}^M
-
E_B{}^N\partial_N E_A{}^M
+
\partial^M E_A{}^P E_B{}^Q\eta_{PQ}
+
\Omega_{AB}{}^C E_C{}^M .
\end{equation}
Here \(E_A\) denotes any of the frame elements \(H_{\hat I}\), \(H_{\bar I}\), or \(E_{\pm\alpha_c}\). We define
\begin{equation}\label{eq: gkk_bracket_definition}
    [E_A,E_B] \equiv \tilde{\mathcal{L}}_{E_A}E_B \, .
\end{equation}
Deformed brackets such as \eqref{eq: gkk_bracket_definition} can be considered in the broader setting of twisted C-bracket
structures in gauged DFT. Related algebraic structures for $O(D,D+n)$ gauged DFT have
recently been studied from the viewpoint of extended doubled algebroids in \cite{Mori:2024}.

Related deformations of the generalized Lie derivative were introduced in
\cite{Cagnacci:2017, Fraiman:2018} to account for the cocycle factors of the current algebra. In the present construction this role is played by the $\star$-product phases
$\tilde f$ associated with the retained GKK labels. The tensor \(\Omega_{AB}{}^C\) is taken to vanish whenever one or more of the indices is a Cartan index. When all three indices are momentum-carrying, we define
\begin{equation}
\Omega_{AB}{}^C
=
(-1)^n
\tilde f_{\mathbb K_C\mathbb K_B\mathbb K_A}
\delta_{\mathbb K_C,\mathbb K_A+\mathbb K_B},
\label{eq: gkk_omega_definition}
\end{equation}
Here $n$ counts negative-root labels among $A,B,C$, and
$\tilde f_{\mathbb K_C\mathbb K_B\mathbb K_A}$ is defined in \eqref{eq: phaseGral}. For related appearances of current-algebra structures in heterotic gauged DFT, see for example \cite{Hatsuda:2023}.

At an enhancement point, the retained GKK labels are mapped to roots of the enhanced gauge algebra. For two charged modes whose labels satisfy \(\mathbb P_c=\mathbb P_a+\mathbb P_b\), the phase \eqref{eq: phaseGral} is instrumental in supplying the correct sign for Cartan--Weyl structure constants. With our conventions,
\begin{equation}
\label{eq: structureConsts}
    f_{\alpha_a\alpha_b}{}^{\alpha_c}
    =
    (-1)^n
    \tilde f_{\mathbb K_c\mathbb K_b\mathbb K_a},
    \qquad
    \alpha_c=\alpha_a+\alpha_b \, .
\end{equation}
Let us once more stress that equality \eqref{eq: kLEqualsRoot} holds only at the enhancement point. However, as $\tilde{f}$ is calculated solely with retained GKK labels $\mathbb P$, its value does not change with displacements around moduli space, and so $\tilde f_{\mathbb K_c\mathbb K_b\mathbb K_a} = \tilde f_{\mathbb \alpha_c\mathbb \alpha_b\mathbb \alpha_a}$ in \eqref{eq: structureConsts}. Thus the \(\Omega\)-term \eqref{eq: gkk_omega_definition} is the off-enhancement continuation of the charged-root part of the enhanced gauge algebra. The remaining Cartan--root brackets are instead produced by the ordinary generalized Lie derivative terms.

The nonzero brackets arising from this proposal are
\begin{align}
\left[E_{\alpha_a},E_{\alpha_b}\right]
&=
(-1)^n
\tilde f_{\mathbb K_c\mathbb K_b\mathbb K_a}
\delta_{\mathbb K_c,\mathbb K_a+\mathbb K_b}
E_{\alpha_c},
\label{eq: gkk_algebra_summary_1}
\\
\left[E_{\alpha_c},E_{-\alpha_c}\right]
&=
\kappa_L^{(\alpha_c)\hat I}H_{\hat I}
-
k_R^{(\alpha_c)\bar I}H_{\bar I},
\label{eq: gkk_algebra_summary_2}
\\
\left[H_{\hat I},E_{\alpha_c}\right]
&=
\kappa_L^{(\alpha_c)\hat I}E_{\alpha_c},
\label{eq: gkk_algebra_summary_3}
\\
\left[H_{\bar I},E_{\alpha_c}\right]
&=
k_R^{(\alpha_c)\bar I}E_{\alpha_c}\, ,
\label{eq: gkk_algebra_summary_4}
\end{align}
where we have used $\alpha_{c}$ as a label inherited from enhancement. For the case of the bracket between two different momentum-carrying frame elements, \(E_{\alpha_a}\) and \(E_{\alpha_b}\), with \(\alpha_{a}\neq \pm \alpha_{b}\), the first three terms in \eqref{eq: gkk_LieDerivativeAlgebra} vanish: the two frame elements have nonzero entries in different orthogonal two-planes, and there are no derivatives along these directions. The nonzero bracket is therefore encoded in the \(\Omega\)-term, which yields \eqref{eq: gkk_algebra_summary_1}.
Relations \eqref{eq: gkk_algebra_summary_2}, \eqref{eq: gkk_algebra_summary_3} and \eqref{eq: gkk_algebra_summary_4} can be obtained by direct substitution from the usual generalized Lie derivative, namely the first three terms in \eqref{eq: gkk_LieDerivativeAlgebra}. 

These brackets hold at generic moduli. At enhancement, \(k_R=0\) and \(\kappa_L=\alpha\), recovering the usual gauge algebra \cite{Aldazabal:2017, Aldazabal:2017b}.

\subsection{Jacobi identity away from enhancement points}

It is possible to check that the bracket defined above satisfies the Jacobi identity away from the enhancement point. For three generators \(X,Y,Z\), we use
\begin{equation}
    [X,[Y,Z]]+[Y,[Z,X]]+[Z,[X,Y]]=0 \, .
    \label{eq: jacobi_identity}
\end{equation}
The point is not that the modes remain massless away from the enhancement locus, but rather that the algebraic relations obtained in \eqref{eq: gkk_algebra_summary_1}--\eqref{eq: gkk_algebra_summary_4} continue to satisfy \eqref{eq: jacobi_identity} when the same GKK lattice labels are retained.

As a representative example, consider
\begin{equation}
    X=E_{\alpha_a},\qquad Y=E_{\alpha_b},\qquad Z=H_{\hat I}.
\end{equation}
Using \eqref{eq: gkk_algebra_summary_1} and \eqref{eq: gkk_algebra_summary_3}, the three terms in \eqref{eq: jacobi_identity} are
\begin{align}
[E_{\alpha_a},[E_{\alpha_b},H_{\hat I}]]
&=
-\kappa _L^{(\alpha_b)\hat I}
(-1)^n
\tilde f_{\mathbb K_c\mathbb K_b\mathbb K_a}
\delta_{\mathbb K_c,\mathbb K_a+\mathbb K_b}
E_{\alpha_c}\, ,
\label{eq: jacobi_example_1}
\\
[E_{\alpha_b},[H_{\hat I},E_{\alpha_a}]]
&=
-\kappa_L^{(\alpha_a)\hat I}
(-1)^n
\tilde f_{\mathbb K_c\mathbb K_b\mathbb K_a}
\delta_{\mathbb K_c,\mathbb K_a+\mathbb K_b}
E_{\alpha_c}\, ,
\label{eq: jacobi_example_2}
\\
[H_{\hat I},[E_{\alpha_a},E_{\alpha_b}]]
&=
\kappa_L^{(\alpha_c)\hat I}
(-1)^n
\tilde f_{\mathbb K_c\mathbb K_b\mathbb K_a}
\delta_{\mathbb K_c,\mathbb K_a+\mathbb K_b}
E_{\alpha_c}\ , 
\label{eq: jacobi_example_3}
\end{align}
where in \eqref{eq: jacobi_example_2} we used the antisymmetry of the phase \eqref{eq: phaseGral} under exchange of the last two entries, $\tilde f_{\mathbb K_c\mathbb K_a\mathbb K_b} = - \tilde f_{\mathbb K_c\mathbb K_b\mathbb K_a} $. If \(\mathbb K_c\neq \mathbb K_a+\mathbb K_b\), all three terms vanish because of the Kronecker delta. If \(\mathbb K_c=\mathbb K_a+\mathbb K_b\), their sum is proportional to
\begin{equation}
    -\kappa_L^{(\alpha_b)\hat I}
    -
    \kappa_L^{(\alpha_a)\hat I}
    +
    \kappa_L^{(\alpha_c)\hat I} \, .
\end{equation}
This vanishes because the left and right momenta are linear functions of the underlying lattice vector, and thus
\begin{equation}
    \mathbb K_c=\mathbb K_a+\mathbb K_b
    \qquad\Longrightarrow\qquad
    \kappa_L^{(\alpha_c)}=\kappa_L^{(\alpha_a)}+\kappa_L^{(\alpha_b)}
\end{equation}
at any point in moduli space. The same argument applies to the right Cartan generators, with \(\kappa_L\) replaced by \(k_R\).

Other choices of generators are analogous. In particular, when brackets of the form \([E_{\alpha},E_{-\alpha}]\) appear, the relevant contractions are made with the extended metric \eqref{eq: eta_LR} and can be rewritten in terms of the underlying momentum and winding data. Thus the cancellations required by the Jacobi identity are controlled by the retained GKK labels and the \(\star\)-product phases, not by the special condition \(k_R=0\). We therefore obtain a consistent algebraic structure for the retained modes away from the enhancement point.

\subsection{Rotated frame and retained algebraic structure}
\label{subsec: retained_algebraic_structure}

The algebraic relations \eqref{eq: gkk_algebra_summary_1}--\eqref{eq: gkk_algebra_summary_4} are written in terms of the left and right momenta \(\kappa_L,k_R\), and therefore their coefficients appear to depend on the moduli. But this dependence is merely a result of the choice of Cartan basis. The underlying labels of the retained modes are the integer GKK data
\begin{equation}
    \mathbb P^B=(P^I,p_m,\tilde p^m),
\end{equation}
which are independent of the point in moduli space. The left and right momenta are obtained from these labels by a moduli-dependent \(O(r_L,r)\) rotation,
\begin{equation}
    \mathbb L^P(\Phi)=\mathcal R^P{}_{B}(\Phi)\,\mathbb P^B \, ,
\end{equation}
where $\mathcal{R}(\Phi)$ is, for the zero-Wilson-line backgrounds considered in our explicit examples, the moduli-dependent matrix
\begin{equation}\label{eq: rotacion}
    \mathcal{R}(\Phi) =  \begin{pmatrix}
        \delta^{I}_{J} & 0 & 0 \\
        0 & g^{mn} & \frac{1}{2}\left(\delta^{m}_{n} - g^{mr}b_{rn} \right) \\
        0 & g^{mn} & -\frac{1}{2}\left(\delta^{m}_{n} + g^{mr}b_{rn} \right)  \\
    \end{pmatrix} \, .
\end{equation}
More details about this change of basis can be found in Appendix \ref{ap: HeteroticStringBasics}.
Equivalently, the coefficients appearing in the Cartan brackets can be collected into the vector defined in \eqref{eq: Kgral},
\begin{equation}
    \mathbb K^S(\mathbb P;\Phi)
    =
    \left(K^{Q}_{L},k_L^{m}(\mathbb P;\Phi),k_R^{\bar I}(\mathbb P;\Phi)\right),
\end{equation}
which is linear in the retained GKK label:
\begin{equation}
    \mathbb K^S(\mathbb P;\Phi)
    =
    \mathcal U^S{}_{B}(\Phi)\,\mathbb P^B .
    \label{eq:K_U_P}
\end{equation}
Here \(\mathcal U(\Phi)\) includes both the moduli-dependent rotation from lattice data to left/right momenta \eqref{eq: rotacion}, and the change to the root basis used for the enhanced algebra. Explicitly, $\mathcal{U}^S{}_{B}(\Phi) = \bm{\alpha}^{S}\,_{P}\mathcal{R}^P{}_{B}(\Phi)$, where
\begin{equation}
    \bm{\alpha} = 
 \begin{pmatrix}
        \tilde{\alpha}^{Q}\,_{I} & 0 & 0 \\
        0 & \alpha^{m}\,_{k} & 0 \\     
        0 & 0  & \alpha^{\bar{I}}\,_{k} \\
    \end{pmatrix} \, .
\end{equation}
On the upper-left block, the above defined matrix contains $\Lambda_{16}$ roots enumerated by $I$ and with Euclidean component $Q$, and the will-be roots of the enhancement point $\alpha^{m}\,_{k}$ in the remaining left and right blocks. Away from the enhancement point, the latter are not roots of the gauge algebra of the massless states, but remain the fixed basis inherited from the enhancement point.
In terms of structure, $\bm{\alpha}^{S}\,_{P}$ has a row index $S$ which tells us about the Euclidean root component, and a column index $P$ labelling the simple root.

Let us also collect the Cartan frame elements as
\begin{equation}
    \mathbb H^S=(H^{I},H^{m},H^{\bar I})\, .
\end{equation}
Then the last three brackets in \eqref{eq: gkk_algebra_summary_1}--\eqref{eq: gkk_algebra_summary_4} can be written compactly as
\begin{align}
    [E_{\alpha_c},E_{-\alpha_c}]
    &=
    \mathbb K_c^S\eta_{SR}\mathbb H^R,
    \label{eq:pre_rot_EE}
    \\
    [\mathbb H^S,E_{\alpha_c}]
    &=
    \mathbb K_c^S E_{\alpha_c}\, ,
    \label{eq:pre_rot_HE}
\end{align}
with contractions performed using the metric \eqref{eq: eta_LR}.
Using \eqref{eq:K_U_P}, we define a rotated Cartan basis by
\begin{equation}
    \mathcal H_B(\Phi)
    \equiv
    \mathcal U^S{}_{B}(\Phi)\,\eta_{SR}\,\mathbb H^R .
    \label{eq:rotated_cartan_definition}
\end{equation}


Equivalently, since the lattice index is $B=(I,{}_n,{}^{n})$, and $\mathbb P^B=(P^I,p_n,\tilde p^{\,n})$, the rotated Cartan generators naturally decompose as $\mathcal H_B = \left( \mathcal H_I,\mathcal H^n,\mathcal H_n
    \right)$. For vanishing Wilson lines, as is our consideration, the heterotic component is simply $\mathcal H_I = \tilde\alpha^{Q}{}_{I}H_Q$, while the compact toroidal components are
\begin{equation}
    \begin{pmatrix}
        \mathcal{H}^{n} \\
        \mathcal{H}_{n}
    \end{pmatrix}
    =
    \begin{pmatrix}
        g^{nk}
        \left(
        (\alpha_L)_{k}{}^{m}H_{m}
        -
        (\alpha_R)_{k}{}^{\bar I}H_{\bar I}
        \right)
        \\
        \frac{1}{2}
        \left[
        \left(
        (\alpha_L)_{n}{}^{m}
        -
        g^{kr}b_{rn}(\alpha_L)_{k}{}^{m}
        \right)H_{m}
        +
        \left(
        (\alpha_R)_{n}{}^{\bar I}
        +
        g^{kr}b_{rn}(\alpha_R)_{k}{}^{\bar I}
        \right)H_{\bar I}
        \right]
    \end{pmatrix}\, .
    \label{eq:rotated_cartan_explicit}
\end{equation}
The placement of the indices follows the ordering of the lattice vector \(\mathbb P^B\): \(\mathcal H_I\) is conjugate to the heterotic lattice component \(P^I\), \(\mathcal H^n\) to the momentum component \(p_n\), and \(\mathcal H_n\) to the winding component \(\tilde p^{\,n}\).

In this basis,
\begin{equation}
    [E_{\alpha_c},E_{-\alpha_c}]
    =
    \mathbb P_c^B \mathcal H_B .
    \label{eq:rotated_EE}
\end{equation}

The action of the rotated Cartan generators on the momentum-carrying frame elements is obtained in the same way:
\begin{equation}
\begin{aligned}
   \relax[\mathcal{ H}_B,E_{\alpha_c}]
    &=
    \mathcal U^S{}_{B}\eta_{SR}
    [\mathbb H^R,E_{\alpha_c}]
    \\
    &=
    \mathcal U^S{}_{B}\eta_{SR}
    \mathcal U^R{}_{C}\mathbb P_c^C
    E_{\alpha_c}.
\end{aligned}
\end{equation}
The combination
\begin{equation}
    \tilde\eta_{BC}
    \equiv
    \mathcal U^S{}_{B}\eta_{SR}\mathcal U^R{}_{C}
    \label{eq: lattice_metric_tilde}
\end{equation}
is the rotated invariant metric on the GKK lattice. In the basis
\(\mathbb P^B=(P^I,p_m,\tilde p^m)\), it takes the standard Narain form
\begin{equation}
    \tilde\eta_{BC}
    =
    \begin{pmatrix}
        \delta_{IJ} & 0 & 0 \\
        0 & 0 & \delta_m{}^n \\
        0 & \delta^m{}_n & 0
    \end{pmatrix}.
    \label{eq: narain_metric_lattice_basis}
\end{equation}
Thus
\begin{equation}
    [\mathcal H_B,E_{\alpha_c}]
    =
    \tilde\eta_{BC}\mathbb P_c^C E_{\alpha_c}.
    \label{eq: rotated_HE}
\end{equation}

It is then natural to rewrite the momentum-carrying bracket directly in terms of the underlying GKK labels,
\begin{equation}
    [E_{\alpha_a},E_{\alpha_b}]
    =
    (-1)^n
    \tilde f_{\mathbb P_c\mathbb P_b\mathbb P_a}
    \delta_{\mathbb P_c,\mathbb P_a+\mathbb P_b}
    E_{\alpha_c}\, .
    \label{eq: rotated_EE_different}
\end{equation}
Notice that this relabelling does not affect the definition in \eqref{eq: phaseGral}.
The algebraic structure of the retained modes therefore takes the moduli-independent form
\begin{align}
    [E_{\alpha_a},E_{\alpha_b}]
    &=
    (-1)^n
    \tilde f_{\mathbb P_c\mathbb P_b\mathbb P_a}
    \delta_{\mathbb P_c,\mathbb P_a+\mathbb P_b}
    E_{\alpha_c},
    \label{eq: retained_algebra_1}
    \\
    [E_{\alpha_c},E_{-\alpha_c}]
    &=
    \mathbb P_c^B\mathcal H_B,
    \label{eq: retained_algebra_2}
    \\
    [\mathcal H_B,E_{\alpha_c}]
    &=
    \tilde\eta_{BC}\mathbb P_c^C E_{\alpha_c}.
    \label{eq: retained_algebra_3}
\end{align}
The continuous moduli have been absorbed into the definition of the rotated Cartan generators \(\mathcal H_B(\Phi)\), while the structure constants are controlled by the GKK lattice labels, the invariant Narain metric, and the \(\star\)-product phases. At the enhancement point, this reduces to the usual Cartan--Weyl algebra of the enhanced left gauge group. Away from that point, the corresponding fields are generically massive, but the algebraic structure associated with the retained GKK modes remains well defined and, when written in this form, remains exactly the same as it was at the enhancement point itself.

\subsection{Example: Algebra around an
\texorpdfstring{$\mathrm{SU}(3)_L$}{SU(3)L} enhancement point} \label{sec: SU3_AlgebraAwayEnhPoint}

Let us illustrate the previous construction for the modes retained around an \(\mathrm{SU(3)}_L\) enhancement point, as analysed in Section \ref{sec: su3example}. We use the same conventions for roots and GKK labels collected in Appendix \ref{ap: SU3facts}. The retained momentum-carrying frame elements are $ E_{\pm \alpha_1},\, E_{\pm \alpha_2},\, E_{\pm \alpha_3}$, with $ \alpha_3=\alpha_1+\alpha_2$. At the enhancement point, \(k_R=0\) and \(\kappa_L=\alpha\), and the algebra
\eqref{eq: gkk_algebra_summary_1}--\eqref{eq: gkk_algebra_summary_4} reduces to the usual Cartan--Weyl form,
\begin{align}
    [E_{\alpha_a},E_{\alpha_b}]
    &=
    (-1)^n
    \tilde f_{\mathbb K_c\mathbb K_b\mathbb K_a}
    \delta_{\mathbb K_c,\mathbb K_a+\mathbb K_b}
    E_{\alpha_c},
    \label{eq:su3_enhanced_1}
    \\
    [E_{\alpha_c},E_{-\alpha_c}]
    &=
    \alpha_c^i H_i,
    \label{eq:su3_enhanced_2}
    \\
    [H_i,E_{\pm\alpha_c}]
    &=
    \pm \alpha_c^i E_{\pm\alpha_c}.
    \label{eq:su3_enhanced_3}
\end{align}
Thus the retained frame elements reproduce the \(\mathrm{SU(3)}_L\) algebra at the enhancement point. While keeping the same GKK labels, we now move away from this particular choice of $\Phi$. In this example the enhancement comes from the compact \(T^2\) momentum--winding sector, so in the Cartan basis introduced in \eqref{eq:rotated_cartan_definition}, and for clarity, we only display the compact components \((\mathcal H^n,\mathcal H_n)\) of the full rotated Cartan vector \(\mathcal H_B=(\mathcal H_I,\mathcal H^n,\mathcal H_n)\). Under these considerations, the Cartan generators are $\mathcal H_B = \left( \mathcal H^1,\mathcal H^2,\mathcal H_1,\mathcal H_2\right)$, and the bracket \eqref{eq: retained_algebra_2} gives
\begin{align}
    [E_{\alpha_1},E_{-\alpha_1}]
    &=
    \mathcal H^1+\mathcal H_1,
    \label{eq:su3_rotated_1}
    \\
    [E_{\alpha_2},E_{-\alpha_2}]
    &=
    -\mathcal H^1+\mathcal H^2+\mathcal H_2,
    \label{eq:su3_rotated_2}
    \\
    [E_{\alpha_3},E_{-\alpha_3}]
    &=
    \mathcal H^2+\mathcal H_1+\mathcal H_2 \, ,
    \label{eq:su3_rotated_3}
\end{align}
while the action of the rotated Cartan generators \eqref{eq: retained_algebra_3} is equivalently written as
\begin{equation}
    [\mathcal H^n,E_{\alpha}]
    =
    \tilde p^{(\alpha)n}E_{\alpha},
    \qquad
    [\mathcal H_n,E_{\alpha}]
    =
    p^{(\alpha)}_n E_{\alpha},
    \label{eq:su3_rotated_cartan_action}
\end{equation}
with the corresponding signs reversed for \(E_{-\alpha}\). Therefore the structure is still controlled by the same integer GKK labels, even though the left and right momenta are no longer those of massless gauge bosons.

It is useful to perform one further change of Cartan basis. We define
\begin{align}
    \tilde{\mathcal H}_1
    &=
    \frac{1}{\sqrt{2}}
    \left(
    \mathcal H^1+\mathcal H_1
    \right),
    \label{eq:su3_final_cartan_1}
    \\
    \tilde{\mathcal H}_2
    &=
    \sqrt{\frac{2}{3}}
    \left(
    -\frac{1}{2}\mathcal H^1
    +\frac{1}{2}\mathcal H_1
    +\mathcal H^2
    +\mathcal H_2
    \right),
    \label{eq:su3_final_cartan_2}
    \\
    \tilde{\mathcal H}_{1R}
    &=
    \mathcal H^1-\mathcal H_1-\mathcal H^2,
    \label{eq:su3_right_cartan_1}
    \\
    \tilde{\mathcal H}_{2R}
    &=
    \mathcal H^1-\mathcal H_1-\mathcal H_2.
    \label{eq:su3_right_cartan_2}
\end{align}
In this basis, the nontrivial brackets become
\begin{align}
    [E_{\alpha_a},E_{\alpha_b}]
    &=
    (-1)^n
    \tilde f_{\mathbb P_c\mathbb P_b\mathbb P_a}
    \delta_{\mathbb P_c,\mathbb P_a+\mathbb P_b}
    E_{\alpha_c},
    \label{eq:su3_final_algebra_1}
    \\
    [E_{\alpha_c},E_{-\alpha_c}]
    &=
    \alpha_c^1 \tilde{\mathcal H}_1
    +
    \alpha_c^2 \tilde{\mathcal H}_2,
    \label{eq:su3_final_algebra_2}
    \\
    [\tilde{\mathcal H}_i,E_{\pm\alpha_c}]
    &=
    \pm \alpha_c^i E_{\pm\alpha_c},
    \label{eq:su3_final_algebra_3}
\end{align}
while
\begin{equation}
    [\tilde{\mathcal H}_{1R},E_{\pm\alpha_c}]
    =
    [\tilde{\mathcal H}_{2R},E_{\pm\alpha_c}]
    =
    0 \, .
    \label{eq:su3_right_cartans_commute}
\end{equation}
In this rotated form we now use the underlying GKK labels \(\mathbb P\), rather than the moduli-dependent momenta \(\mathbb K(\mathbb P;\Phi)\), as detailed in the previous section. Thus, after the rotated change of basis, the retained algebra separates into the usual \(\mathrm{SU(3)}\) Cartan--Weyl algebra together with two commuting abelian generators. Away from the enhancement point, this should not be interpreted as an unbroken non-Abelian gauge symmetry of massless fields. Rather, it shows that the GKK modes retained around the \(\mathrm{SU(3)}_L\) point continue to carry the same algebraic structure, with the extra abelian directions corresponding to the right-sector Cartan generators.

\section{Summary and Outlook} \label{sec: Remarks_and_Outlook}

Throughout this work, we revisited the interpolating action proposed in \cite{Aldazabal:2018uzm}, which is a truncated field theory description of heterotic toroidal compactifications, written in terms of doubled internal coordinates, and with a background-independent form. This construction, written in $d$ spacetime dimensions, captures symmetry enhancement-breaking effects by preserving fields with $\bar{N}=0$ and $N=0,1$ that can be understood as modes of a GKK mode-expansion in the internal double torus. Different subsets of these become massless at special points in moduli space. The target space of this construction displays a non-commutative nature, introduced by a $\star$-product that encodes stringy KK momenta and winding modes. This product can be viewed as a doubled-coordinate implementation of the cocycle phases of compact string vertex operators, related to the non-commutativity of closed string zero modes \cite{Sakamoto:1989,Freidel:2017a,Freidel:2017b}.

Building on these results, we presented candidate transformations for the scalar and vector fields relevant to the enhancement sector in the interpolating action, written in the unexpanded double-coordinate formulation and making use of the $\star$-product. These proposals successfully reproduce ordinary non-Abelian gauge transformations at enhancement points in moduli space after a GKK mode-expansion is performed, in the same spirit that the derivative structures in \eqref{eq: DmuMjUplift}, \eqref{eq: DmuHkjUplift}, \eqref{eq: FmunuChargedUplift} and \eqref{eq: FmunuCartanUplift} were shown in \cite{Aldazabal:2018uzm} to reproduce the expected covariant derivatives. We illustrated this extension by further elaborating on the $T^{2}$-compactification -- leading to an $\mathrm{SU(3)}$ enhancement -- studied in the original work \cite{Aldazabal:2018uzm}, and also considered an enhancement to $\mathrm{SU(4)}$, stemming from a $T^{3}$ compactification, and further demonstrating the robustness of the candidate transformations. These variations also display the feature of being enough, just like the derivative proposals, to determine the physical combinations of $N=1$ fields that complete the representations of the enhanced gauge groups. In particular, the two examples analyzed in Section \ref{subsec: Examples} suggest a normalization pattern for the physical combinations of $SU(r+1)$, $N=1$ massive fields, expressible in terms of the stringy phases $\tilde{f}$ and the matrix elements of the relevant generators. This evidence merits further investigation. Our examples fall within the standard $A_r\simeq \mathfrak{su}(r+1)$ enhancement points of toroidal compactifications, but they do not exhaust the possible enhancement patterns, which more generally depend on the choice of Narain lattice point and may involve other simply-laced algebras or products thereof \cite{Narain:1986,Narain:1987,Giveon:1994, Font:2020rsk}.

The core question addressed in this work is whether these $\star$-product transformations constitute, or are at least part of, an underlying hidden symmetry structure of the interpolating action. In this direction, we showed that the unexpanded derivatives of scalar fields $D_{\mu}M_{\cal J}$ and $D_{\mu}\mathcal{H}_{\mathcal{KJ}}$ do transform covariantly. A key point in this analysis is, however, that covariance is not apparent from a naive term comparison between the two natural computations, i.e. varying the derivative term by term, $D_\mu\delta$, and assuming a covariant transformation law for the derivative, $\delta D_\mu$.
To arrive at the result, we must keep track of the origin of each $\star$-product factor, the allowed GKK support of each field, and the projection onto physical representations after mode expansion. The comparison between the term-by-term ($D_{\mu}\delta$) variation and the assuming-covariance ($\delta D_{\mu}$) variation shows that all terms can be matched, sometimes only after using the representation-dependent constraints imposed by momentum conservation. 

Finally, by leveraging generalized-frame elements associated with the Cartan directions and with the momentum-carrying modes that become charged generators at an enhancement point, we isolated the algebraic structure carried by the retained GKK labels themselves.
This construction is closely related to the tangent-space enlargement used in DFT descriptions of gauge enhancement at fixed points in moduli space \cite{Aldazabal:2016,Aldazabal:2017,Aldazabal:2017b,Cagnacci:2017}.
The resulting brackets reproduce the Cartan--Weyl algebra at the enhancement point, while away from it the same underlying GKK labels continue to define a consistent, preserved algebraic structure. The Jacobi identity follows from the linearity of the map from GKK labels to left/right momenta and from the $\star$-product phase properties. By rotating the Cartan basis, the apparent moduli dependence of the Cartan brackets can be absorbed into the definition of the Cartan generators, leaving a moduli-independent algebra written in terms of the underlying lattice labels $\mathbb P^B$ and the Narain metric.  In this sense, the rotation plays a role analogous to a choice of frame: the moduli-dependent left/right momentum components are traded for generators adapted to the moduli-independent GKK lattice data.

Overall, these results sharpen the interpretation of the interpolating action given in \cite{Aldazabal:2018uzm}. Previous DFT descriptions of enhancement were naturally adapted to a chosen enhancement point, often through an enlargement of the tangent space or through fluxes that reproduce the enhanced gauge algebra at that point
\cite{Aldazabal:2016,Aldazabal:2017,Aldazabal:2017b,Cagnacci:2017}. The present analysis suggests that, in the interpolating formulation, the same algebraic data are instead encoded in the GKK mode labels and in the $\star$-product phases. The non-commutative $\star$-product, required to reproduce cubic string-amplitude data and the signs of the enhanced gauge algebra \cite{Aldazabal:2018uzm}, also allows us to organize candidate field transformations, covariant derivative structures, and a preserved algebra of GKK labels away from the enhancement locus. 

Several limitations remain. The construction is intrinsically truncated: a complete string-theoretic description would require the full tower of oscillator modes and spins. In particular, the sector under consideration should not be understood as closed under the full string interactions, as operator products or string three-point couplings can in general involve states with higher oscillator number, higher spin, or higher $\alpha'$ corrections \cite{Bedoya:2014,Hohm:2015} that lie outside the construction. Thus the truncation is best viewed as the first step of a possible level-by-level organization of the string spectrum, rather than as a complete effective theory by itself. Moreover, while the transformations proposed here reproduce the expected gauge transformations after mode expansion and on the enhancement locus, and make the relevant derivative structures covariant even in the unexpanded formulation, a complete proof of invariance of the full truncated action remains to be established. The examples studied provide nontrivial evidence, but a systematic treatment of arbitrary enhancement groups, representations, and Wilson-line backgrounds is still open. For an in-depth classification of gauge-symmetry-enhancement patterns in heterotic toroidal compactifications, see for example \cite{Fraiman:2018, Font:2020rsk}.

Natural extensions of our results include proving closure of the proposed transformations, checking the full variation of the truncated action, and systematically extending the construction to higher oscillator levels, where higher-spin
fields and higher $\alpha'$ corrections are also expected to be relevant \cite{Green:1987,Bedoya:2014,Hohm:2015,Eloy:2020,Hronek:2022, Baron:2018, HsiaKamalWulff2025, Ciafardini:2024ujx}. One possible systematic route is an iterative Noether procedure, in which higher-oscillator fields are added level by level and their interactions and gauge transformations are deformed order by order to cancel the variation of the action, with residual terms indicating which additional sectors are required for consistency. This has a parallel in
Exceptional Field Theory, where gauge-algebra closure requires a
tensor hierarchy~\cite{Aldazabal:2013TensorHierarchy}. The problem of closure is naturally related to homotopy-algebraic formulations of gauge symmetries, and in particular to $L_\infty$ descriptions of gauged DFT and enhanced DFT \cite{Hohm:2017pnh, Lescano:2021}.  Another natural direction is the inclusion of fermions, either through a supersymmetric
extension of the present construction or through its relation to heterotic DFT descriptions \cite{Hohm:2011a,Jeon:2011c,Jeon:2012,Lescano:2021b}.
It would also be important to derive the observed normalization of physical massive fields in a representation-independent way, and to formulate the preserved GKK algebra for general enhancement groups and Wilson-line backgrounds. 
Clarifying the precise sense in which the unexpanded doubled-coordinate formulation is
background independent, and how it relates to string-field-theoretic approaches to
target-space duality \cite{Kugo:1992md,Hohm:2010}, remain open questions. Recent developments in duality-covariant heterotic
geometry may provide useful language for this problem \cite{Hassler:2024curv}.
Finally, the connection with gauged DFT deserves a more systematic treatment. More
ambitiously, one may ask whether the retained-frame construction admits an
infinite-dimensional completion, in which all GKK labels that can participate in
enhancement phenomena are treated as frame-like directions. Such a formulation would
provide a unified counterpart of the finite tangent-space enlargements used at
individual enhancement points.

\newpage

\appendix
\section{Some Heterotic string basics}\label{ap: HeteroticStringBasics}

We summarize here some string-theory ingredients needed in the body of the article. We mainly focus on the \(\mathrm{Spin(32)}\) heterotic string. For a heterotic string compactified to \(d\) space-time dimensions, let $r=10-d$ be the number of torus-compact directions. We collect the moduli-independent GKK data into the lattice vector
\begin{equation}\label{eq:gkk_lattice_vector}
    \mathbb P^B
    =
    \left(P^I,p_m,\tilde p^{\,m}\right),
    \qquad
    I=1,\ldots,16,
    \qquad
    m=1,\ldots,r .
\end{equation}
Here \(P^I\) denotes the heterotic gauge-lattice component, while \(p_m\) and \(\tilde p^{\,m}\) are the KK momenta and winding numbers, respectively. The corresponding moduli-dependent left- and right-moving momentum coordinates are collected into
\begin{equation}\label{eq: generalizedmomentum}
    \mathbb L^P(\mathbb P;\Phi)
    =
    \left(L_L^I,l_L^m,l_R^m\right),
\end{equation}
where \(L_L^I\) are the heterotic left-moving components and \(l_L^m,l_R^m\) are the compact left- and right-moving components. These coordinates define a vector on the Narain lattice \(\Gamma_{26-d,10-d}\), of signature \((26-d,10-d)\).

Writing \(\Phi=(g,b,A)\) for the moduli, with \(g_{mn}\) the internal metric, \(b_{mn}\) the antisymmetric tensor and \(A_m^I\) the Wilson lines, the components of \(\mathbb L^P(\mathbb P;\Phi)\) are
\begin{equation}\label{eq: leftrightmomenta}
\begin{aligned}
L_L^I
&=
P^I + R A_n^I \tilde p^{\,n},
\\
l_L^m
&=
\frac{\sqrt{\alpha'}}{2}
\left[
\frac{\tilde p^{\,m}}{\tilde R}
+
2g^{mn}
\left(
\frac{p_n}{R}
-
\frac{1}{2}b_{nr}\frac{\tilde p^{\,r}}{\tilde R}
\right)
-
P^I A_I^m
-
\frac{R}{2}A_I^m A_n^I\tilde p^{\,n}
\right],
\\
l_R^m
&=
\frac{\sqrt{\alpha'}}{2}
\left[
-\frac{\tilde p^{\,m}}{\tilde R}
+
2g^{mn}
\left(
\frac{p_n}{R}
-
\frac{1}{2}b_{nr}\frac{\tilde p^{\,r}}{\tilde R}
\right)
-
P^I A_I^m
-
\frac{R}{2}A_I^m A_n^I\tilde p^{\,n}
\right].
\end{aligned}
\end{equation}
Equivalently,
\begin{equation}\label{eq:narain_rotation}
    \mathbb L^P(\mathbb P;\Phi)
    =
    \mathcal R^P{}_{B}(\Phi)\,\mathbb P^B .
\end{equation}
The rotation matrix corresponding to \eqref{eq: leftrightmomenta} is
\begin{equation}\label{eq:general_rotation_matrix}
\mathcal R(\Phi)
=
\begin{pmatrix}
\delta^I{}_J
&
0
&
R A_n^I
\\[0.4em]
-\frac{\sqrt{\alpha'}}{2}A_J^m
&
\frac{\sqrt{\alpha'}}{R}g^{mn}
&
\frac{\sqrt{\alpha'}}{2\tilde R}
\left(
\delta^m{}_n
-
g^{mr}b_{rn}
-
\frac{\alpha'}{2}A_I^m A_n^I
\right)
\\[0.4em]
-\frac{\sqrt{\alpha'}}{2}A_J^m
&
\frac{\sqrt{\alpha'}}{R}g^{mn}
&
-\frac{\sqrt{\alpha'}}{2\tilde R}
\left(
\delta^m{}_n
+
g^{mr}b_{rn}
+
\frac{\alpha'}{2}A_I^m A_n^I
\right)
\end{pmatrix}.
\end{equation}
Thus \(\mathbb P\) labels the GKK mode and is independent of the point in moduli space, while \(\mathbb L(\mathbb P;\Phi)\) gives the corresponding moduli-dependent momentum coordinates.

The notation \(\mathbb L^P\) should not be confused with the Euclidean root-space quantities used in the Cartan--Weyl algebra. When a root-space basis is chosen, one further changes basis according to
\begin{equation}\label{eq:K_alpha_L_appendix}
    \mathbb K^S(\mathbb P;\Phi)
    =
    \bm{\alpha}^S{}_{P}\,
    \mathbb L^P(\mathbb P;\Phi)
    =
    \mathcal U^S{}_{B}(\Phi)\,\mathbb P^B .
\end{equation}
Here \(\bm{\alpha}\) is the root-basis matrix introduced in Section \ref{sec: hidden_algebra_GKK}. In this notation, \(\mathbb K = (\kappa_{L}, k_{R}) \equiv (K_{L}, k_{L}, k_{R})\) gives the components that enter the Cartan--Weyl brackets, while \(\mathbb L\) denotes left/right Narain momentum coordinates.

In the examples considered in the main text we set the Wilson lines to zero and fix the reference scale of the torus parametrization by
\begin{equation}\label{eq: self_dual_radius_choice}
    \tilde R=\frac{\alpha'}{R},
    \qquad
    R=\sqrt{\alpha'} .
\end{equation}
Here $R$, typically referred to as the compactification radius, sets a common length scale, and $\widetilde R$ is its dual. The torus geometry is specified by the metric $g_{mn}$, which encodes the lengths and angles of the torus cycles in these units, while $b_{mn}$ specifies the antisymmetric
Kalb--Ramond background entering the momentum-winding spectrum.

The mass formulas for string states are
\begin{equation}\label{eq: LRstringmasses}
\begin{aligned}
   \frac{\alpha'}{2}m_L^2
   &=
   \frac{1}{2}\kappa_L^2 + N - 1,
   \\
   \frac{\alpha'}{2}m_R^2
   &=
   \frac{1}{2}k_R^2 + \bar N .
\end{aligned}
\end{equation}
Here \(N=N_B\), while $\bar N=\bar N_B+\bar N_F+\bar E_0$ . The quantities \(N_B\) and \(\bar N_B\) are the left- and right-moving bosonic oscillator numbers, \(\bar N_F\) is the right-moving fermionic oscillator number, and
\[
    \bar E_0=
    \begin{cases}
        -\frac{1}{2}, & \text{NS sector},\\
        0, & \text{R sector}.
    \end{cases}
\]
The level-matching condition is \(m_L^2=m_R^2\), or equivalently
\begin{equation}\label{LMCwidings}
    \frac{1}{2}\mathbb K^2
    \equiv
    \frac{1}{2}\left(\kappa_L^2-k_R^2\right)
    =
    \tilde p^{\,m}p_m+\frac{1}{2}P^2
    =
    1-N+\bar N .
\end{equation}

In our discussion we restrict to
\[
    \bar N_B=0,
    \qquad
    \bar N_F=\frac{1}{2},
    \qquad
    \bar E_0=-\frac{1}{2},
\]
so that \(\bar N=0\). The charged-vector sector corresponds to \(N=0\), and therefore satisfies
\[
    \mathbb K^2=2 .
\]
Massless charged vectors occur at enhancement points, where
\[
    k_R(\mathbb P;\Phi_0)=0,
    \qquad
    \frac{1}{2}\kappa_L(\mathbb P;\Phi_0)^2=1 .
\]
Equivalently, after the change to the root-space basis described in \eqref{eq:K_alpha_L_appendix}, the corresponding left-moving momentum is identified with a root of the enhanced algebra\footnote{
With our conventions, the roots are normalized as \(\alpha^2=2\). In the examples, the dimensionless coordinates of the left-moving momenta in the chosen simple-root basis coincide, at the enhancement point, with the coordinates of roots or weights of the corresponding representation.
}.

As is well known, there are \(10-d+16\) left gauge bosons: \(16\) Cartan gauge bosons of the original heterotic gauge algebra, associated with vertex-operator factors of the form $\partial_z Y^I \tilde\psi^\mu$, together with \(10-d\) left gauge bosons coming from the compact metric and antisymmetric tensor components, associated with $\partial_z Y^m \tilde\psi^\mu $. The \(10-d\) right-moving combinations, associated with $\partial_z X^\mu \tilde\psi^m, \, m=1,\ldots,10-d$, generate the right Abelian group. These Cartan-sector states have vanishing GKK label, $ \mathbb P=0$, and hence $\kappa_L=0, \, k_R=0$.

\section{The \texorpdfstring{$\bm{\star}$-product}{star-product}}\label{ap: star_product}

\subsection{Definition}

A $\star$-product of this type was proposed in \cite{Freidel:2017a} in order to incorporate, in a doubled-coordinate description, the cocycle phases of compact string vertex operators. We first recall the bosonic part of the construction, namely the part involving only KK momenta and windings along the compact torus. Starting from the full GKK label used in Appendix \ref{ap: HeteroticStringBasics}, $\mathbb P^B=(P^I,p_m,\tilde p^{\,m})$, we denote by $\check{\mathbb P}=(p_m,\tilde p^{\,m})$ its $O(r,r)$ compact torus component. The corresponding doubled compact coordinate is $\check{\mathbb Y}=(y^m,\tilde y_m)$.

As in the main text, we label modes by the moduli-dependent momenta $\check{\mathbb K}=\check{\mathbb K}(\check{\mathbb P};\Phi)$. The $\star$-product phase, however, is computed from the underlying lattice data $\check{\mathbb P}$.

For two fields expanded as
\begin{equation}
    \phi_i(x,\check{\mathbb Y})
    =
    \sum_{\check{\mathbb K}_i}
    \phi_i^{(\check{\mathbb K}_i)}(x)
    e^{i\check{\mathbb K}_i\cdot\check{\mathbb Y}},
    \quad
    \check{\mathbb K}_i=\check{\mathbb K}(\check{\mathbb P}_i;\Phi),
\end{equation}
the $\star$-product is defined by
\begin{equation}
\label{eq: starproductFourier}
\begin{aligned}
(\phi_1\star\phi_2)(x,\check{\mathbb Y})
&=
\sum_{\check{\mathbb K}_1,\check{\mathbb K}_2}
e^{i\pi p_{1m}\tilde p_2^{\,m}}
\phi_1^{(\check{\mathbb K}_1)}(x)
\phi_2^{(\check{\mathbb K}_2)}(x)
e^{i\check{\mathbb K}_{12}\cdot\check{\mathbb Y}},
\end{aligned}
\end{equation}
where $p_{1m}$ and $\tilde p_2^{\,m}$ are read from the underlying labels $\check{\mathbb P}_1$ and $\check{\mathbb P}_2$, and
\begin{equation}
    \check{\mathbb K}_{12}
    =
    \check{\mathbb K}(\check{\mathbb P}_1+\check{\mathbb P}_2;\Phi)
    =
    \check{\mathbb K}_1+\check{\mathbb K}_2 \, .
\end{equation}

The phase in \eqref{eq: starproductFourier} makes the product generically non-commutative. Its origin can be traced to the non-commutativity of the compact string coordinate zero modes; see for instance \cite{Sakamoto:1989,Freidel:2017b}.

\subsection{Associativity and projected products}

The $\star$-product in \eqref{eq: starproductFourier} is associative on the full Fourier-expanded field space. Indeed,
\begin{equation}
\label{eq: assocStarProduct}
\begin{aligned}
\left((\phi_1\star\phi_2)\star\phi_3\right)(x,\check{\mathbb Y})
&=
\sum_{\check{\mathbb K}_1,\check{\mathbb K}_2,\check{\mathbb K}_3}
e^{i\pi(p_{1m}+p_{2m})\tilde p_3^{\,m}}
e^{i\pi p_{1m}\tilde p_2^{\,m}}
\phi_1^{(\check{\mathbb K}_1)}(x)
\phi_2^{(\check{\mathbb K}_2)}(x)
\phi_3^{(\check{\mathbb K}_3)}(x)
e^{i\check{\mathbb K}_{123}\cdot\check{\mathbb Y}}
\\
&=
\sum_{\check{\mathbb K}_1,\check{\mathbb K}_2,\check{\mathbb K}_3}
e^{i\pi p_{1m}(\tilde p_2^{\,m}+\tilde p_3^{\,m})}
e^{i\pi p_{2m}\tilde p_3^{\,m}}
\phi_1^{(\check{\mathbb K}_1)}(x)
\phi_2^{(\check{\mathbb K}_2)}(x)
\phi_3^{(\check{\mathbb K}_3)}(x)
e^{i\check{\mathbb K}_{123}\cdot\check{\mathbb Y}}
\\
&=
\left(\phi_1\star(\phi_2\star\phi_3)\right)(x,\check{\mathbb Y}) .
\end{aligned}
\end{equation}

Here, we have defined
\begin{equation}
    \check{\mathbb K}_{123}
    =
    \check{\mathbb K}(\check{\mathbb P}_1+\check{\mathbb P}_2+\check{\mathbb P}_3;\Phi)
    =
    \check{\mathbb K}_1+\check{\mathbb K}_2+\check{\mathbb K}_3 \, .
\end{equation}
This proof is unaffected if the fields themselves have Fourier support only on modes satisfying the level-matching condition appropriate to their oscillator number, provided the product is still evaluated as a product on the full Fourier-expanded field space. In other words, one may restrict the fields appearing in the product, but one must not project the intermediate products onto a smaller set of modes and still expect associativity.

To make this distinction explicit, let $\Pi_{\mathcal S}$ denote a projection onto some restricted set of GKK labels or field components, and define a projected product by
\begin{equation}
    \phi\star_{\mathcal S}\psi
    \equiv
    \Pi_{\mathcal S}(\phi\star\psi).
\end{equation}
In general,
\begin{equation}
    (\phi_1\star_{\mathcal S}\phi_2)\star_{\mathcal S}\phi_3
    \neq
    \phi_1\star_{\mathcal S}(\phi_2\star_{\mathcal S}\phi_3).
\end{equation}
Thus the failure is not a failure of the $\star$-product itself, but of the product obtained after imposing intermediate projections.

This point appears explicitly in the comparison between the term-by-term and assuming-covariance calculations of Section \ref{sec: covariance}. Consider, for instance, the term
\begin{equation}
\label{eq: ex_ApB_a}
 \delta D_{\mu}M_{\mathcal K\bar J}\big|_{\mathrm{t.b.t.}}
 \supset
 -g\,\partial_{\mathcal K}(\delta A_{\mu})\star M_{\bar J}
 \supset
 -ig\,(\partial_{\mathcal K}\lambda\star A_{\mu})\star M_{\bar J}.
\end{equation}

In the following example, the superscripts $\alpha_i$ and $\Lambda_s$ denote the enhancement-point left momenta $\check{\mathbb K}$ of the corresponding modes; the underlying GKK labels are implicit. We focus on the variation of the mode with weight $\Lambda_2$ in the $\bm 6_{(-\frac{2}{3},-\frac{1}{3})}$ representation of $\mathrm{SU(3)}$. After mode expansion, and keeping only the labels relevant to the support argument, \eqref{eq: ex_ApB_a} gives
\begin{equation}
\label{eq: ex_ApB_b}
    \begin{aligned}
     \delta D_{\mu}M_{\mathcal K\bar J}^{(\Lambda_{2})}\big|_{\mathrm{t.b.t.}}
     &\supset
     -g\left[
     \partial_{\mathcal K}(\delta A_{\mu}^{(\alpha_{1})})M_{\bar J}^{(\Lambda_{3})}
     +
     \partial_{\mathcal K}(\delta A_{\mu}^{(-\alpha_{1})})M_{\bar J}^{(\Lambda_{1})}
     \right]
     \\
     &\supset
     -ig\left[
     (\partial_{\mathcal K}\lambda^{(\alpha_{3})}A_{\mu}^{(-\alpha_{2})})M_{\bar J}^{(\Lambda_{3})}
     +
     (\partial_{\mathcal K}\lambda^{(-\alpha_{3})}A_{\mu}^{(\alpha_{2})})M_{\bar J}^{(\Lambda_{1})}
     \right.
     \\
     &\quad \left.
     +
     (\partial_{\mathcal K}\lambda^{(-\alpha_{2})}A_{\mu}^{(\alpha_{3})})M_{\bar J}^{(\Lambda_{3})}
     +
     (\partial_{\mathcal K}\lambda^{(\alpha_{2})}A_{\mu}^{(-\alpha_{3})})M_{\bar J}^{(\Lambda_{1})}
     \right] .
    \end{aligned}
\end{equation}
The terms in \eqref{eq: ex_ApB_b}  are allowed in this bracketing because the first product is a product of two adjoint modes which can be combined to obtain the adjoint modes in the first line. For example,
\begin{equation}
    \alpha_3-\alpha_2=\alpha_1,
    \qquad
    \alpha_1+\Lambda_3=\Lambda_2,
\end{equation}
and similarly for the other terms.

On the other hand, in the assuming-covariance calculation one encounters the structure
\begin{equation}
\label{eq: ex_ApB_c}
 \delta D_{\mu}M_{\mathcal K\bar J}\big|_{\mathrm{a.c.}}
 \supset
 -\partial_{\mathcal K}\lambda\star D_{\mu}M_{\bar J}
 \supset
 -ig\,\partial_{\mathcal K}\lambda\star(A_{\mu}\star M_{\bar J}) \, ,
\end{equation}
apparently equivalent to \eqref{eq: ex_ApB_a}. For the same external mode $\Lambda_2$ as was considered in \eqref{eq: ex_ApB_b}, the first step gives
\begin{equation}
 \delta D_{\mu}M_{\mathcal K\bar J}^{(\Lambda_{2})}\big|_{\mathrm{a.c.}}
 \supset
 -\partial_{\mathcal K}\lambda^{(\alpha_{1})}D_{\mu}M_{\bar J}^{(\Lambda_{3})}
 -
 \partial_{\mathcal K}\lambda^{(-\alpha_{1})}D_{\mu}M_{\bar J}^{(\Lambda_{1})}\, .
\end{equation}
However, the particular contribution $-ig\,\partial_{\mathcal K}\lambda\star(A_{\mu}\star M_{\bar J})$ in \eqref{eq: ex_ApB_c}
does not produce the terms in \eqref{eq: ex_ApB_b} once the intermediate product $A_\mu\star M_{\bar J}$ is required to lie in the allowed $N=0$ $M_{\bar J}$-sector of the representation. For example,
\begin{equation}
    -\alpha_2+\Lambda_3=\Lambda_5,
\end{equation}
but $\Lambda_5$ is an $N=1$ state in the $\bm 6_{(-\frac{2}{3},-\frac{1}{3})}$, not an $N=0$ mode of $M_{\bar J}$. Other terms either give intermediate labels outside the representation or labels belonging to the wrong oscillator sector. Thus the contribution is removed by the intermediate projection, even though the unprojected triple product has the same total GKK label $\Lambda_2$.

This illustrates the precise sense in which associativity can fail after projection: the full $\star$-product is associative, but the product obtained by restricting intermediate states to a fixed representation or field sector is not. Consequently, in the covariance checks one cannot freely reassociate products after mode expansion while simultaneously imposing representation-level projections at each intermediate step. This illustrates why the process of mode-expansion becomes necessary to put terms into correspondence, as was done in Section \ref{sec: covariance}.

\subsection{Heterotic extension}

We now extend the product to include the heterotic gauge-lattice degrees of freedom. The underlying GKK label is
\begin{equation}
    \mathbb P=(P^I,p_m,\tilde p^{\,m}),
    \qquad
    I=1,\ldots,16,
    \qquad
    m=1,\ldots,r .
\end{equation}
The moduli-dependent left and right momentum coordinates \(\mathbb L(\mathbb P;\Phi)\) are obtained from this label by the rotation described in Appendix \ref{ap: HeteroticStringBasics}. The \(\star\)-product phase, however, is naturally written in terms of the moduli-independent GKK data \(\mathbb P\).

The heterotic gauge momenta \(P^I\) may be interpreted as originating from a sixteen-dimensional torus \cite{Giveon:1994}. Introducing auxiliary momenta and windings \((p_I,\tilde p^{\,I})\) in these directions, and imposing the chiral condition \(P_R^I=0\), one finds
\begin{equation}
    \tilde p_1^{\,I}p_{2I}
    =
    \frac12 P_1^I E_{IJ}P_2^J,
    \qquad
    E_{IJ}=G_{IJ}+B_{IJ}\, .
\end{equation}
Here \(G_{IJ}\) is the Cartan metric of the gauge lattice, and \(B_{IJ}\) is chosen so as to reproduce the usual cocycle ordering of the gauge-lattice vertex operators. For the \(\mathrm{Spin(32)}\) lattice, this can be implemented by taking \(B_{IJ}=G_{IJ}=-B_{JI}\) for \(I>J\).

We define the bilinear phase
\begin{equation}
\label{eq: heterotic_star_phase}
    \Theta(\mathbb P_1,\mathbb P_2)
    =
    p_{1m}\tilde p_2^{\,m}
    +
    \frac12 P_1^I E_{IJ}P_2^J .
\end{equation}
The heterotic \(\star\)-product is then obtained from \eqref{eq: starproductFourier} by replacing
\begin{equation}
    p_{1m}\tilde p_2^{\,m}
    \quad\longrightarrow\quad
    \Theta(\mathbb P_1,\mathbb P_2).
\end{equation}
Equivalently,
\begin{equation}
\label{eq: heterotic_star_product}
\begin{aligned}
(\phi_1\star\phi_2)(x,\mathbb Y)
&=
\sum_{\mathbb K_1,\mathbb K_2}
e^{i\pi\Theta(\mathbb P_1,\mathbb P_2)}
\phi_1^{(\mathbb K_1)}(x)
\phi_2^{(\mathbb K_2)}(x)
e^{i\mathbb K(\mathbb P_1+\mathbb P_2;\Phi)\cdot\mathbb Y}.
\end{aligned}
\end{equation}

\subsection{Phases and structure constants}

The elementary phase in the $\star$-product is the two-field cocycle phase. For an output two-field product with associated mode $\mathbb{K}_{1}$ and underlying GKK label $\mathbb{P}_{1}$,  i.e. selecting a single mode from the full GKK expansion \eqref{eq: heterotic_star_product}, the corresponding mode coefficients are
\begin{equation}
\label{eq:star_product_mode_K}
    (\phi_2\star\phi_3)^{(\mathbb K_1)}(x)
    =
    \sum_{\mathbb P_2+\mathbb P_3=\mathbb P_1}
    e^{i\pi\Theta(\mathbb P_2,\mathbb P_3)}
    \phi_2^{(\mathbb K_2)}(x)
    \phi_3^{(\mathbb K_3)}(x)\, .
\end{equation}

The phases denoted by $\tilde f$ in the main text, and defined in \eqref{eq: phaseGral}, are a convenient rewriting of this two-field phase, with the output mode placed as the first index,

\begin{equation*}
 \tilde{f}_{\mathbb{K}
_1 
\mathbb {K}_2
\mathbb{K}_3}=e^{i{\pi } 
{p}_1\cdot \tilde p_2}e^{i{\pi } 
({p}_1+{p}_2)\cdot \tilde p_3}\equiv\pm 1\, .
\end{equation*}
In the sector considered in this paper, $\bar N=0$, and the level-matching condition gives $    \Theta(\mathbb P_1,\mathbb P_1) = \frac{1}{2}\mathbb K_1^2 = 1-N_1$ ,
where $N_1$ is the left oscillator number of the output mode. Bearing this in mind, we find a convenient rewrite of \eqref{eq: heterotic_star_phase} as 
\begin{equation}
\label{eq:ftilde_definition_from_binary_product}
    e^{i\pi\Theta(\mathbb P_2,\mathbb P_3)}
    =
    e^{i\pi(1-N_1)}
    \tilde f_{\mathbb K_1\mathbb K_2\mathbb K_3}.
\end{equation}
Equivalently,
\begin{equation}
\label{eq: ftilde_binary_phase}
    \tilde f_{\mathbb K_1\mathbb K_2\mathbb K_3}
    =
    e^{i\pi(1-N_1)}
    e^{i\pi\Theta(\mathbb P_2,\mathbb P_3)} \, .
\end{equation}
Using $\mathbb P_1=\mathbb P_2+\mathbb P_3$ and the bilinearity of $\Theta$, this can also be written as
\begin{equation}
\label{eq: three_field_phase_appendix}
    \tilde f_{\mathbb K_1\mathbb K_2\mathbb K_3}
    =
    e^{i\pi\Theta(\mathbb P_1,\mathbb P_2)}
    e^{i\pi\Theta(\mathbb P_1+\mathbb P_2,\mathbb P_3)} \, .
\end{equation}

In the main text, labels such as $\mathbb K_i$, $\alpha_i$ or $\Lambda_s$ are used to identify modes. The underlying GKK labels are always implicit. Thus phases such as $\tilde f_{\mathbb K_1\mathbb K_2\mathbb K_3}\, , \tilde f_{\alpha_c\alpha_b\alpha_a}$ and $\tilde f_{\Lambda_s\alpha_l\Lambda_r}$ are functions of the corresponding underlying GKK labels, not of the root or weight vectors alone. For the purely gauge-lattice part, the cocycle phase $\epsilon(P_1,P_2) = e^{i\pi\frac12 P_1^I E_{IJ}P_2^J}$ provides the standard cocycle ordering of the heterotic charged operators. Here $P_1^I$ and $P_2^I$ denote only the heterotic gauge-lattice components of the full GKK labels. In particular, for gauge-lattice roots $P_1$ and $P_2$ such that $P_3=P_1+P_2$ is also a root, the charged operator algebra takes the schematic form
\begin{equation}
    [E_{P_1},E_{P_2}]
    =
    \epsilon(P_1,P_2)E_{P_3},
\end{equation}
up to the conventional normalization of the step generators; see for instance \cite{Green:1987}.

The same mechanism applies to the enhancement algebras generated by compact momentum and winding modes. At an enhancement point $\Phi_0$, the relevant GKK labels $\mathbb P_c$ are mapped to roots after the change to the root-space basis:
\begin{equation}
    k_R(\mathbb P_c;\Phi_0)=0,
    \qquad
    \kappa_L(\mathbb P_c;\Phi_0)=\alpha_c .
\end{equation}
Equivalently, the root-space vector $\mathbb K(\mathbb P_c;\Phi_0)$ has left component $\alpha_c$ and vanishing right component. The derivatives acting on the corresponding Fourier modes reproduce the Cartan action, while the cocycle phases $\tilde f$ participate in reproducing the signs of the charged-root structure constants. With the conventions used in the main text,
\begin{equation}
\label{eq: ctesFaseTildef}
    f_{\alpha_a,\alpha_b}{}^{\alpha_c}
    =
    (-1)^n
    \tilde f_{\alpha_c\alpha_b\alpha_a},
    \qquad
    \alpha_c=\alpha_a+\alpha_b,
\end{equation}
where $n$ is the number of negative roots among $\alpha_{a}\, ,\alpha_{b}$ and $\alpha_{c}$, up to the normalization convention for the step generators. This is the sense in which the non-commutative $\star$-product already contains the charged-root part of the enhanced non-Abelian algebra before a conventional gauge-theory basis is chosen.

\section{Lie algebras in the Cartan-Weyl basis}
We collect here the Lie-algebra conventions used throughout this work. We work in a Cartan--Weyl basis, with Cartan generators \(T_i\) and step generators \(T_{\pm\alpha}\) associated with the roots of the enhanced algebra. Specifically, the algebraic relations read
\begin{equation}\label{eq: CW_algebra}
    \begin{aligned}
        \left[T_{i}, T_{\alpha}  \right]&= \alpha^i T_\alpha \\
        \left[T_\alpha, T_\beta \right]  &= f_{\alpha  \, \beta}{}^{\alpha + \beta} T_{\alpha + \beta}\\
        \left[ T_\alpha, T_{-\alpha}\right] &=  \alpha^m T_m \, ,
    \end{aligned}
\end{equation}
where the second bracket is understood to vanish whenever \(\alpha+\beta\) is nor a root nor zero.
The following subsections contain the explicit \(\mathrm{SU(3)}\) and \(\mathrm{SU(4)}\) root conventions, enhancement-point GKK labels, representation data and generator matrices used in our examples.

\subsection{Some useful \texorpdfstring{$\mathrm{SU(3)}$}{SU(3)} expressions} \label{ap: SU3facts}
Here we collect some relevant $\mathrm{SU(3)}$ conventions used in the examples in Sections \ref{sec: su3example} and \ref{sec: SU3_AlgebraAwayEnhPoint}. The eight $\mathrm{SU(3)}$ generators are denoted by the Cartan generators  $T_1,T_2$ 
and the step raising (lowering) generators $T_\alpha$ ($T_{-\alpha}$) with $\alpha\in\{\alpha_1,\alpha_2,\alpha_3\}$, $(\alpha_3=\alpha_1+\alpha_2)$. In particular, they must satisfy \eqref{eq: CW_algebra}.
We choose the simple root basis $\lbrace \alpha_1,\alpha_2 \rbrace$   with $\mathbb{R}^2$ coordinates 
\begin{equation}
\label{eq: simplesu3roots}
    \alpha_1=
    \left(\sqrt{2},0\right),
    \qquad
    \alpha_2=
    \left(-\frac{1}{\sqrt{2}},\sqrt{\frac{3}{2}}\right)\, .
\end{equation}
The corresponding fundamental weights are
\begin{equation}
\label{eq:su3_fundamental_weights}
    \omega_1=
    \left(
    \frac{1}{\sqrt{2}},
    \frac{1}{2}\sqrt{\frac{2}{3}}
    \right),
    \qquad
    \omega_2=
    \left(
    0,
    \sqrt{\frac{2}{3}}
    \right),
\end{equation}
where $\omega_j$ are dual to the simple roots,  $\omega_j. \alpha^m=\delta_j ^m$.

For the \(\mathrm{SU(3)}_L\) enhancement point considered in our examples, we take vanishing Wilson lines and choose the torus metric and antisymmetric tensor
\begin{equation}
   g =
   \begin{pmatrix}
        2 & -1 \\
        -1 & 2
    \end{pmatrix},
    \qquad
   b =
   \begin{pmatrix}
        0 & -1 \\
        1 & 0
    \end{pmatrix}.
    \label{eq:su3_moduli_appendix}
\end{equation}
With the radius conventions used in Appendix \ref{ap: HeteroticStringBasics}, the left and right coordinates of momenta in the simple-root basis are
\begin{equation}
    \begin{aligned}
        l^{1}_{L}
        &=
        \frac{1}{3}
        \left(
        2p_{1}+p_{2}+\tilde p^{1}+\tilde p^{2}
        \right),
        &
        l^{1}_{R}
        &=
        \frac{1}{3}
        \left(
        2p_{1}+p_{2}-2\tilde p^{1}+\tilde p^{2}
        \right),
        \\
        l^{2}_{L}
        &=
        \frac{1}{3}
        \left(
        p_{1}+2p_{2}-\tilde p^{1}+2\tilde p^{2}
        \right),
        &
        l^{2}_{R}
        &=
        \frac{1}{3}
        \left(
        p_{1}+2p_{2}-\tilde p^{1}-\tilde p^{2}
        \right).
    \end{aligned}
    \label{eq:su3_left_right_momenta_appendix}
\end{equation}
The GKK labels that become the six charged \(\mathrm{SU(3)}_L\) generators at this point are
\begin{equation}
    \mathbb P_{\pm\alpha_1}
    =
    \pm(1,0;1,0),
    \qquad
    \mathbb P_{\pm\alpha_2}
    =
    \pm(-1,1;0,1),
    \qquad
    \mathbb P_{\pm\alpha_3}
    =
    \pm(0,1;1,1).
    \label{eq:su3_root_gkk_labels_appendix}
\end{equation}
Indeed, substituting these labels in \eqref{eq:su3_left_right_momenta_appendix} gives
\begin{equation}
    (l_R^1,l_R^2)=(0,0),
    \qquad
    (l_L^1,l_L^2)
    =
    \pm(1,0),\quad
    \pm(0,1),\quad
    \pm(1,1),
\end{equation}
which are precisely the coordinates of
\(\pm\alpha_1,\pm\alpha_2,\pm\alpha_3\) in the simple-root basis. They satisfy
\begin{equation}
    l_L^m g_{mn} l_L^n
    =
    2\tilde p^{\,m}p_m
    =
    2,
\end{equation}
and label the six charged massless vectors of \(\mathrm{SU(3)}_L\). Together with the heterotic gauge-lattice roots, this gives the enhanced gauge group
\begin{equation}
    \mathrm{SO(32)}_L\times \mathrm{SU(3)}_L\times \mathrm{U(1)^2}_R
\end{equation}
at the chosen point.

The GKK labels, left-momentum coordinates, Dynkin labels and oscillator numbers for the $\bm{8}$ adjoint representation and the $\bm{6}$ symmetric representation are detailed in tables \ref{tab: Lambdas_SU3_8} and \ref{tab: Lambdas_SU3_6}, respectively. The subscripts denote the right-moving momentum chosen for the representation.

\begin{table}[ht]
\small{
\centering
\caption{KK momenta, winding numbers and weights of $\mathrm{SU(3)}$ representations.}
\label{tab: Lambdas_SU3_8_6}
\begin{subtable}[t]{0.48\linewidth}
\centering
\caption{$\bm{8}_{(0,0)}$ representation.}
\label{tab: Lambdas_SU3_8}
\renewcommand{\arraystretch}{2}
\begin{adjustbox}{width=\linewidth}
\begin{tabular}{| c | c | c | c |}
\hline
  Dynkin label & $\left(p_{1}, p_{2}; \tilde{p}^{1}, \tilde{p}^{2}\right)$ & $\left(l^{1}_{L}, l^{2}_{L} \right)$ & $N$  \\
\hline
$\pm (1,1)$ & $\pm (0,1;1,1)$ & $\pm(1,1) \equiv \pm \alpha_{3}$ & $0$\\
\hline
$\pm (2,-1)$ & $\pm (1,0;1,0)$ & $\pm(1,0) \equiv \pm \alpha_{1}$ & $0$\\
\hline
$\pm (-1,2)$ & $\pm (-1,1;0,1)$ & $\pm(0,1) \equiv \pm \alpha_{2}$ & $0$\\
\hline
$2 \times (0,0)$ & $\pm (0,0;0,0)$ & $\pm(0,0)$ & $1$\\
\hline
\end{tabular}
\end{adjustbox}
\end{subtable}
\hfill
\begin{subtable}[t]{0.48\linewidth}
\centering
\caption{$\bm{6}_{(-\frac{2}{3}, -\frac{1}{3})}$ representation.}
\label{tab: Lambdas_SU3_6}
\renewcommand{\arraystretch}{2}
\begin{adjustbox}{width=\linewidth}
\begin{tabular}{| c | c | c | c |}
\hline
 r $\equiv$ Dynkin label & $\left(p_{1}, p_{2}; \tilde{p}^{1}, \tilde{p}^{2}\right)$ & $\left(l^{1}_{L}, l^{2}_{L} \right)$ & $N$  \\
\hline
$1 \equiv (2,0)$ & $(0,1;2,1)$ & $\left(\frac{4}{3}, \frac{2}{3} \right)$ & $0$\\
\hline
$2 \equiv (0,1)$ & $(-1,1;1,1)$ & $\left(\frac{1}{3}, \frac{2}{3} \right)$ & $1$\\
\hline
$3 \equiv (-2,2)$ & $(-2,1;0,1)$ & $\left(-\frac{2}{3}, \frac{2}{3} \right)$ & $0$\\
\hline
$4 \equiv (1,-1)$ & $(0,0;1,0)$ & $\left(\frac{1}{3}, -\frac{1}{3} \right)$ & $1$\\
\hline
$5 \equiv (-1,0)$ & $(-1,0;0,0)$ & $\left(-\frac{2}{3}, -\frac{1}{3} \right)$ & $1$\\
\hline
$6 \equiv (0,-2)$ & $(0,-1;0,-1)$ & $\left(-\frac{2}{3}, -\frac{4}{3} \right)$ & $0$\\
\hline
\end{tabular}
\end{adjustbox}
\end{subtable}
}
\end{table}

Satisfying \(T_{-\alpha}=(T_\alpha)^t\), the generators for the \(\bm 6\) representation of \(\mathrm{SU(3)}\) read

{\tiny{
\begin{equation}\label{eq: Generators6SU3}
\begin{aligned}
T_1
&=
\begin{pmatrix}
\sqrt{2} & 0 & 0 & 0 & 0 & 0\\
0 & 0 & 0 & 0 & 0 & 0\\
0 & 0 & -\sqrt{2} & 0 & 0 & 0\\
0 & 0 & 0 & \frac{1}{\sqrt{2}} & 0 & 0\\
0 & 0 & 0 & 0 & -\frac{1}{\sqrt{2}} & 0\\
0 & 0 & 0 & 0 & 0 & 0
\end{pmatrix}
\qquad
T_2
=
\begin{pmatrix}
\sqrt{\frac{2}{3}} & 0 & 0 & 0 & 0 & 0\\
0 & \sqrt{\frac{2}{3}} & 0 & 0 & 0 & 0\\
0 & 0 & \sqrt{\frac{2}{3}} & 0 & 0 & 0\\
0 & 0 & 0 & -\frac{1}{\sqrt{6}} & 0 & 0\\
0 & 0 & 0 & 0 & -\frac{1}{\sqrt{6}} & 0\\
0 & 0 & 0 & 0 & 0 & -2\sqrt{\frac{2}{3}}
\end{pmatrix}
\\[1.5em]
T_{\alpha_1}
&=
\begin{pmatrix}
0 & \sqrt{2} & 0 & 0 & 0 & 0\\
0 & 0 & \sqrt{2} & 0 & 0 & 0\\
0 & 0 & 0 & 0 & 0 & 0\\
0 & 0 & 0 & 0 & 1 & 0\\
0 & 0 & 0 & 0 & 0 & 0\\
0 & 0 & 0 & 0 & 0 & 0
\end{pmatrix}
\qquad
T_{\alpha_2}
=
\begin{pmatrix}
0 & 0 & 0 & 0 & 0 & 0\\
0 & 0 & 0 & 1 & 0 & 0\\
0 & 0 & 0 & 0 & \sqrt{2} & 0\\
0 & 0 & 0 & 0 & 0 & 0\\
0 & 0 & 0 & 0 & 0 & \sqrt{2}\\
0 & 0 & 0 & 0 & 0 & 0
\end{pmatrix}
\qquad
T_{\alpha_3}
=
\begin{pmatrix}
0 & 0 & 0 & \sqrt{2} & 0 & 0\\
0 & 0 & 0 & 0 & 1 & 0\\
0 & 0 & 0 & 0 & 0 & 0\\
0 & 0 & 0 & 0 & 0 & \sqrt{2}\\
0 & 0 & 0 & 0 & 0 & 0\\
0 & 0 & 0 & 0 & 0 & 0
\end{pmatrix}\, .
\end{aligned}
\end{equation}
}}

\subsection{Some useful \texorpdfstring{$\mathrm{SU(4)}$}{SU(4)} expressions}\label{ap: SU4facts}

Here we collect some relevant $\mathrm{SU(4)}$ conventions used in the example in Section \ref{sec: su4example}.
 The fifteen $\mathrm{SU(4)}$ generators are denoted by the Cartan generators  $T_1,T_2, T_3$ 
and the step raising (lowering) generators $T_\alpha$ ($T_{-\alpha}$) with $\alpha\in\{\alpha_1,\alpha_2,\alpha_3, \alpha_{4}, \alpha_{5}, \alpha_{6} \}$, $(\alpha_4=\alpha_1+\alpha_2,\, \alpha_5=\alpha_2+\alpha_3,\, \alpha_6=\alpha_1+\alpha_2+\alpha_{3} )$. In particular, they must satisfy \eqref{eq: CW_algebra}. We choose the simple root basis $\lbrace \alpha_{1}, \alpha_{2}, \alpha_{3} \rbrace$, with $\mathbb{R}^{3}$ coordinates 
\begin{equation}\label{eq: rootBasis_SU4}
    \alpha_{1} = \left(\sqrt{2}, 0, 0 \right), \qquad \alpha_{2} = \left(-\frac{1}{\sqrt{2}}, \sqrt{\frac{3}{2}}, 0 \right), \qquad \alpha_{3} = \left(0, -\sqrt{\frac{2}{3}}, \frac{2}{\sqrt{3}} \right)\, .
\end{equation}
The corresponding fundamental weights are
\begin{equation}\label{eq: weightBasis_SU4}
    \omega_{1} = \left(\frac{1}{\sqrt{2}}, \frac{1}{\sqrt{6}}, \frac{1}{2\sqrt{3}} \right), \qquad \omega_{2} = \left(0,\sqrt{\frac{2}{3}}, \frac{1}{\sqrt{3}} \right), \qquad \omega_{3} = \left(0, 0, \frac{\sqrt{3}}{2} \right) \, ,
\end{equation}
where \(\omega_j\) are dual to the simple roots, $\omega_j. \alpha^m=\delta_j ^m$.

For the \(\mathrm{SU(4)}_L\) enhancement point considered in Section \ref{sec: su4example}, we take vanishing Wilson lines and choose the torus metric and antisymmetric tensor
\begin{equation}
   g =
   \begin{pmatrix}
        2 & -1 & 0 \\
        -1 & 2 & -1 \\
        0 & -1 & 2
    \end{pmatrix},
    \qquad
   b =
   \begin{pmatrix}
        0 & -1 & 0 \\
        1 & 0 & -1 \\
        0 & 1 & 0
    \end{pmatrix}.
    \label{eq:su4_moduli_appendix}
\end{equation}
With the radius conventions used in Appendix \ref{ap: HeteroticStringBasics}, namely \(\tilde R=\alpha'/R\) and \(R=\sqrt{\alpha'}\), the left and right momenta have coordinates
\begin{equation}\label{eq: LRmomenta_SU4}
    \begin{aligned}
        l^{1}_{L}
        &=
        \frac{1}{4}
        \left(
        3p_{1}+2p_{2}+p_{3}
        +\tilde p^{1}+\tilde p^{2}+\tilde p^{3}
        \right),
        &
        l^{1}_{R}
        &=
        \frac{1}{4}
        \left(
        3p_{1}+2p_{2}+p_{3}
        -3\tilde p^{1}+\tilde p^{2}+\tilde p^{3}
        \right),
        \\
        l^{2}_{L}
        &=
        \frac{1}{2}
        \left(
        p_{1}+2p_{2}+p_{3}
        -\tilde p^{1}+\tilde p^{2}+\tilde p^{3}
        \right),
        &
        l^{2}_{R}
        &=
        \frac{1}{2}
        \left(
        p_{1}+2p_{2}+p_{3}
        -\tilde p^{1}-\tilde p^{2}+\tilde p^{3}
        \right),
        \\
        l^{3}_{L}
        &=
        \frac{1}{4}
        \left(
        p_{1}+2p_{2}+3p_{3}
        -\tilde p^{1}-\tilde p^{2}+3\tilde p^{3}
        \right),
        &
        l^{3}_{R}
        &=
        \frac{1}{4}
        \left(
        p_{1}+2p_{2}+3p_{3}
        -\tilde p^{1}-\tilde p^{2}-\tilde p^{3}
        \right)\, .
    \end{aligned}
\end{equation}
The GKK labels that become the twelve charged \(SU(4)_L\) generators at this point are
\begin{equation}\label{eq: KKyW_SU4}
    \begin{aligned}
    \mathbb P_{\pm\alpha_1}
    &=
    \pm(1,0,0;1,0,0),
    &
    \mathbb P_{\pm\alpha_2}
    &=
    \pm(-1,1,0;0,1,0),
    &
    \mathbb P_{\pm\alpha_3}
    &=
    \pm(0,-1,1;0,0,1),
    \\
    \mathbb P_{\pm\alpha_4}
    &=
    \pm(0,1,0;1,1,0),
    &
    \mathbb P_{\pm\alpha_5}
    &=
    \pm(-1,0,1;0,1,1),
    &
    \mathbb P_{\pm\alpha_6}
    &=
    \pm(0,0,1;1,1,1).
    \end{aligned}
\end{equation}
Indeed, substituting these labels in \eqref{eq: LRmomenta_SU4} gives
\begin{equation}
    (l_R^1,l_R^2,l_R^3)=(0,0,0)\, , 
\end{equation}
together with
\begin{equation}
\begin{aligned}
    (l_L^1,l_L^2,l_L^3) =&
    \pm(1,0,0),\quad
    \pm(0,1,0),\quad
    \pm(0,0,1)\, ,\\
    &\pm(1,1,0),\quad
    \pm(0,1,1),\quad
    \pm(1,1,1) \, .\\ 
\end{aligned}
\end{equation}
These are precisely the coordinates of
\(\pm\alpha_1,\pm\alpha_2,\pm\alpha_3,\pm\alpha_4,\pm\alpha_5,\pm\alpha_6\)
in the simple-root basis. They satisfy
\begin{equation}
    l_L^m g_{mn} l_L^n
    =
    2\tilde p^{\,m}p_m
    =
    2,
\end{equation}
and label the twelve charged massless vectors of \(SU(4)_L\). Together with the heterotic gauge-lattice roots, this gives the enhanced gauge group
\begin{equation}
    \mathrm{SO(32)}_L\times \mathrm{SU(4)}_L\times \mathrm{U(1)^3}_R
\end{equation}
at the chosen point.

The weights, coordinates in the root basis, oscillator number, as well as KK momenta and winding numbers, for the $\bm{15}$ adjoint and $\bm{10}$ symmetric representations are detailed in tables \ref{tab: LambdasKKyW_SU4_15} and \ref{tab: LambdasKKyW_SU4_10}, respectively. The subscripts indicate right charge in the simple root basis.

\begin{table}[h]
\small{
\centering
\caption{ KK momenta, winding numbers and weights for $\mathrm{SU(4)}$ representations.}
\label{tab: LambdasKKyW_SU4_15_10}
\begin{subtable}[t]{0.48\linewidth}
\centering
\caption{$\bm{15}_{(0,0,0)}$ representation.}
\label{tab: LambdasKKyW_SU4_15}
\renewcommand{\arraystretch}{2}
\begin{adjustbox}{width=\linewidth}
\begin{tabular}{| c | c | c | c |}
\hline
Dynkin label & $(p_{n};\Tilde{p}^{n})$ & $(l^{1}_{L}, l^{2}_{L}, l^{3}_{L})$ & $N$   \\
\hline
$\pm (2,-1,0)$ & $\pm(1,0,0;1,0,0)$  & $\pm \alpha_{1} \equiv \pm (1,0,0)$ &  $0$ \\
\hline
$\pm(-1,2,-1)$& $\pm (-1, 1, 0; 0, 1, 0)$ & $\pm \alpha_{2} \equiv \pm (0,1,0)$  &  $0$ \\
\hline
$\pm(0,-1,2)$& $\pm (0, -1, 1; 0, 0, 1)$ & $\pm \alpha_{3} \equiv \pm (0,0,1)$  &  $0$ \\
\hline
$\pm (1,1,-1)$& $\pm (0, 1, 0; 1, 1, 0)$& $\pm \alpha_{4} \equiv \pm(1,1,0)$  &  $0$ \\
\hline
$\pm (-1,1,1)$& $\pm (-1, 0, 1; 0, 1, 1)$ & $\pm \alpha_{5} \equiv \pm(0,1,1)$  &  $0$ \\
\hline
$\pm (1,0,1)$ & $\pm (0,0,1;1,1,1)$ & $\pm \alpha_{6} \equiv \pm(1,1,1)$  &  $0$ \\
\hline
$3 \times (0,0,0)$ & $\pm (0,0,0;0,0,0)$ & $\pm (0,0,0)$ & $1$ \\
\hline
\end{tabular}
\end{adjustbox}
\end{subtable}
\hfill
\begin{subtable}[t]{0.48\linewidth}
\centering
\caption{$\bm{10}_{(-\frac{1}{2},0,\frac{1}{2})}$ representation.}
\label{tab: LambdasKKyW_SU4_10}
\renewcommand{\arraystretch}{2}
\begin{adjustbox}{width=\linewidth}
\begin{tabular}{| c | c | c | c |}
\hline
 r $\equiv$ Dynkin label &$(p_{n};\Tilde{p}^{n})$ & $(l^{1}_{L}, l^{2}_{L}, l^{3}_{L})$ & $N$   \\
\hline
$1 \equiv (2,0,0)$ &  $(0, 1, 1, 2, 1, 0)$& $\left(  \frac{3}{2}, 1, \frac{1}{2} \right)$  &$0$ \\
\hline
$2 \equiv (0,1,0)$ &$(-1,1,1,1,1,0)$& $\left(  \frac{1}{2}, 1, \frac{1}{2} \right)$  & $1$ \\
\hline
$3 \equiv (-2,2,0)$ & $(-2,1,1,0,1,0)$ & $\left( -\frac{1}{2}, 1, \frac{1}{2} \right)$  &  $0$ \\
\hline
$4 \equiv (1,-1,1)$ & $(0,0,1,1,0,0)$ & $\left(  \frac{1}{2}, 0, \frac{1}{2} \right)$  & $1$  \\
\hline
$5 \equiv (-1,0,1)$ & $(-1,0,1,0,0,0)$ & $\left(  -\frac{1}{2}, 0, \frac{1}{2} \right)$  & $1$\\
\hline
$6 \equiv (0,-2,2)$ & $(0,-1,1,0,-1,0)$ &$\left( - \frac{1}{2}, -1, \frac{1}{2} \right)$  &  $0$ \\
\hline
$7 \equiv (1,0,-1)$ & $(0,1,0,1,0,-1)$ & $\left(  \frac{1}{2}, 0, -\frac{1}{2} \right)$  &  $1$ \\
\hline
$8 \equiv (-1,1,-1)$ & $(-1,1,0,0,0,-1)$ &$\left( - \frac{1}{2}, 0, -\frac{1}{2} \right)$  &  $1$ \\
\hline
$9 \equiv (0,-1,0)$ & $(0,0,0,0,-1,-1)$ & $\left(  -\frac{1}{2}, -1, -\frac{1}{2} \right)$  & $1$ \\
\hline
$10 \equiv (0,0,-2)$ & $(0,1,-1,0,-1,-2)$ & $\left(  -\frac{1}{2}, -1, -\frac{3}{2} \right)$  &  $0$ \\
\hline
\end{tabular}
\end{adjustbox}
\end{subtable}
}
\end{table}

Satisfying $T_{-\alpha}={(T_{\alpha})}^t$, generators for the $10$ dimensional representation of $\mathrm{SU(4)}$ read
{\tiny{
\begin{equation}\label{eq: Generators10SU4}
\begin{alignedat}{2}
T_1 \; &= \;
\begin{adjustbox}{width=0.51\textwidth, keepaspectratio}
\(
\begin{pmatrix}
\sqrt{2} & 0 & 0 & 0 & 0 & 0 & 0 & 0 & 0 & 0 \\
0 & 0 & 0 & 0 & 0 & 0 & 0 & 0 & 0 & 0 \\
0 & 0 & -\sqrt{2} & 0 & 0 & 0 & 0 & 0 & 0 & 0 \\
0 & 0 & 0 & \frac{1}{\sqrt{2}} & 0 & 0 & 0 & 0 & 0 & 0 \\
0 & 0 & 0 & 0 & -\frac{1}{\sqrt{2}} & 0 & 0 & 0 & 0 & 0 \\
0 & 0 & 0 & 0 & 0 & 0 & 0 & 0 & 0 & 0 \\
0 & 0 & 0 & 0 & 0 & 0 &\frac{1}{\sqrt{2}} & 0 & 0 & 0 \\
0 & 0 & 0 & 0 & 0 & 0 & 0 & -\frac{1}{\sqrt{2}} & 0 & 0 \\
0 & 0 & 0 & 0 & 0 & 0 & 0 & 0 & 0 & 0 \\
0 & 0 & 0 & 0 & 0 & 0 & 0 & 0 & 0 & 0
\end{pmatrix}
\)
\end{adjustbox}
\quad
&T_{\alpha_1} \;=\;
\begin{adjustbox}{max width=0.4\textwidth, keepaspectratio}
\(
\begin{pmatrix}
0 & \sqrt{2} & 0 & 0 & 0 & 0 & 0 & 0 & 0 & 0\\
0 & 0 & \sqrt{2} & 0 & 0 & 0 & 0 & 0 & 0 & 0\\
0 & 0 & 0 & 0 & 0 & 0 & 0 & 0 & 0 & 0\\
0 & 0 & 0 & 0 & 1 & 0 & 0 & 0 & 0 & 0\\
0 & 0 & 0 & 0 & 0 & 0 & 0 & 0 & 0 & 0\\
0 & 0 & 0 & 0 & 0 & 0 & 0 & 0 & 0 & 0\\
0 & 0 & 0 & 0 & 0 & 0 & 0 & 1 & 0 & 0\\
0 & 0 & 0 & 0 & 0 & 0 & 0 & 0 & 0 & 0\\
0 & 0 & 0 & 0 & 0 & 0 & 0 & 0 & 0 & 0\\
0 & 0 & 0 & 0 & 0 & 0 & 0 & 0 & 0 & 0
\end{pmatrix}
\)
\end{adjustbox}\\[1.5em]
T_2 \; &= \;
\begin{adjustbox}{max width=0.54\textwidth}
\(
\begin{pmatrix}
\sqrt{\frac{2}{3}} & 0 & 0 & 0 & 0 & 0 & 0 & 0 & 0 & 0 \\
0 & \sqrt{\frac{2}{3}} & 0 & 0 & 0 & 0 & 0 & 0 & 0 & 0 \\
0 & 0 & \sqrt{\frac{2}{3}} & 0 & 0 & 0 & 0 & 0 & 0 & 0 \\
0 & 0 & 0 & -\frac{1}{\sqrt{6}} & 0 & 0 & 0 & 0 & 0 & 0 \\
0 & 0 & 0 & 0 & -\frac{1}{\sqrt{6}} & 0 & 0 & 0 & 0 & 0 \\
0 & 0 & 0 & 0 & 0 & -2\sqrt{\frac{2}{3}} & 0 & 0 & 0 & 0 \\
0 & 0 & 0 & 0 & 0 & 0 & \frac{1}{\sqrt{6}} & 0 & 0 & 0 \\
0 & 0 & 0 & 0 & 0 & 0 & 0 & \frac{1}{\sqrt{6}} & 0 & 0 \\
0 & 0 & 0 & 0 & 0 & 0 & 0 & 0 & -\sqrt{\frac{2}{3}} & 0 \\
0 & 0 & 0 & 0 & 0 & 0 & 0 & 0 & 0 & 0
\end{pmatrix}
\)
\end{adjustbox}
\quad
&T_{\alpha_2} \;=\;
\begin{adjustbox}{max width=0.4\textwidth, keepaspectratio}
\(
\begin{pmatrix}
0 & 0 & 0 & 0 & 0 & 0 & 0 & 0 & 0 & 0\\
0 & 0 & 0 & 1 & 0 & 0 & 0 & 0 & 0 & 0\\
0 & 0 & 0 & 0 & \sqrt{2} & 0 & 0 & 0 & 0 & 0\\
0 & 0 & 0 & 0 & 0 & 0 & 0 & 0 & 0 & 0\\
0 & 0 & 0 & 0 & 0 & \sqrt{2} & 0 & 0 & 0 & 0\\
0 & 0 & 0 & 0 & 0 & 0 & 0 & 0 & 0 & 0\\
0 & 0 & 0 & 0 & 0 & 0 & 0 & 0 & 0 & 0\\
0 & 0 & 0 & 0 & 0 & 0 & 0 & 0 & 1 & 0\\
0 & 0 & 0 & 0 & 0 & 0 & 0 & 0 & 0 & 0\\
0 & 0 & 0 & 0 & 0 & 0 & 0 & 0 & 0 & 0
\end{pmatrix}
\)
\end{adjustbox}\\[1.5em]
T_3 \; &= \;
\begin{adjustbox}{max width=0.54\textwidth}
\(
\begin{pmatrix}
\frac{1}{\sqrt{3}} & 0 & 0 & 0 & 0 & 0 & 0 & 0 & 0 & 0 \\
0 & \frac{1}{\sqrt{3}} & 0 & 0 & 0 & 0 & 0 & 0 & 0 & 0 \\
0 & 0 & \frac{1}{\sqrt{3}} & 0 & 0 & 0 & 0 & 0 & 0 & 0 \\
0 & 0 & 0 & \frac{1}{\sqrt{3}} & 0 & 0 & 0 & 0 & 0 & 0 \\
0 & 0 & 0 & 0 & \frac{1}{\sqrt{3}} & 0 & 0 & 0 & 0 & 0 \\
0 & 0 & 0 & 0 & 0 & \frac{1}{\sqrt{3}} & 0 & 0 & 0 & 0 \\
0 & 0 & 0 & 0 & 0 & 0 & -\frac{1}{\sqrt{3}} & 0 & 0 & 0 \\
0 & 0 & 0 & 0 & 0 & 0 & 0 & -\frac{1}{\sqrt{3}} & 0 & 0 \\
0 & 0 & 0 & 0 & 0 & 0 & 0 & 0 & -\frac{1}{\sqrt{3}} & 0 \\
0 & 0 & 0 & 0 & 0 & 0 & 0 & 0 & 0 & -\sqrt{3}
\end{pmatrix}
\)
\end{adjustbox}
\quad
&T_{\alpha_3} \;=\;
\begin{adjustbox}{max width=0.4\textwidth, keepaspectratio}
\(
\begin{pmatrix}
0 & 0 & 0 & 0 & 0 & 0 & 0 & 0 & 0 & 0\\
0 & 0 & 0 & 0 & 0 & 0 & 0 & 0 & 0 & 0\\
0 & 0 & 0 & 0 & 0 & 0 & 0 & 0 & 0 & 0\\
0 & 0 & 0 & 0 & 0 & 0 & 1 & 0 & 0 & 0\\
0 & 0 & 0 & 0 & 0 & 0 & 0 & 1 & 0 & 0\\
0 & 0 & 0 & 0 & 0 & 0 & 0 & 0 & \sqrt{2} & 0\\
0 & 0 & 0 & 0 & 0 & 0 & 0 & 0 & 0 & 0\\
0 & 0 & 0 & 0 & 0 & 0 & 0 & 0 & 0 & 0\\
0 & 0 & 0 & 0 & 0 & 0 & 0 & 0 & 0 & \sqrt{2}\\
0 & 0 & 0 & 0 & 0 & 0 & 0 & 0 & 0 & 0
\end{pmatrix}
\)
\end{adjustbox}\\[1.5em]
T_{\alpha_{4}} \; &= \;
\begin{adjustbox}{width=0.35\textwidth}
\(
\begin{pmatrix}
0 & 0 & 0 & \sqrt{2} & 0 & 0 & 0 & 0 & 0 & 0\\
0 & 0 & 0 & 0 & 1 & 0 & 0 & 0 & 0 & 0\\
0 & 0 & 0 & 0 & 0 & 0 & 0 & 0 & 0 & 0\\
0 & 0 & 0 & 0 & 0 & \sqrt{2} & 0 & 0 & 0 & 0\\
0 & 0 & 0 & 0 & 0 & 0 & 0 & 0 & 0 & 0\\
0 & 0 & 0 & 0 & 0 & 0 & 0 & 0 & 0 & 0\\
0 & 0 & 0 & 0 & 0 & 0 & 0 & 0 & 1 & 0\\
0 & 0 & 0 & 0 & 0 & 0 & 0 & 0 & 0 & 0\\
0 & 0 & 0 & 0 & 0 & 0 & 0 & 0 & 0 & 0\\
0 & 0 & 0 & 0 & 0 & 0 & 0 & 0 & 0 & 0\\
\end{pmatrix}
\)
\end{adjustbox}
\hfill
&T_{\alpha_5} \;=\;
\begin{adjustbox}{width=0.35\textwidth, keepaspectratio}
\(
\begin{pmatrix}
0 & 0 & 0 & 0 & 0 & 0 & 0 & 0 & 0 & 0\\
0 & 0 & 0 & 0 & 0 & 0 & 1 & 0 & 0 & 0\\
0 & 0 & 0 & 0 & 0 & 0 & 0 & \sqrt{2} & 0 & 0\\
0 & 0 & 0 & 0 & 0 & 0 & 0 & 0 & 0 & 0\\
0 & 0 & 0 & 0 & 0 & 0 & 0 & 0 & 1 & 0\\
0 & 0 & 0 & 0 & 0 & 0 & 0 & 0 & 0 & 0\\
0 & 0 & 0 & 0 & 0 & 0 & 0 & 0 & 0 & 0\\
0 & 0 & 0 & 0 & 0 & 0 & 0 & 0 & 0 & \sqrt{2}\\
0 & 0 & 0 & 0 & 0 & 0 & 0 & 0 & 0 & 0\\
0 & 0 & 0 & 0 & 0 & 0 & 0 & 0 & 0 & 0\\
\end{pmatrix}
\)
\end{adjustbox}\\[1.5em]
T_{\alpha_{6}} \; &= \;
\begin{adjustbox}{width=0.35\textwidth, keepaspectratio}
\(
\begin{pmatrix}
0 & 0 & 0 & 0 & 0 & 0 & \sqrt{2} & 0 & 0 & 0\\
0 & 0 & 0 & 0 & 0 & 0 & 0 & 1 & 0 & 0\\
0 & 0 & 0 & 0 & 0 & 0 & 0 & 0 & 0 & 0\\
0 & 0 & 0 & 0 & 0 & 0 & 0 & 0 & 1 & 0\\
0 & 0 & 0 & 0 & 0 & 0 & 0 & 0 & 0 & 0\\
0 & 0 & 0 & 0 & 0 & 0 & 0 & 0 & 0 & 0\\
0 & 0 & 0 & 0 & 0 & 0 & 0 & 0 & 0 & \sqrt{2}\\
0 & 0 & 0 & 0 & 0 & 0 & 0 & 0 & 0 & 0\\
0 & 0 & 0 & 0 & 0 & 0 & 0 & 0 & 0 & 0\\
0 & 0 & 0 & 0 & 0 & 0 & 0 & 0 & 0 & 0\\
\end{pmatrix} \, .
\)
\end{adjustbox}
\end{alignedat}
\end{equation}
}}

 \section*{Acknowledgements}

We thank Chris Hull, Mariana Graña, Walter Baron, Diego Marqués, and Bernardo Fraiman for useful discussions.
G. A. acknowledges support by a Simons targeted grant to Instituto Balseiro. G. A.
thanks  ICTP and IFT UAM-CSIC via the Centro de Excelencia Severo Ochoa for hospitality and support. The work of G. A. and L.M.C. was  partially supported by the PICT-2020-01760 grant. L. M. C. is grateful to ICTP for hospitality and support.

\bibliographystyle{JHEP}
\bibliography{aacm2026}

@article{Aldazabal:2018uzm,
    author = "Aldazabal, G. and Andr\'es, E. and Mayo, M. and Penas, V.",
    title = "{Symmetry enhancement interpolation, non-commutativity and Double Field Theory}",
    eprint = "1805.10306",
    archivePrefix = "arXiv",
    primaryClass = "hep-th",
    doi = "10.1007/JHEP03(2019)012",
    journal = "JHEP",
    volume = "03",
    pages = "012",
    year = "2019"
}

@article{Kugo:1992md,
    author = "Kugo, Taichiro and Zwiebach, Barton",
    title = "{Target space duality as a symmetry of string field theory}",
    eprint = "hep-th/9201040",
    archivePrefix = "arXiv",
    reportNumber = "YITP-K-961, IASSNS-HEP-92-3, MIT-CTP-2058",
    doi = "10.1143/ptp/87.4.801",
    journal = "Prog. Theor. Phys.",
    volume = "87",
    pages = "801--860",
    year = "1992"
}

@article{Siegel:1993th,
    author = "Siegel, W.",
    title = "{Superspace duality in low-energy superstrings}",
    eprint = "hep-th/9305073",
    archivePrefix = "arXiv",
    reportNumber = "ITP-SB-93-28",
    doi = "10.1103/PhysRevD.48.2826",
    journal = "Phys. Rev. D",
    volume = "48",
    pages = "2826--2837",
    year = "1993"
}

@article{Hull:2009mi,
    author = "Hull, Chris and Zwiebach, Barton",
    title = "{Double Field Theory}",
    eprint = "0904.4664",
    archivePrefix = "arXiv",
    primaryClass = "hep-th",
    reportNumber = "IMPERIAL-TP-2009-CH-02, MIT-CTP-4031",
    doi = "10.1088/1126-6708/2009/09/099",
    journal = "JHEP",
    volume = "09",
    pages = "099",
    year = "2009"
}

@article{Aldazabal:2013sca,
    author = "Aldazabal, Gerardo and Marqués, Diego and Núñez, Carmen",
    title = "{Double Field Theory: A Pedagogical Review}",
    eprint = "1305.1907",
    archivePrefix = "arXiv",
    primaryClass = "hep-th",
    doi = "10.1088/0264-9381/30/16/163001",
    journal = "Class. Quant. Grav.",
    volume = "30",
    pages = "163001",
    year = "2013"}

@article{Hohm:2013bwa,
author = {Hohm, Olaf and L\"ust, Dieter and Zwiebach, Barton},
    title = "{The Spacetime of Double Field Theory: Review, Remarks, and Outlook}",
    eprint = "1309.2977",
    archivePrefix = "arXiv",
    primaryClass = "hep-th",
    reportNumber = "MIT-CTP-4494, LMU-ASC-59-13, MPP-2013-241",
    doi = "10.1002/prop.201300024",
    journal = "Fortsch. Phys.",
    volume = "61",
    pages = "926--966",
    year = "2013"
}

@article{Aldazabal:2016,
    author = "Aldazabal, G. and Gra{\~n}a, M. and Iguri, S. and Mayo, M. and N{\'u}{\~n}ez, C. and Rosabal, J. A.",
    title = "{Enhanced gauge symmetry and winding modes in Double Field Theory}",
    eprint = "1510.07644",
    archivePrefix = "arXiv",
    primaryClass = "hep-th",
    doi = "10.1007/JHEP03(2016)093",
    journal = "JHEP",
    volume = "03",
    pages = "093",
    year = "2016"
}

@article{Aldazabal:2017,
    author = "Aldazabal, G. and Andr{\'e}s, E. and Mayo, M. and Rosabal, J. A.",
    title = "{Gauge symmetry enhancing-breaking from a Double Field Theory perspective}",
    eprint = "1704.04427",
    archivePrefix = "arXiv",
    primaryClass = "hep-th",
    doi = "10.1007/JHEP07(2017)045",
    journal = "JHEP",
    volume = "07",
    pages = "045",
    year = "2017"
}

@article{Cagnacci:2017,
    author = "Cagnacci, Y. and Gra{\~n}a, M. and Iguri, S. and N{\'u}{\~n}ez, C.",
    title = "{The bosonic string on string-size tori from double field theory}",
    eprint = "1704.04242",
    archivePrefix = "arXiv",
    primaryClass = "hep-th",
    doi = "10.1007/JHEP06(2017)005",
    journal = "JHEP",
    volume = "06",
    pages = "005",
    year = "2017"
}

@article{Aldazabal:2017b,
    author = "Aldazabal, G. and Andr{\'e}s, E. and Mayo, M. and Penas, V.",
    title = "{Double Field Theory description of Heterotic gauge symmetry enhancing-breaking}",
    eprint = "1708.07148",
    archivePrefix = "arXiv",
    primaryClass = "hep-th",
    doi = "10.1007/JHEP10(2017)046",
    journal = "JHEP",
    volume = "10",
    pages = "046",
    year = "2017"
}

@article{Fraiman:2018,
    author = "Fraiman, Bernardo and Gra{\~n}a, Mariana and N{\'u}{\~n}ez, Carmen A.",
    title = "{A new twist on heterotic string compactifications}",
    eprint = "1805.11128",
    archivePrefix = "arXiv",
    primaryClass = "hep-th",
    doi = "10.1007/JHEP09(2018)078",
    journal = "JHEP",
    volume = "09",
    pages = "078",
    year = "2018"
}

@article{Mohaupt:1992jf,
    author = {Mohaupt, Thomas},
    title = {Critical Wilson lines in toroidal compactifications of heterotic strings},
    eprint = {hep-th/9209101},
    archivePrefix = {arXiv},
    doi = {10.1142/S0217751X93001429},
    journal = {Int. J. Mod. Phys. A},
    volume = {8},
    pages = {3529--3552},
    year = {1993}
}

@article{Font:2020rsk,
    author = {Font, Anamar{\'\i}a and Fraiman, Bernardo and
              Gra{\~n}a, Mariana and N{\'u}{\~n}ez, Carmen A. and
              Parra De Freitas, H{\'e}ctor},
    title = {Exploring the landscape of heterotic strings on {$T^d$}},
    eprint = {2007.10358},
    archivePrefix = {arXiv},
    primaryClass = {hep-th},
    doi = {10.1007/JHEP10(2020)194},
    journal = {JHEP},
    volume = {10},
    pages = {194},
    year = {2020}
}

@article{Hohm:2010,
    author = "Hohm, Olaf and Hull, Chris and Zwiebach, Barton",
    title = "{Background independent action for double field theory}",
    eprint = "1003.5027",
    archivePrefix = "arXiv",
    primaryClass = "hep-th",
    doi = "10.1007/JHEP07(2010)016",
    journal = "JHEP",
    volume = "07",
    pages = "016",
    year = "2010"
}

@article{Hohm:2011a,
    author = "Hohm, Olaf and Kwak, Seung Ki",
    title = "{Double Field Theory Formulation of Heterotic Strings}",
    eprint = "1103.2136",
    archivePrefix = "arXiv",
    primaryClass = "hep-th",
    doi = "10.1007/JHEP06(2011)096",
    journal = "JHEP",
    volume = "06",
    pages = "096",
    year = "2011"
}

@article{Hohm:2015,
    author = "Hohm, Olaf and Sen, Ashoke and Zwiebach, Barton",
    title = "{Heterotic Effective Action and Duality Symmetries Revisited}",
    eprint = "1411.5696",
    archivePrefix = "arXiv",
    primaryClass = "hep-th",
    doi = "10.1007/JHEP02(2015)079",
    journal = "JHEP",
    volume = "02",
    pages = "079",
    year = "2015"
}

@article{Bedoya:2014,
    author = "Bedoya, Oscar A. and Marqu{\'e}s, Diego and N{\'u}{\~n}ez, Carmen",
    title = "{Heterotic $\alpha'$-corrections in Double Field Theory}",
    eprint = "1407.0365",
    archivePrefix = "arXiv",
    primaryClass = "hep-th",
    doi = "10.1007/JHEP12(2014)074",
    journal = "JHEP",
    volume = "12",
    pages = "074",
    year = "2014"
}

@article{Freidel:2017a,
    author = "Freidel, Laurent and Leigh, Robert G. and Minic, Djordje",
    title = "{Intrinsic non-commutativity of closed string theory}",
    eprint = "1706.03305",
    archivePrefix = "arXiv",
    primaryClass = "hep-th",
    doi = "10.1007/JHEP09(2017)060",
    journal = "JHEP",
    volume = "09",
    pages = "060",
    year = "2017"
}

@article{Freidel:2017b,
    author = "Freidel, Laurent and Leigh, Robert G. and Minic, Djordje",
    title = "{Noncommutativity of closed string zero modes}",
    eprint = "1707.00312",
    archivePrefix = "arXiv",
    primaryClass = "hep-th",
    doi = "10.1103/PhysRevD.96.066003",
    journal = "Phys. Rev. D",
    volume = "96",
    number = "6",
    pages = "066003",
    year = "2017"
}

@article{Sakamoto:1989,
  author = {M. Sakamoto},
  title = {A Physical Interpretation of Cocycle Factors in Vertex Operator Representations},
  journal = {Phys. Lett. B},
  volume = {231},
  pages = {258},
  year = {1989},
  doi = {10.1016/0370-2693(89)90210-4},
}

@article{Narain:1987,
  author = {K. S. Narain and M. H. Sarmadi and E. Witten},
  title = {A Note on Toroidal Compactification of Heterotic String Theory},
  journal = {Nucl. Phys. B},
  volume = {279},
  pages = {369},
  year = {1987},
  doi = {10.1016/0550-3213(87)90001-0},
}

@article{Narain:1986,
  author = {K. S. Narain},
  title = {New Heterotic String Theories in Uncompactified Dimensions $<$ 10},
  journal = {Phys. Lett. B},
  volume = {169},
  pages = {41--46},
  year = {1986},
  doi = {10.1016/0370-2693(86)90682-9},
}

@article{Giveon:1994,
  author = {A. Giveon and M. Porrati and E. Rabinovici},
  title = {Target space duality in string theory},
  journal = {Phys. Rept.},
  volume = {244},
  pages = {77},
  year = {1994},
  eprint = {hep-th/9401139},
}

@article{Aldazabal:2011,
    author = "Aldazabal, G. and Baron, W. and Marqu{\'e}s, D. and N{\'u}{\~n}ez, C.",
    title = "{The effective action of Double Field Theory}",
    eprint = "1109.0290",
    archivePrefix = "arXiv",
    primaryClass = "hep-th",
    doi = "10.1007/JHEP11(2011)052",
    journal = "JHEP",
    volume = "11",
    pages = "052",
    year = "2011",
    note = "{Erratum: JHEP 11 (2011) 109}"
}

@article{Geissbuhler:2013,
    author = "Geissb{\"u}hler, David and Marqu{\'e}s, Diego and N{\'u}{\~n}ez, Carmen and Penas, Victor",
    title = "{Exploring Double Field Theory}",
    eprint = "1304.1472",
    archivePrefix = "arXiv",
    primaryClass = "hep-th",
    doi = "10.1007/JHEP06(2013)101",
    journal = "JHEP",
    volume = "06",
    pages = "101",
    year = "2013"
}

@article{Hohm:2013b,
  author = {O. Hohm and H. Samtleben},
  title = "{Gauge theory of {Kaluza-Klein} and winding modes}",
  journal = {Phys. Rev. D},
  volume = {88},
  pages = {085005},
  year = {2013},
  doi = {10.1103/PhysRevD.88.085005},
  eprint = "1307.0039",
  archivePrefix = "arXiv",
  primaryClass = "hep-th",
}

@article{Jeon:2011c,
    author = "Jeon, Imtak and Lee, Kanghoon and Park, Jeong-Hyuck",
    title = "{Incorporation of fermions into double field theory}",
    eprint = "1109.2035",
    archivePrefix = "arXiv",
    primaryClass = "hep-th",
    doi = "10.1007/JHEP11(2011)025",
    journal = "JHEP",
    volume = "11",
    pages = "025",
    year = "2011"
}

@article{Jeon:2012,
    author = "Jeon, Imtak and Lee, Kanghoon and Park, Jeong-Hyuck",
    title = "{Supersymmetric Double Field Theory: Stringy Reformulation of Supergravity}",
    eprint = "1112.0069",
    archivePrefix = "arXiv",
    primaryClass = "hep-th",
    doi = "10.1103/PhysRevD.85.081501",
    journal = "Phys. Rev. D",
    volume = "85",
    pages = "081501",
    year = "2012",
    note = "{Erratum: Phys. Rev. D 86 (2012) 089903}"
}

@book{Green:1987,
    author = "Green, Michael B. and Schwarz, John H. and Witten, Edward",
    title = "{Superstring Theory}",
    series = "Cambridge Monographs on Mathematical Physics",
    publisher = "Cambridge University Press",
    address = "Cambridge",
    year = "1987",
    note = "{Volumes 1 and 2}"
}

@article{Lescano:2021,
    author = "Lescano, Eric and Mayo, Mart{\'i}n",
    title = "{Gauged Double Field Theory as an $L_{\infty}$ algebra}",
    eprint = "2103.07361",
    archivePrefix = "arXiv",
    primaryClass = "hep-th",
    doi = "10.1007/JHEP06(2021)058",
    journal = "JHEP",
    volume = "06",
    pages = "058",
    year = "2021"
}

@article{Hohm:2017pnh,
    author = "Hohm, Olaf and Zwiebach, Barton",
    title = "{$L_{\infty}$ Algebras and Field Theory}",
    eprint = "1701.08824",
    archivePrefix = "arXiv",
    primaryClass = "hep-th",
    doi = "10.1002/prop.201700014",
    journal = "Fortsch. Phys.",
    volume = "65",
    number = "3-4",
    pages = "1700014",
    year = "2017"
}

@article{Lescano:2021b,
    author = "Lescano, Eric and N{\'u}{\~n}ez, Carmen A. and Rodr{\'i}guez, Jes{\'u}s A.",
    title = "{Supersymmetry, T-duality and heterotic $\alpha'$-corrections}",
    eprint = "2104.09545",
    archivePrefix = "arXiv",
    primaryClass = "hep-th",
    doi = "10.1007/JHEP07(2021)092",
    journal = "JHEP",
    volume = "07",
    pages = "092",
    year = "2021"
}

@article{Eloy:2020,
    author = "Eloy, Camille and Hohm, Olaf and Samtleben, Henning",
    title = "{Duality Invariance and Higher Derivatives}",
    eprint = "2004.13140",
    archivePrefix = "arXiv",
    primaryClass = "hep-th",
    doi = "10.1103/PhysRevD.101.126018",
    journal = "Phys. Rev. D",
    volume = "101",
    number = "12",
    pages = "126018",
    year = "2020"
}

@article{Hronek:2022,
    author = "Hronek, Stanislav and Wulff, Linus and Zacar{\'i}as, Salom{\'o}n",
    title = "{The $\alpha'^2$ correction from double field theory}",
    eprint = "2206.10640",
    archivePrefix = "arXiv",
    primaryClass = "hep-th",
    doi = "10.1007/JHEP11(2022)090",
    journal = "JHEP",
    volume = "11",
    pages = "090",
    year = "2022"
}

@article{Grana:2012,
    author = "Gra{\~n}a, Mariana and Marqu{\'e}s, Diego",
    title = "{Gauged Double Field Theory}",
    eprint = "1201.2924",
    archivePrefix = "arXiv",
    primaryClass = "hep-th",
    doi = "10.1007/JHEP04(2012)020",
    journal = "JHEP",
    volume = "04",
    pages = "020",
    year = "2012"
}

@article{Hatsuda:2023,
    author = "Hatsuda, Machiko and Mori, Haruka and Sasaki, Shin and Yata, Masaya",
    title = "{Gauged Double Field Theory, Current Algebras and Heterotic Sigma Models}",
    eprint = "2212.06476",
    archivePrefix = "arXiv",
    primaryClass = "hep-th",
    doi = "10.1007/JHEP05(2023)220",
    journal = "JHEP",
    volume = "05",
    pages = "220",
    year = "2023"
}

@article{Hassler:2024curv,
    author = "Hassler, Falk and Osten, David and Sakatani, Yuho",
    title = "{Duality covariant curvatures for the heterotic string}",
    eprint = "2412.17893",
    archivePrefix = "arXiv",
    primaryClass = "hep-th",
    doi = "10.1007/JHEP09(2025)031",
    journal = "JHEP",
    volume = "09",
    pages = "031",
    year = "2025"
}

@article{Mori:2024,
    author = "Mori, Haruka and Sasaki, Shin",
    title = "{Extended Doubled Structures of Algebroids for Gauged Double Field Theory}",
    eprint = "2402.03895",
    archivePrefix = "arXiv",
    primaryClass = "hep-th",
    doi = "10.1007/JHEP06(2024)096",
    journal = "JHEP",
    volume = "06",
    pages = "096",
    year = "2024"
}

@article{Baron:2018,
    author = "Baron, Walter H. and Lescano, Eric and Marqu{\'e}s, Diego",
    title = "{The generalized Bergshoeff-de Roo identification}",
    eprint = "1810.01427",
    archivePrefix = "arXiv",
    primaryClass = "hep-th",
    doi = "10.1007/JHEP11(2018)160",
    journal = "JHEP",
    volume = "11",
    pages = "160",
    year = "2018"
}

@article{Coimbra:2014qaa,
  author        = {Coimbra, Andr{\'e} and Minasian, Ruben and
                   Triendl, Hagen and Waldram, Daniel},
  title         = {Generalised geometry for string corrections},
  journal       = {JHEP},
  volume        = {11},
  pages         = {160},
  year          = {2014},
  doi           = {10.1007/JHEP11(2014)160},
  eprint        = {1407.7542},
  archivePrefix = {arXiv},
  primaryClass  = {hep-th}
}

@article{Aldazabal:2013TensorHierarchy,
  author = {Aldazabal, G. and Gra{\~n}a, M. and
            Marqu{\'e}s, D. and Rosabal, J. A.},
  title = {{The gauge structure of Exceptional Field Theories
            and the tensor hierarchy}},
  journal = {JHEP},
  volume = {04},
  pages = {049},
  year = {2014},
  doi = {10.1007/JHEP04(2014)049},
  eprint = {1312.4549},
  archivePrefix = {arXiv},
  primaryClass = {hep-th}
}

@article{HsiaKamalWulff2025,
  author        = {Hsia, Steven Weilong and Kamal, Ahmed Rakin and Wulff, Linus},
  title         = {No manifest {T} duality at order $\alpha'^3$},
  journal       = {Phys. Rev. D},
  volume        = {111},
  number        = {6},
  pages         = {L061904},
  year          = {2025},
  doi           = {10.1103/PhysRevD.111.L061904},
  eprint        = {2411.15302},
  archivePrefix = {arXiv},
  primaryClass  = {hep-th}
}

@article{Ciafardini:2024ujx,
  author        = {Ciafardini, Marco and Marqu{\'e}s, Diego
                   and N{\'u}{\~n}ez, Carmen A.
                   and Grau, Agustina Pereyra},
  title         = {Hidden symmetries from extra dimensions},
  journal       = {JHEP},
  volume        = {02},
  pages         = {072},
  year          = {2025},
  doi           = {10.1007/JHEP02(2025)072},
  eprint        = {2410.07325},
  archivePrefix = {arXiv},
  primaryClass  = {hep-th}
}

\end{document}